\documentclass[12pt,a4paper]{article}
\pdfoutput=1
\usepackage{graphicx}
\usepackage[height=8.85in, width=6.45in]{geometry}
\usepackage{lipsum}
\usepackage{tikz}
\usepackage{tikz-cd}
\usepackage{braids}
\usepackage{amsmath,amsfonts,amssymb}
\usepackage{slashed,stmaryrd}
\usepackage{enumitem}
\allowdisplaybreaks

\usepackage{amsthm}
\usepackage{youngtab}
\usepackage{jheppub}

\theoremstyle{definition}

\def\Z{\mathbb{Z}}
\def\mod{\text{mod~}}
\graphicspath{ {./Figs/} }
\usepackage{xspace}
\usepackage{amscd,amsxtra,mathrsfs,graphics,graphicx,amsthm,epsfig,bm,longtable,float,color,tikz,mathtools,xfrac,footnote,rotating,lscape,wrapfig}
\usepackage{simpler-wick}
\usepackage[makeroom]{cancel}
\usepackage[utf8]{inputenc}
\usepackage[T1]{fontenc}
\usepackage{bm}
\usepackage{physics}
\usepackage{caption}
\usepackage{subcaption,booktabs}
\usetikzlibrary{arrows}
\usepackage{empheq}
\usepackage{array}   % for \newcolumntype macro
\newcolumntype{C}{>{$}c<{$}} % math-mode version of "l" column type

\def\nn{\nonumber}
\def\SO{\text{SO}}

\def\SU{\text{SU}}

\def\U{\text{U}(1)}

\def\Spin{\text{Spin}}
\def\Spinc{\text{Spin}_c}

\def\UU{\text{U}}

\def\Tr{\text{~Tr~}}

\def\ee{\text{e}}
\def\dd{\text{d}}

\def\R{\mathbb{R}}

\def\CP{\mathbb{CP}}

\newcommand{\beq}{\begin{equation}}
\newcommand{\eeq}{\end{equation}}

\newcommand{\avg}[1]{\left\langle#1\right\rangle}

\title{Remarks on QCD$_4$ with fundamental and adjoint matter}
\date{}
\author{Mohamed M. Anber, Nakarin Lohitsiri, Tin Sulejmanpasic}
\affiliation{Department of Mathematical Sciences, Durham University}
\emailAdd{mohamed.anber@durham.ac.uk}
\emailAdd{nakarin.lohitsiri@durham.ac.uk}
\emailAdd{tin.sulejmanpasic@durham.ac.uk}

\abstract{We study 4-dimensional $\SU(N)$ gauge theory with one adjoint Weyl fermion
  and fundamental matter -- either bosonic or fermionic. Symmetries, their 't Hooft anomalies, and the Vafa-Witten-Weingarten theorems strongly constrain the possible bulk phases. The first part of the paper is dedicated to a single fundamental fermion or boson. As long as the adjoint Weyl fermion is massless, this theory always possesses a $\Z_{2N}^\chi$ chiral symmetry, which breaks spontaneously, supporting $N$ vacua and domain walls between them for a generic mass of the matter fields. We argue, however, that the domain walls generically undergo a phase transition, and we establish the corresponding 3d gauge theories on the walls. The rest of the paper is dedicated to studying the multi-flavor fundamental matter. Here, the phases crucially depend on the ratio of the number of colors and the number of fundamental flavors. We also discuss the limiting scenarios of heavy adjoint and fundamentals, which align neatly with our current understanding of QCD and ${\cal N}=1$ super Yang-Mills theory.
}

\begin{document}
\maketitle

\section{Introduction}
\label{sec:intro}

Quantum field theories (QFTs), fundamental or emergent, are important in modern physics. They are the cornerstone of the Standard Model of particles. In statistical mechanics, they are an efficient way to capture universal features of phase transitions. At the same time, in condensed matter, they are useful to parametrize the important, low-energy features of various materials. 

Non-abelian gauge theories are the most infamous of QFTs. They describe Quantum Chromodynamics (QCD). Quarks, labeled by 3 colors, interact with non-abelian gauge fields to produce a theory that makes sense as a genuine quantum field theory (i.e., has a continuum limit) and, in addition, has highly nontrivial features in the infrared, such as chiral symmetry breaking and confinement which have yet to have a complete theoretical understanding.

This is why understanding QCD-like theories has a special place in theoretical physics. Most recently, a popular method of analyzing such theories was pioneered by the discovery of novel generalized global symmetries and 't Hooft anomalies \cite{Kapustin:2014gua,Gaiotto:2014kfa,Gaiotto:2017yup}, leading to an avalanche of discoveries (see, e.g., \cite{Gaiotto:2017tne,Komargodski:2017dmc,Komargodski:2017smk,Komargodski:2017keh,Sulejmanpasic:2018upi,Tanizaki:2018xto,Tanizaki:2018wtg,Karasik:2019bxn,Anber:2018iof,Anber:2018jdf,Anber:2019nfu,Anber:2021lzb,Sulejmanpasic:2020zfs,Smith:2021vbf} for an incomplete list). In this paper, we will be concerned with one such theory, namely the $\SU(N)$ gauge theory with either fermionic or bosonic quarks (i.e., matter in the fundamental representation), supplemented by a single Weyl fermion in the adjoint representation of the $\SU(N)$ gauge group. In some sense, the theory in question is halfway between QCD (the version of our theory without the adjoint Weyl fermion) and ${\cal N}=1$ Super Yang-Mills theory (SYM) (our theory without fundamental matter). 

We will mostly be concerned with a massless adjoint Weyl fermion so that there is always a $\mathbb Z_{2N}^\chi$ discrete chiral symmetry\footnote{We use the superscript $\chi$ in $\mathbb Z_{2N}^{\chi}$ to distinguish the discrete chiral symmetry from other discrete groups that appear in the text. As we shall discuss, if the fundamental matter is fermionic and massless, the discrete chiral symmetry enhances to $\U_\chi$ chiral symmetry.}. The fundamental matter (bosonic or fermionic) can be massless or massive. Let us send the mass of the fundamental matter to infinity. The theory becomes super Yang-Mills, for which many things are known. The theory breaks its chiral $\mathbb Z_{2N}^\chi$ symmetry down to $\mathbb Z_2$ fermion number symmetry, leading to $N$ degenerate supersymmetric vacua separated by domain walls that support a topological quantum field theory (TQFT). In this paper, we use the existence of a novel anomaly involving the baryon symmetry and the methods of effective field theory to show that such a phase persists in the bulk at any finite fundamental mass. 

In particular, consider the simplest case of one adjoint Weyl fermion and one massive fundamental Dirac fermion. The global symmetries are only the $\mathbb Z_{2N}^\chi$ chiral symmetry, acting on the adjoint Weyl fermions, and the $\U_B$ baryon symmetry, acting on the fundamental fermion. We will show that there is a mixed anomaly between the two symmetries for any mass of the fundamentals, indicating that either $\U_B$ or $\mathbb Z_{2N}^\chi$ is spontaneously broken\footnote{The anomaly manifests itself as follows. An insertion of the minimal $\U_B$ flux will activate the color `t Hooft fluxes. This happens because $\U_B$ is the quotient of the $\U_q$ quark symmetry by the center of the gauge group $\mathbb Z_N$, i.e., $\U_B\cong \U_q/\mathbb Z_N$. Consistency of the cocycle condition in the presence of the $\U_B$ background requires we also activate the 't Hooft flux of the center of $SU(N)$. When this happens, however, the instanton number becomes fractionally quantized, and $\mathbb Z_{2N}^\chi$  is explicitly broken to $\mathbb Z_{2}$ by the presence of the $\U_B$ flux. This is similar to \cite{Gaiotto:2017tne,Anber:2019nze}.}. The Euclidean path integral measure of this system is positive definite, and one can apply the Vafa-Witten theorem \cite{Vafa:1983tf} to conclude that $\U_B$ cannot be spontaneously broken. Hence, anomaly matching conditions demand that $\mathbb Z_{2N}^\chi$ be spontaneously broken. In addition, $\mathbb Z_{2N}^{\chi}$ has a mixed anomaly with gravity; thus, even if the Vafa-Witten theorem did not hold (e.g., Vafa-Witten is applied to a fermionic fundamental matter, but as we will see, the same anomalies apply to bosonic fundamental matter) the breaking of $\U_B$ would not saturate all the anomalies. Therefore, introducing massive fundamental fermions does not change the bulk phase of SYM: the theory still has $N$ vacua connected via domain walls. While there is no phase transition in bulk as we vary the mass of the fundamentals, we shall argue that phase transitions will occur on the domain walls. At $m=0$, the $\mathbb Z_{2N}^\chi$ symmetry enhances to $\U_\chi$, and the domain walls melt away, leaving a Goldstone boson for the spontaneously broken $\U_\chi$. 

If we replace a fundamental fermion with a fundamental scalar, a similar conclusion can be drawn\footnote{There is a slight caveat to this statement.}: no phase transition in the bulk takes place as the fundamental mass squared is driven from large and positive (where the boson decouples) to large and negative (where the boson condenses).  Yet, a transition on the domain wall is expected to occur.

We also discuss theories with more fundamental flavors and establish analogous anomalies. For $N_f$ fundamental fermions, matching the anomalies in the IR happen via one of two channels depending on the number of flavors: either the theory breaks its symmetries spontaneously, or it flows to a conform field theory (CFT). These scenarios are summarized in Figures \ref{fig:one-ferm-phase} and \ref{fig:mult-ferm-phases} for the single flavor and multi flavors, respectively. 

We briefly give an incomplete review of the literature on the mixed representation QCD. Of course,  4d super QCD is the most well-known mixed-representation QCD-like theory. Thanks to holomorphy, a lot is known about the IR phases of these theories, which, by now, is textbook material. The recent work \cite{Bashmakov:2018ghn} is relevant to our studies, which analyzed domain walls in super QCD. The mixed fundamental/adjoint representation was also analyzed in the context of the weakly coupled $R^3\times S^1$ and adiabatic continuity \cite{Poppitz:2009tw}. In \cite{Anber:2017pak}, theories with adjoint and higher mixed-representations were studied on $R^3\times S^1$, and in  \cite{Anber:2019nfu}, $SU(6)$ with fermions in the adjoint and the $3$-index antisymmetric mixed representations were studied in 4d and on $R^3\times S^1$.   In \cite{DeGrand:2016pgq}, a systematic analysis of chiral perturbation theory was done for arbitrarily mixed representations, while in \cite{Ayyar:2017qdf,Cossu:2019hse} lattice simulations with fermion fundamental and anti-symmetric sextet representation of $\SU(4)$ gauge theory was done. More relevant for our work is \cite{Bergner:2020mwl}, where the mixed representation of fundamental/adjoint fermions was simulated.

%%%%%%%%%%%%%%%%%%%%%%%%%%%%%%%%%%%%%%%%%%%%%%%%%%
\subsection{A warmup: Higgs phase of $\SU(2)$ gauge theory with a single scalar}
\label{A warmup: Higgs phase of SU(2) gauge theory with a single scalar}\label{sec:warmup}
%%%%%%%%%%%%%%%%%%%%%%%%%%%%%%%%%%%%%%%%%%%%%%%%%

The various anomalies involving the chiral symmetry $\Z_{2N}^\chi$, like the mixed $\Z_{2N}^\chi$--gravitational anomaly, or the pure $\Z_{2N}^{\chi}$ anomaly, are purely due to the adjoint fermions. But there exists an anomaly involving $\Z_{2N}^\chi$ and $\U_B$ -- the baryon symmetry carried by the fundamental matter. Such an anomaly exists regardless of whether the fundamental matter is bosonic or fermionic. This seems surprising at first, as even if the fundamental matter is fermionic, there are no triangles involving $\U_B$ and the chiral symmetry. When the fundamental matter is bosonic, the anomaly seems even stranger still. We will illustrate this anomaly by a simple example of an $\SU(2)$ gauge theory with one fundamental scalar $\Phi$ and one adjoint Weyl fermion $\lambda$. The theory has a $\mathbb Z^{\chi}_4$ discrete chiral symmetry, whose $\mathbb Z_2$ subgroup is the fermion number. It acts on $\lambda$ as
\beq
\lambda\rightarrow i\lambda\;.
\eeq
There is also a $\U$ symmetry acting on the scalar $\phi\rightarrow e^{i\alpha}\phi$. The $\mathbb Z_2$ subgroup of $\U$ is, however, the center of $\SU(2)$ gauge group, so one can view the global symmetry as $\U/\mathbb Z_2\cong \U$. We distinguish between two normalizations. First, we define $\U_B$ as the baryon symmetry, i.e., a symmetry under which the smallest charge of the baryon is unity. This symmetry group is related to the quark symmetry $\U_q$, under which the quark has the unit charge, as follows $\U_B=\U_q/\mathbb Z_2$ (or for general $\SU(N)$ as $\U_B=\U_q/\mathbb Z_N$). The scalar then transforms as $\phi\rightarrow e^{i\alpha/2}\phi$ under $\U_B$, where now $\alpha\sim \alpha+2\pi$. 

If we condense the scalars, they fully Higgs the $\SU(2)$ gauge group, leaving only three ungapped free adjoint Weyl fermions associated with the algebra of $\SU(2)$. We can write the Weyl fermion as $\lambda=\lambda^a\frac{\tau^a}{2}$, where summation over $a=1,2,3$ is implied and where $\tau^a$ are the Pauli matrices. The fermions $\lambda^a$ are, however, not gauge invariant, and the correct gauge invariant operators corresponding to the three Weyl fermions are
\begin{align}\label{eq:xi123}
\nonumber
\xi_1=&\phi^*_{A} {\lambda^A}_B \phi^B\;,\\
\xi_2=&\phi^{A}\epsilon_{AB}{\lambda^B}_{C}\phi^C\;,\\
\xi_3=&{\phi^*}_{A}{\lambda^{A}}_B\epsilon^{BC}\phi_C^*\;,
\nonumber
\end{align}
where we explicitly wrote out fundamental indices $A,B,C=1,2$ of $\SU(2)$ ($\epsilon_{AB}$ being the totally anti-symmetric tensor). We enumerate the charges of the three Weyl fermions under the $\mathbb Z_4$ chiral symmetry and $\U_B$ symmetry\footnote{\label{enhanced symmetry}The model enjoys an enhanced flavor symmetry, called custodial symmetry, because of the pseudo-reality of the gauge group. The true flavor symmetry is actually $\SO(3)$ with $\U_B$ as its subgroup. Since we have a triplet of Weyl fermions, two of which are charged as $\pm1$ under the $\U_B$, the three naturally fit into the triplet of $\SO(3)$ flavor group.} in Table~\ref{tab:xi123}.

% Requires the booktabs if the memoir class is not being used
\begin{table}[htbp]
   \centering
   %\topcaption{Table captions are better up top} % requires the topcapt package
   \begin{tabular}{ccc} % Column formatting, @{} suppresses leading/trailing space
 %     \toprule
         	    	& $\mathbb Z_4^{\chi}$ 	& $\U_B$\\
      \midrule
      $\xi_1$    	& 1 				& 0 \\
      $\xi_2$    	& 1  				& 1 \\
      $\xi_3$      & 1	  			& -1 \\
      \bottomrule
   \end{tabular}
   \caption{The charges of the three fermions \eqref{eq:xi123} of the IR phase.}
   \label{tab:xi123}
\end{table}

The effective theory is hence a theory of three Weyl fermions $\xi_{1,2,3}$ with a given charge assignment under the global symmetries given in Table~\ref{tab:xi123}. It is now straightforward to see that there is a standard triangle $\U_B-\mathbb Z_4^\chi$ anomaly. Here we found an anomaly using the IR theory\footnote{Note however that the $\mathbb Z_4^{\chi}$ enhances to $\U_\chi$ symmetry because all mass terms involving $\xi_{1,2,3}$ either preserve all of $\U_\chi$ or keep only $\mathbb Z_2\subset\mathbb Z_4^{\chi}$. }, but since the anomaly is RG invariant, it implies the same anomaly exists in the UV theory. Later, we will see more formally how this occurs in a more general setting.

At first, this anomaly sounds odd, as the baryon-number carrying fundamental scalar $\Phi$ naively has nothing to do with the chirally charged $\lambda$.  Further, it generalizes to any fundamental matter charged under the baryon number in the presence of a massless adjoint Weyl fermion. In addition the scalar charged under the baryon number can be made arbitrarily massive, which goes against the traditional lore that such fields can participate in the anomaly. However in recent years there are many examples of this type. In \cite{Komargodski:2017dmc,Komargodski:2017smk} a mixed anomaly between $\U$-topological symmetry and $\text{SO}(3)$ (or more generally $\text{PSU}(N)$) flavor symmetry was utilized in Abelian-Higgs models. The scalars can be massive, and the anomaly persists. In \cite{Gaiotto:2017tne} an anomaly between $T$ and vector-like flavor symmetries was also used in QCD in 4d, again for any mass of the flavor multiplet. In all these cases (including the one we discuss here) the decoupling limit results in a theory with a 1-form symmetry, while the anomaly involving flavor transmutes to an anomaly involving the 1-form symmetry.  

In our theory, a way to understand this anomaly from the UV point of view is to start with a massless adjoint fermion and a fundamental scalar coupled to the $\SU(N)$ gauge fields. Let the scalar have a very large positive mass. Then, the theory has a mixed anomaly between the discrete $\Z_{2N}^\chi$ chiral symmetry and an emergent  $\Z_N^{[1]}$ $1$-form center symmetry (the $1$-form symmetry becomes exact in the infinite mass limit) \cite{Gaiotto:2014kfa} (see also \cite{Gaiotto:2017yup}). Now, lower the mass of the scalar or even take it to be negative. The presence of the scalar breaks the $ 1$-form symmetry explicitly, but it introduces a $\U_B$ 0-form symmetry. As we will see the anomaly involving the $\Z_N^{[1]}$ 1-form symmetry transmutes into the anomaly involving the $\U_B$ baryon symmetry. Roughly speaking the background gauge fields for both the $\U_B$ as well as the 1-form $\Z_N^{[1]}$ symmetry activates 't Hooft fluxes, which causes the anomaly to manifest itself. We shall see the details of this anomaly in the bulk of the paper using different methods.

\begin{table}[htbp]
   \centering
   %\topcaption{Table captions are better up top} % requires the topcapt package
   \begin{tabular}{ccc} % Column formatting, @{} suppresses leading/trailing space
 %     \toprule
         	    	& $\U_\chi/\Z_4^\chi$ 	& $\U_B$\\
      \midrule
      $\lambda$		& $1$				&$0$\\
      $\phi$ 			& $-1$				& $\frac{1}{2}$\\
      $\psi$    		& $-2$				& $\frac{1}{2}$ \\
      $\tilde\psi$    	& $-2$  				& $-\frac{1}{2}$ \\
      \bottomrule
   \end{tabular}\qquad\qquad \begin{tabular}{ccc} % Column formatting, @{} suppresses leading/trailing space
 %     \toprule
         	    	& $\U_\chi/\Z_4^\chi$ 	& $\U_B$\\
      \midrule
      $\chi_0$     	 	& $-1$	  			& $1$ \\
      $\chi_1$ 		& $-1$				& $-1$	\\
      $\chi_2 $  		& $-3$				& $0$	\\
      \bottomrule
   \end{tabular}
   \caption{Left: The charges of the fundamental fields in the fermion-Higgs $\SU(2)$ model. Right: The charges of the composite fermions in the Higgs phase.}
   \label{tab:chi012}
\end{table}

To see that the anomaly is there also for the theory with fermions, we can introduce a fundamental Dirac fermion $\Psi$ consisting of two Weyl fermions $\psi$ and $\tilde{\psi}$ in the fundamental and the anti-fundamental representation respectively. Further we postulate a Yukawa coupling
\begin{equation}
\mathcal{L}_{\text{Yukawa}}= \phi^\dagger\lambda \psi+c.c.\;.
\end{equation}
If we add a mass to the scalars, we decouple them and the model reduces to an $\SU(2)$ gauge theory with a single fundamental Dirac fermion. On the other hand condensing the scalar and Higgsing the theory will reveal a composite fermion phase with the required mixed $\Z_4^\chi$-$\U_B$ anomaly, as we will see in a moment. Since the symmetries in these two limits are unchanged, the anomaly structure has to be the same.

We can also add a mass term $m\tilde\psi\psi+c.c.$. The symmetries of the model involve the $\U_B$ symmetry and, if $m=0$, the $\U_\chi$ chiral symmetry. When the mass $m\ne0$ the chiral symmetry reduces back to $\Z_4^\chi$. Table \ref{tab:chi012} on the left summarizes the charges of the fundamental fields under these symmetries\footnote{Note that since there is no Yukawa coupling for $\tilde\psi$, it seems that there is an additional baryon symmetry rotating $\tilde\psi$ only. This symmetry however is anomalous. To make it anomaly-free we must either rotate the adjoint $\lambda$ by an appropriate phase, or $\psi$ by an appropriate phase or a combination of both. It is easy to see that this results in two $\U$ symmetries, which we choose to label as $\U_B$ and $\U_\chi$ with fields charged as in Table \ref{tab:chi012}.  }. 

Now let us condense the scalar $\phi$ like before, higgsing the gauge fields entirely. In the gauge where $\phi=\begin{pmatrix}
v\\
0
\end{pmatrix}$, the fermions $\psi^A$ and $v{\lambda^1}_A+m\tilde\psi_A$, while fermions ${\lambda^2}_1$ and $m{\lambda^1}_A-v\tilde \psi_{A}$ remain massless, 3 fermions in total. The massless fermions can be written in a gauge invariant form as
\begin{align}\nonumber
\chi_0=&\phi^A\epsilon_{AB}{\lambda^{B}}_C\phi^C\\
\chi_1=&m\phi^*_A{\lambda^A}_C\epsilon^{CD}\phi^*_D-|\phi|^2\tilde\psi_A\epsilon^{AB}\phi^*_B\\
\nonumber
\chi_2=&m\phi^*_A{\lambda^A}_B\phi^B-|\phi|^2\tilde\psi_A\phi^A
\end{align}
We summarize the charges under the global symmetries of these fields in the right of Table \ref{tab:chi012}. As we will see these composite fermions precisely correspond to the proposal. Further we can send $m\to\infty$, and then we restore the single Higgs model where $\eta_{0,1,2}\to \xi_{2,3,1}$\footnote{Note that in the model without the (anti-)fundamental fermions we defined the scalar field not to transform under the chiral symmetry $\U_\chi$, so the identification of charges of Table \ref{tab:xi123} and \ref{tab:chi012} should be made up to this redefinition}. We will have more to say about this model in Sec.~\ref{sec:one-ferm-IR-phase}.

Finally let us briefly add that we could also see the anomaly in $\SU(N)$ gauge theory with fermionic flavors. One would then couple $N-1$ fundamental Higgs fields, with appropriate Yukawas designed to preserve the global symmetries of the theory, and condense them in such a way to Higgs the gauge fields completely. The resulting phase is always a phase of free fermions, and the anomalies become manifest. We will not pursue this in details

\subsection{Outline}
\label{sec:outline}

The rest of the paper is organized as follows. In \S~\ref{sec:one-Dirac-SUN-symmetry}, we study in detail the theory with a single fundamental Dirac fermion in the presence of a massless adjoint Weyl. We identify the faithful global symmetry and the associated 't Hooft anomalies constraining the IR phase and its domain wall theory. In  \S~\ref{sec:one-fund-scal}, we repeat the analysis with a single fundamental scalar that replaces the single fundamental fermion. We generalize the story in \S~\ref{sec:multi-flavor-fermion} to several fundamental Dirac fermions and a single adjoint Weyl fermion and elaborate on their anomalies. We generalize the analysis to multi-fundamental scalars in \S~\ref{sec:mult-scal-flav}. We conclude and outline some possible future directions in \S~\ref{sec:conclusions}. The appendices summarize some points used in different parts of the paper. In Appendix \ref{app:measure-positive-definite}, we show that the fermionic measure is positive definite. In Appendix \ref{app:spectral}, we review the relation between spectral flow and the index theorem. Finally, the $3$-loop $\beta$-function used in some of our analyses in \S~\ref{sec:multi-flavor-fermion} is displayed in Appendix \ref{beta function section}.

%%%%%%%%%%%%%%%%%%%%%%%%%%%%%%%%%%%%%%%%%%%%%%%%%%%%%%
\section{Theory with one fundamental fermion: symmetries, anomalies, and the phase diagram}
\label{sec:one-Dirac-SUN-symmetry}
%%%%%%%%%%%%%%%%%%%%%%%%%%%%%%%%%%%%%%%%%%%%%%%%%%%%%%

In this section, we study the symmetries and anomalies of  4d $\SU(N)$
gauge theory coupled to one Dirac fermion in the fundamental
representation and one Majorana fermion in the adjoint
representation. Note that we can alternatively view the Majorana
fermion in the adjoint representation as a Weyl fermion, but not both
simultaneously. In the rest of the paper, we will take the adjoint
fermion to be Weyl for convenience of the analysis. 

\subsection{The massive case}
\label{sec:one-Dirac-massive}

Denote the $\SU(N)$ gauge field that couples to the fundamental
fermions by $a$ and its counterpart for the adjoint fermions by
$a_{adj}$. Then the action of the fermionic sector is given by
\begin{equation}
  S = \int \dd^4x\, \left(i \bar{\lambda}(\slashed{\partial} -i \slashed{a}_{adj})\lambda + i \bar{\Psi} (\slashed{\partial}-i \slashed{a} -m)\Psi\right)\,.
\end{equation}
Here, $\lambda$ is a left-handed Weyl fermion in the adjoint
representation of $\SU(N)$. $\Psi$ is a Dirac fermion in the fundamental
representation of $\SU(N)$. In the chiral basis, we decompose $\Psi$
as
\begin{equation*}
  \Psi = \begin{pmatrix} \psi \\ \tilde{\psi}^{\dagger}\end{pmatrix}\,,
  \end{equation*}
  where we use the notation that both $\psi$ and $\tilde{\psi}$ are left-handed Weyl fermions in the fundamental and anti-fundamental representations of the gauge group.
 
The theory has a classical $\U_\chi$ chiral symmetry acting on the adjoint:
\begin{equation}
\lambda\rightarrow e^{i \alpha}\lambda\;,
\end{equation}
which reduces to $\mathbb Z_{2N}^{\chi}$ discrete chiral symmetry by the ABJ anomaly. Further, there is
a $\U$ symmetry acting on the fundamental quark:
\begin{equation}
\Psi\rightarrow e^{i \beta}\Psi\;.
\end{equation}
We will call this symmetry $\U_q$, where $q$ stands for \emph{quark}.
Note that if $\beta \in \frac{2\pi }{N}\Z$, we can absorb the symmetry in the
$\SU(N)$ gauge transformation so that the true global symmetry group is
$\U_B\cong\U_q/\Z_N$ -- the Baryon number symmetry. Under $\U_B$, the smallest charge of the baryon is unity. Thus, the faithful global symmetry is\footnote{We can also use the cocycle conditions as a systematic way to find the faithful symmetries. See \S~\ref{sec:multi-flavor-fermion}.}
\begin{equation}
G^{\text{Global}}= \frac{\U_q}{\mathbb Z_N}\times \Z_{2N}^{\chi}\;.
\end{equation}
The fermion content is summarised in terms of their representations
under various symmetries in Table \ref{tab:massive-fermion-reps}
below.
\begin{table}[h]
  \centering
  \begin{tabular}{c||c|cc}
    & $\SU(N)$ & $\U_q$ & $\Z_{2N}^{\chi}$\\
    \hline
    $\psi$ & ${\tiny\yng(1)}$ & $+1$ & $0$ \\
    $\tilde{\psi}$ & ${\tiny \overline{\yng(1)}}$ & $-1$ & $0$ \\
    $\lambda$ & $\textbf{adj}$ & $0$ & $+1$
  \end{tabular}
  \caption{Representations of the fermions under various symmetry groups in the massive case}
  \label{tab:massive-fermion-reps}
\end{table}

Finally, the action of the $\Z_2$ subgroup of the symmetry $\U_B\times \Z_{2N}^{\chi}$:
\begin{equation}
\mathbb Z_2: (\lambda,\Psi)\rightarrow (-\lambda,-\Psi)\;,
\end{equation}
coincides with the action of the $\mathbb Z_2^F$ fermion number. Thus,
there is a mixing between the $\Spin$ spacetime symmetry and the $\Z_{2N}^{\chi}$ into a $\Spin$-$\Z_{2N}^{\chi}$
$:= (\Spin \times \Z_{2N}^{\chi})/\Z_2^F$ structure. We can then use this structure to
define the theory on some orientable manifolds that are non-spin by
turning on a non-trivial $\Z_N\cong \Z_{2N}^{\chi}/\Z_2^F$ bundle on such a
manifold whose obstruction to lifting to a $\Z_{2N}^{\chi}$ bundle is
precisely the second Stiefel--Whitney class of the
manifold\footnote{Note that this only works for orientable manifolds
  $M$ with non-vanishing $H^1(M;\Z_N)$ because otherwise, there can be
  no non-trivial $\Z_N$ bundle in the first place.}.

To study the anomalies, let us couple a background gauge field
$\mathcal{A}_q$ to the $\U_q$ symmetry by promoting
\begin{equation}
  \slashed{\partial}-i \slashed{a}\rightarrow \slashed{\partial}-i \slashed{a}-i \slashed{\mathcal{A}}_q \textbf{1}_N=\slashed{\partial}-i \slashed{A}\;,\nn
\end{equation}
where $A$ is a $\UU(N)$ gauge field whose traceless part is
dynamical. We write $F$ for the field strength of $A$ and
$\mathcal{F}_q$ for the field strength of $\mathcal{A}_q$. Note that
$\tr F =\tr (\mathcal F_q \textbf{1}_N) =N \mathcal F_q$ is quantized in integer units of $2\pi$. Hence it follows $\mathcal{F}_q$ can be fractionally quantized in units
$2\pi/N$. This is because $\U_q$ is not the proper global symmetry (i.e. there are no gauge invariant operators with the unit charge under $\U_q$) but $\U_B$ is. We will also make use of the properly quantized baryon gauge
field $A_B= N\mathcal{A}_q$, and its curvature $F_B=N\mathcal{F}_q$.

Now, applying the
$\mathbb Z_{2N}^{\chi}: \lambda\rightarrow e^{i\frac{\pi k}{N}}\lambda$ symmetry transformation, the
action changes as
\begin{equation}
\Delta S = i  \frac{k}{4\pi}\int \tr\left[ \left(F-\frac{\tr F}{N}\right)\wedge \left(F-\frac{\tr F}{N}\right)\right]\,,
\end{equation}
where the trace is taken in the fundamental representation. In the above result the traceless part is subtracted, because the adjoint field $\lambda$ cannot see the trace of $F$. Now we write
\begin{multline}\label{eq:Z2N_anomaly}
\frac{k}{4\pi}\int \tr \left[\left(F-\frac{\tr F}{N} \right)\wedge \left(F-\frac{\tr F}{N}\right)\right]=\frac{k}{4\pi}\int \tr F\wedge F -\frac{k}{4\pi N}\int F_B\wedge F_B\;.
\end{multline}
The first term on the RHS is integer-quantized on a spin manifold. So
the partition function changes under the $\Z_{2N}^{\chi}$
transformation as
\begin{align}\label{eq:Z2N_anomaly1}
{\cal Z} \rightarrow {\cal Z} \exp[{\frac{-i k}{4\pi N}\int F_B\wedge F_B}]\;.
\end{align}
Recall that $F_B$ is a properly quantized $\U$ field strength; on a
closed spin manifold, we have
\begin{equation}
\frac{1}{2}\int \frac{F_B}{2\pi} \wedge \frac{F_B}{2\pi} \in \Z\;.
\end{equation}
We can then see that the above phase is nontrivial for
$k=1,2,\dots, N-1$. Thus, there is a mixed anomaly between the
$\Z_{2N}^{\chi}$ discrete chiral symmetry and the $\U_B$ baryon symmetry\footnote{Notice that when $k=N$, the phase is trivial, meaning that
  the fermion number symmetry is not anomalous with the $\U_B$ baryon
  symmetry.}.  As we mentioned in \S~\ref{A warmup: Higgs phase of SU(2) gauge theory with a single scalar}, the reader might not feel at ease about a mixed anomaly between $\mathbb Z^{\chi}_{2N}$ and $\U_B$ since none of the two symmetries couples to the two fermion species simultaneously, i.e., one does not see such an anomaly from triangle diagrams\footnote{However, recall that this is precisely how we observed the anomaly in \S~\ref{sec:warmup}, by going to a Higgs phase and establishing the fermion content in the IR.}. Another equivalent way to obtain the anomaly is to realize that a minimal flux for $A_B$ field induces a 't Hooft flux for the color fields, which makes the color topological charge fractional, $Q_c\in \mathbb Z/N$, and thus leads to a reduction of the chiral symmetry to $\Z_2^\chi$.  Such anomaly mechanism, where putting background fields for a global flavor symmetry forces the instanton number to be fractional, has been observed before in \cite{Gaiotto:2017tne,Shimizu:2017asf,Tanizaki:2018wtg}, or more closely related to our setup, in \cite{Anber:2019nze,Anber:2021iip} where they were dubbed baryon-color-flavor (BCF) and color-flavor-$\U$ (CFU) anomalies. We will discuss this point of view in \S~\ref{sec:multi-flavor-fermion} in more detail.

The theory also exhibits a mixed $\mathbb Z_{2N}^{\chi}$-gravitational anomaly. Under a $\mathbb Z_{2N}^{\chi}$ rotation, the partition function transforms as
\begin{eqnarray}
{\cal Z}\rightarrow {\cal Z} \exp\left[-i \pi \frac{k(N^2-1)}{ N}\int_{ M}\frac{p_1}{24}\right]\,,
\end{eqnarray}
where $p_1\equiv -\frac{1}{8\pi^2}\mbox{tr}R\wedge R$ is the first Pontryagin class of the tangent bundle, $R$ is the curvature $2$-form, and the integral is taken on a closed  4-manifold $ M$. Notice that on a spin manifold $\int_{M}p_1\in 48\mathbb Z$.

Finally, there is a nonperturbative discrete anomaly arising from the symmetry group
$\Z_{2N}^{\chi}$. Recall that, because of the quotient $\Z_2^F$ between the
spin group and the $\Z_{2N}^{\chi}$ symmetry group, we can define a theory
on a non-spin manifold as long as the manifold admits a
$\Spin\text{-}{\Z_{2N}^{\chi}}$ structure. The anomalies in this structure is
classified by the cobordism group
\cite{Hsieh:2018ifc,Delmastro:2022pfo}
\begin{equation}
\mho^6_{\Spin\text{-}{\Z_{2N}^{\chi}}} \cong \text{Hom} \left( \Omega^{\Spin\text{-}{\Z_{2N}^{\chi}}}_5,\U \right) \cong \Z_a \times \Z_b\;,
\end{equation}
where
\begin{equation}
  \begin{split}
    a &=
    \begin{cases}
      24 N, &\quad N = 0 ~\mod 6,\\
      8 N, &\quad N = 0 ~\mod 2 \;\text{and}\; N \ne 0 ~\mod 3,\\
      3 N, &\quad N = 0 ~\mod 3 \; \text{and}\; N \ne 0 ~\mod 2,\\
      N, &\quad \text{else}
    \end{cases},\\
    b &=
    \begin{cases}
      N/6, &\quad N = 0 ~\mod 6,\\
      N/2, &\quad N = 0 ~\mod 2\; \text{and}\; N \ne 0 ~\mod 3,\\
      N/3, &\quad N = 0 ~\mod 3\; \text{and}\; N \ne 0 ~\mod 2,\\
      N, &\quad \text{else}
    \end{cases}\;.
  \end{split}
\end{equation}
For a Weyl fermion of charge $q ~\mod 2N$ under $\Z_{2N}^{\chi}$, the anomaly
is given by a pair of indices
$\left( \nu_a, \nu_b \right)\in \Z_a \times \Z_b$ where $\nu_a$ and
$\nu_b$ are explicitly given by \cite{Hsieh:2018ifc,Davighi:2020uab}
\footnote{We thank Joe Davighi for working out the general expression
  for $\nu_b$ with one of the authors.}
\begin{equation}
  \begin{split}
    \nu_a &=  \frac{a}{48N} \left( \left( 2N^2+N+1 \right)q^3-(N+3)q \right) ~\mod a,\\
    \nu_b &=
    \begin{cases}
      \frac{b}{4N} \left( \left( N+1 \right)\left( 2N+1 \right)q^3-(N+1)q \right) ~\mod b , &\quad N = 0 ~\mod 6, N/6 \in 2\Z,\\
      \frac{b}{4N}\left( \left( 10N^2+3N+5 \right)q^3-\left( 5N+17 \right)q \right) ~\mod b, &\quad N = 0 ~\mod 6, N/6\in 2\Z+1,\\
      \frac{b}{4N} \left( \left( 2N^2-N+1 \right)q^3-(N+5)q \right), &\quad N = 2 ~\mod 4, N \ne 0 ~\mod 3,\\
      \frac{b}{4N} \left( \left( N+1 \right)\left( 2N+1 \right)q^3-(N+1)q \right) ~\mod b, &\quad N = 0 ~\mod 4, N \ne 0 ~\mod 3,\\
      & \text{or } N = 0 ~\mod 3, N\neq 0 ~\mod 2 \\
      \frac{b}{2N} \left( N q^3 + q \right) ~\mod b, &\quad \text{else}
    \end{cases}
   \end{split}
\end{equation}
For instance, when $N=3$, $\mho^6_{\Spin^{\Z_6^{\chi}}}\cong \Z_{a=9}$. The anomaly from our adjoint fermion is then
\begin{equation}
\nu_9 = -1 ~\mod 9 \in \Z_9\;.\nn
\end{equation}

As we shall see this anomaly can be saturated by $\mathbb Z_{2N}^{\chi}$ symmetry breaking (see \S~\ref{sec:one-ferm-IR-phase}).

How are all these anomalies matched in the IR? The most natural way is to spontaneously break the $\Z_{2N}^\chi$ symmetry for any mass of the fundamental matter. We know this is the right answer when the mass of the fundamentals is large enough, but one could speculate some sort of bulk transition for small enough fundamental fermion mass. What could this phase be? In \cite{Cordova:2019bsd, Cordova:2019jqi}, it was shown that unitary and symmetry-preserving TQFTs are excluded in 4d. Another option is to have massless composite fermions. We will in fact propose such fermions for massless fundamental matter which will match all the anomalies. However one can use the argument of Weingarten\footnote{Weingarten considered QCD -- i.e. a theory fundamental fermions only. However the argument is unchanged as the adjoint fermion does not invalidate the positivity of the measure as we show in the Appendix 
\ref{app:measure-positive-definite}.} \cite{Weingarten1983} to show that the meson mass is always smaller than the baryon. Since massless composites must necessarily include baryons to saturate the mixed $\Z_{2N}^\chi-\U_B$ anomaly, it follows that the meson must also be massless, which is not natural unless it is a Goldstone boson. So it seems that $\Z_{2N}^{\chi}$ chiral symmetry must be spontaneously broken. The order parameter of this breaking is the bilinear condensate $\mbox{Tr}\lambda\lambda$. A non-vanishing expectation value of $\mbox{Tr}\lambda\lambda$ breaks $\mathbb Z_{2N}^{\chi}$ down to the fermion number $\mathbb Z_2$ and results in $N$ distinct vacua connected via domain walls. Finally let us briefly consider a scenario where $\Z_{2N}^\chi$ is broken, but $\U_B$ is spontaneously broken as well. However this is prohibited by the Vafa-Witten \cite{Vafa:1983tf} argument.

\subsection{The massless case}
\label{sec:one-Dirac-massless}

When the mass of the fundamental Dirac fermion goes to zero, i.e., $m=0$, the $\U_B$
baryon symmetry remains unchanged, and the theory has an enhanced
chiral (axial) $\U_\chi$ symmetry\footnote{The adjoint/fundamental mixed representation was also studied on $M^4=\R^3\times S^1$ in \cite{Poppitz:2009tw}, which was shown to have a massless goldstone boson associated with the $\U_\chi$ symmetry. Therefore, it was conjectured that the theory with a single fundamental flavor is continuously connected to the theory on $\R^4$ as we decompactify  $S^1$.}. To see this, note that there are more classical
chiral symmetries in the massless case, which now acts on the
fundamental Dirac fermion. The chiral transformation
\begin{equation}
\lambda\rightarrow e^{i \alpha}\lambda, \quad \psi \rightarrow e^{i \beta}\psi, \quad \tilde{\psi} \to e^{i \beta}\tilde{\psi}
\end{equation}
produces a change in the action
\begin{equation}
  \begin{split}
    \Delta S &= i \alpha\frac{2N}{8\pi^2}\int \tr f\wedge f + i 2\beta\frac{1}{8\pi^2}\int \tr f\wedge f\;\\
    &= 2i \frac{(N\alpha+\beta)}{8\pi^2}\int \tr f\wedge f\;,
  \end{split}
\end{equation}
where $f$ is the $\SU(N)$ 2-form field strength. If we set $\beta=-N\alpha$ the action is obviously invariant, hence the transformation
transformation
\begin{equation}\label{eq:chiral-transf}
\begin{split}
\U_\chi:\quad&\lambda\rightarrow e^{i \alpha}\lambda\,,\\
&\psi \rightarrow e^{-i N \alpha} \psi, \quad \tilde{\psi} \to e^{-i N\alpha}\tilde{\psi}\;,
\end{split}
\end{equation}
is a good symmetry even at the quantum level. Thus, the faithful global symmetry is
\begin{equation}
G^{\text{Global}} = \frac{\U_q}{\Z_N} \times \U_\chi\;,
\label{genuine global symmetry for one flavor massless Dirac}
\end{equation}
The fermions transform under $G^{\text{Global}}$ in the
representations given by Table \ref{tab:massless-fermion-reps}.
 \begin{table}[h]
   \centering
  \begin{tabular}{c||c|cc}
    & $\SU(N)$ & $\U_B$ & $\U_\chi$ \\
    \hline
    $\psi$ & ${\tiny\yng(1)}$ & $+1$ & $-N$ \\
    $\tilde{\psi}$ & ${\tiny \overline{\yng(1)}}$ & $-1$ & $-N$  \\
    $\lambda$ & $\textbf{adj}$ & $0$ & $+1$ 
  \end{tabular}
  \caption{Representations of the fermions under various symmetry groups in the massless case.}
  \label{tab:massless-fermion-reps}
\end{table}

The $\U_\chi$ symmetry carries a 't Hooft anomaly.  Indeed, the anomaly is a
consequence of the triangle diagrams containing $\lambda$ and $\Psi$. The anomaly coefficient is given by
\begin{equation}\label{eq:A3}
\mathcal{C}_{A^3} = N(-N)^3+N(-N)^3+(N^2-1)1^3=-2N^4+N^2-1\;,
\end{equation}
where the first two factors come from the fundamental fermions
$\psi, \tilde{\psi}$, and the third one comes from the adjoint fermion
$\lambda$. Note that the fundamentals carry charge $-N$ under the chiral
symmetry, and there are $N$ colors, while the adjoint carries a charge
$1$ and there are $N^2-1$ colors.

In addition to this cubic anomaly of $\U_\chi$, we also have a mixed anomaly between
$\U_\chi$ and $\U_B$. If we were not careful about the modding by $\mathbb Z_N$ in (\ref{genuine global symmetry for one flavor massless Dirac}), we would find that the coefficient of this anomaly comes from the triangle diagrams $\U_\chi\left[\U_q\right]^2$, which yield $-2N^2$. This is the traditional 't Hooft anomaly. However, the modding by $\mathbb Z_N$ in (\ref{genuine global symmetry for one flavor massless Dirac}) refines this anomaly.  To compute it, we put background fields for the $\U_B$ symmetry and perform
the chiral transformation \eqref{eq:chiral-transf}. We then have
\begin{equation}
\label{mixed BCF anomaly}
  \begin{split}
    \Delta S &=  i\frac{2N\alpha}{8\pi^2}\int \tr\left[ \left(F-\frac{\tr F}{N}\right)\wedge \left(F-\frac{\tr F}{N}\right)\right]-i\frac{2N\alpha}{8\pi^2}\int\tr\left[ F\wedge F\right]\\
    &=\frac{-i \alpha}{4\pi^2} \int F_B\wedge F_B\,,
  \end{split}
\end{equation}
where $F_B=N\mathcal F_q$ is the properly normalized baryon background
field. Notice that this is the same as \eqref{eq:Z2N_anomaly} and
\eqref{eq:Z2N_anomaly1} we found previously for $\mathbb Z_{2N}^{\chi}$
chiral symmetry, which is just $\alpha= \frac{\pi k}{N}, k\in\mathbb Z$. The theory also has a mixed $\U_\chi$-gravitational  anomaly. All these are captured by a 5d action\footnote{From the descent formulas, we have that
  $\frac{1}{3!(2\pi)^3}\int_{6d} F_\chi\wedge F_\chi\wedge F_\chi\in \Z$, so that the 5d CS
  theory is $\frac{k}{24 \pi^2} \int A_\chi \wedge F_\chi\wedge F_\chi$ with
  $k\in \mathbb Z$. On the other hand
  $\frac{1}{2!(2\pi)^3}\int_{6d} F_\chi\wedge F_B\wedge F_B\in\Z$ implies that the 5d
  mixed CS theory is given by the action
  $\frac{k}{4\pi^2}\int A_\chi \wedge F_B\wedge F_B$.}
\begin{multline}\label{eq:pert-anom-inflow}
S_{5d}=  i\frac{-2N^4+N^2-1}{24 \pi^2} \int A_\chi\wedge F_\chi\wedge F_\chi+  i\frac{-2}{8\pi^2} \int A_\chi\wedge  F_B\wedge F_B\\
+ i\frac{N^2+1}{24}\int A_\chi\wedge p_1\;,
\end{multline}
where $A_\chi$ and $F_\chi$ are the background gauge field and its field
strength for the $\U_\chi$ chiral symmetry, and
where integral over the first Pontryagin class obeys $\int_{ M} p_1\in 48\mathbb Z$.

%%%%%%%%%%%%%%%%%%%%%
\subsection{The IR phases}
\label{sec:one-ferm-IR-phase}
%%%%%%%%%%%%%%%%%%%%%%

In the infinite mass limit, the Dirac fermion completely decouples,
leaving us with a pure $\mathcal{N}=1$ SYM with gauge group
$\SU(N)$. The discrete chiral symmetry remains. Moreover, there is now
an emergent $\Z_N^{[1]}$ 1-form global symmetry. Therefore, the global
symmetry is given by
\begin{equation}
  \label{eq:infinite-flavour}
  G^{\scriptsize \mbox{Global}}_{m\rightarrow \infty} = \Z_{2N}^{\chi} \times \Z_N^{[1]}\,.
\end{equation}
Of course, the super Yang-Mills and the above symmetry will emerge as long as $m\gg \Lambda$, where $\Lambda$ is the strong scale.

In this case, there is a mixed anomaly between the 0-form and 1-form symmetries in
$  G^{\scriptsize \mbox{Global}}_{m\rightarrow \infty}$ \cite{Gaiotto:2017yup,Komargodski:2017smk}\footnote{To see this, couple a background
gauge field to $\Z_N^{[1]}$. This is a closed 2-cochain $B_2$ with
$\Z_N$ coefficient satisfying the 1-form gauge transformation\footnote{The $\delta$}
\[B_2\mapsto B_2+\dd \lambda^{(1)}\] 
with $\lambda^{(1)}$ a 1-cochain. It, therefore,
defines a class $B\in H^2(\mathcal{M}_4;\Z_N)$, where $\mathcal{M}_4$ is
the 4-dimensional spin manifold on which our theory lives. The theory
now admits a non-trivial $P\SU(N)$ gauge bundles that are not $\SU(N)$
gauge bundles, with the obstruction given by $w_2(P\SU(N)) =
B$. Consequently, the instanton number can now be fractional and
quantized in units of $1/N$. The theta angle is now $2\pi
N$-periodic. As the classical $\U_\chi$ chiral transformation with
parameter $\alpha$ shifts $\theta \mapsto \theta-2N \alpha$, the non-anomalous subgroup is
generated by $\alpha=\pi$, instead of $\alpha=\pi/N$ as we had before when the
periodicity was $2\pi$. It is also simple to see that any local counter terms cannot restore this periodicity. Thus, the $\Z_{2N}^{\chi}$ chiral symmetry is
anomalous, but its $\Z_2^F$ subgroup remains anomaly-free. We
interpret this as a 't Hooft anomaly between the 1-form symmetry and
the chiral symmetry.}. In addition, there is, of course, the mixed gravitational anomaly with $\mathbb Z_{2N}^\chi$ and the nonperturbative $\mathbb [\Z^\chi_{2N}]^3$ anomaly we discussed in \S~\ref{sec:one-Dirac-massive}.

All the anomalies are saturated in the IR by the spontaneous symmetry
breaking of $\Z_{2N}^{\chi}$ down to $\Z_2^F$ via the formation of the bilinear condensate
\begin{equation}
  \label{eq:condensate}
  \expval{\text{Tr}~\lambda\lambda}  = \Lambda^3 \exp \left( \frac{2\pi i k}{N} \right),\; k=0,1,\ldots, N-1\;, 
\end{equation}
where the $N$ vacua are labelled by $k=0,1,\ldots, N-1$.

In the opposite limit, when $m\rightarrow 0$, a priori, we can saturate the anomalies in two ways. One can spontaneously break $\U_\chi$ symmetry or have massless composite fermions. Weingarten's identities exclude the latter
whenever the Euclidean action is positive definite. The fact that our action is positive definite is
demonstrated in Appendix
\ref{app:measure-positive-definite}. So the only alternative is that $\U_\chi$ is spontaneously broken. The IR theory contains a Goldstone boson: a compact scalar $\varphi\sim \varphi+2\pi$ whose effective action is given by 
\begin{equation}\label{eq:goldstone0}
 S_{\text{Goldstone}}\propto \int d^4x\;\Lambda^2(\partial\varphi)^2\;,
\end{equation}
up to an overall multiplicative constant that scales as $N^3$ in the large-$N$ limit (see \S~\ref{sec:large_N}).  The scalar $\varphi$ is associated with the $\tr\lambda\lambda\sim e^{i\varphi}$, and therefore, the operator $e^{i\varphi}$ carries a charge $2$ under $\U_\chi$. 

The above IR action, however, does not reproduce the UV anomalies, in particular the $[\U_\chi]^3$, the $[\U_\chi][\U_B]^2$ and the $\U_\chi$--gravitational anomalies. To fix this problem, we find a set of composite fermions that match anomalies.  We assume that we have $K$ massless fermions with charges $q^\chi_1,q^\chi_2,\dots, q^\chi_K$ under $\U_\chi$  and charges $q^B_1,q^B_2,\dots, q^B_K$ under $\U_B$. Then they match the [$\U_\chi]^3$, mixed $\U_B$ and $\U_\chi$, and the $\U_\chi$-gravitational anomalies if and only if\footnote{In fact, we could replace the second condition $\sum_{k=1}^K q_k^\chi (q_k^B)^2 = -2$ with the traditional 't Hooft anomaly $\U_\chi\left[\U_B\right]^2$, which gives $\sum_{k=1}^K q_k^\chi (q_k^B)^2 = -2N^2$, without affecting any of our conclusions. It was proven in \cite{Anber:2019nze} that if a vector-like theory does not possess a genuine discrete chiral symmetry, massless composites that saturate the traditional 't Hooft anomalies will also match the CFU anomalies. This is easily seen by observing that the only difference between the traditional $\U_\chi\left[\U_B\right]^2$ and the anomaly (\ref{mixed BCF anomaly}) is the different normalizations of the $\U_B$ charges, which account for the multiplicative  $N^2$ factor that appears in the traditional anomaly.}
\begin{align}\label{eq:U1_anom_condition}
\nonumber
&\sum_{k=1}^K \left(q_k^\chi\right)^3= -2N^4+N^2-1\,,\\
&\sum_{k=1}^K q_k^\chi (q_k^B)^2 = -2\,,\\
\nonumber
&\sum_{k=1}^K q_k^\chi = -(N^2+1)\,,
\end{align}
and $q_k^\chi$ need not be distinct. In general, there are multiple ways of satisfying the above conditions for particular $N$. Here, we find a solution that works on all $N$, so it is a natural large $N$ candidate.

First, note the identity
\beq
-2N^4+N^2=\sum_{i=1}^{N}(-2i+1)^3\;.
\eeq
Thus, we satisfy the $[\U_\chi]^3$ anomaly by taking $q_0^\chi=-1$ and $q_i^\chi=-2i+1$ for $i=1,\dots, N$. On the other hand, if we take $q_0^B=1=-q_1^B$, with all other $q_i^B=0,\forall\, 2\le i\le N$, we satisfy the mixed $\U_\chi[\U_B]^2$  anomaly. Finally we also look at the mixed $\U_\chi$-gravitational anomaly, which is given by $-1+\sum_{i=0}^N(-2i+1)=-(N^2+1)$ as it should be. Note that for particular values of $N$, we can match the anomaly with other choices, so the above formula should be interpreted as a natural choice reproducing the smooth large-$N$ limit.

But we do not want the fermions to be gapless; as we pointed out, the Weingarten theorem \cite{Weingarten1983} excludes this scenario. Instead, we want to supplement the Goldstone action \eqref{eq:goldstone0} with fermions that are gapped in the bulk, but otherwise become massless on the vortex worldsheet. To this end, we write bulk fermion mass terms. Such mass terms must respect $U(1)_B$ and can be made $\U_\chi$ invariant by decorating them with the appropriate power of the operator $e^{i\varphi}$. With the charges $q_i^\chi$ and $q_i^B$ we chose above, we can supplement the Goldstone action with 
\beq\label{eq:LGoldstone_fermions}
\mathcal L_f=\sum_{k=0}^N\bar\chi_k i\sigma^\mu\partial_\mu\chi_k+ m_{01}\chi_0\chi_1 e^{-i\varphi \frac{q_0^\chi+q_1^\chi}{2}}+ \sum_{i,j\neq 0,1}^{N}m_{ij}\chi_i\chi_j e^{-i\varphi \frac{q_i^\chi+q_j^\chi}{2}}+\mbox{c.c.}\;.
\eeq
We note that $q_i^\chi$ are odd, so $q_i^\chi+q_j^\chi$ is even for any pair $i,j$. 
The above action should \emph{not} be viewed as an effective theory of the bulk. Indeed the masses of the fermions above are expected to be of order $\Lambda$, which is the UV cutoff of the would-be effective theory. Instead the purpose of including these fermions is four-fold.
\begin{enumerate}
\item The construction explicitly shows that the anomaly can be saturated by the Goldstone boson\footnote{Another way to argue is to write a WZW-like term $(d\varphi+A)\wedge F\wedge F$ in the auxiliary 5d bulk. One can think of this term as stemming from gauging the ``WZW term'' $d\varphi\wedge d^2\varphi\wedge d^2\varphi$. Of course $d^2=0$ on smooth fields $\varphi$, however it does not vanish on singular vortex configurations of $\varphi$, and the anomaly can be interpreted as being saturated by vortex configurations. See also \cite{Cherman:2018jir} for a related discussion.} and shows that the massless composite regime is in the same deformation class as the Goldstone boson. 
\item In a moment we will introduce a mass term for the fundamental fermion breaking the $\U_\chi$ symmetry to $\mathbb Z_{2N}$ explicitly. The $\mathbb Z_{2N}$ will be spontaneously broken, and the discrete vacua will support domain walls. One may wonder whether the inclusion of massive fermions in the bulk would result in light fermions on some domain walls. Indeed since the mass term couples to $\varphi$, this remains an a priori possibility. This is especially important for $T$-preserving domain walls which have 't Hooft anomalies, and it was proposed in \cite{Lohitsiri:2022jyz} that massless fermions do saturate such theories. We will however see that this does not happen in the regime we discuss here (see the next section titled \emph{Domain Walls}). However the bulk fermions will allow us to speculate about this possibility in the regime inaccessible by our analysis. 
\item The vortex of the Goldstone theory carries an anomaly, which is naturally saturated by massless fermions. Inclusion of the massive bulk fermion contains a proposal for this vortex. Indeed the bulk fermion are arbitrarily light near the vortex, and in that sense they are distinguished from other excitations whose mass is of order $\sim \Lambda$,
\item Finally, the fermions in question can be made arbitrarily light by coupling the theory to fundamental Higgs fields in such a way so that the Higgs regime fully breaks the $\SU(N)$ gauge group to nothing. We will explicitly demonstrate this for $\SU(2)$ where only one scalar Higgs is sufficient (see discussion preceding \eqref{eq:eff_SU2_ferm_higgs}). The Higgs regime in this case is characterized by massless composites. Assuming that the transition is 2nd order, the massless fermion phase will transition into a Goldstone phase, with a low energy effective theory given by the Goldstone coupled to light, but massive fermions, of the type we discussed above. 

\end{enumerate}

%\footnote{It is perhaps interesting to note that coupling scalars in the fundamental representation of the gauge group will in general reduce the masses of the composite fermions, and it may even push the model into S-confinement. This model is also interesting for making connections with the Super QCD. We leave the study of this for the future.  }.

Next, let us discuss the IR phase when we introduce a small mass of the fundamental fermion. This explicitly breaks $\U_\chi\rightarrow \mathbb Z_{2N}^\chi$. The most relevant term that achieves the breaking is $\cos (N\varphi)$,  and thus, the Goldstone action needs to be modified to
\begin{equation}\label{eq:Sgoldstone}
S_{\text{Goldstone}} = \int\dd^4x\, \Lambda^2 g \left( m/\Lambda,N \right) \left[\left( \partial\varphi \right)^2 - m\Lambda f \left(m/\Lambda,N\right)\left( \cos N\varphi -1\right)+...\right]\,,
\end{equation}
where $g(0,N)$ and $f(0,N)$ are finite numbers scaling as $\sim N^3$ and $\sim 1$ respectively in the large-$N$ limit (see \S~\ref{sec:large_N}). 
This gives mass to the
Goldstone, as well as lifting the $S^1$ vacuum manifold to $N$
vacua. We hence again have $N$ vacua for small $m$. The domain walls have width $\sim \frac{1}{\sqrt{m\Lambda}}$, and thus, they are thick compared to the strong scale.  

We conclude that in both the limit of small and large $m$, the discrete chiral symmetry $\mathbb Z_{2N}^\chi$ is spontaneously broken to $\mathbb Z_2$. The simplest assumption is that the intermediate regime has no phase transition. 

\begin{figure}[h!]
  \centering
  \includegraphics[scale=0.5]{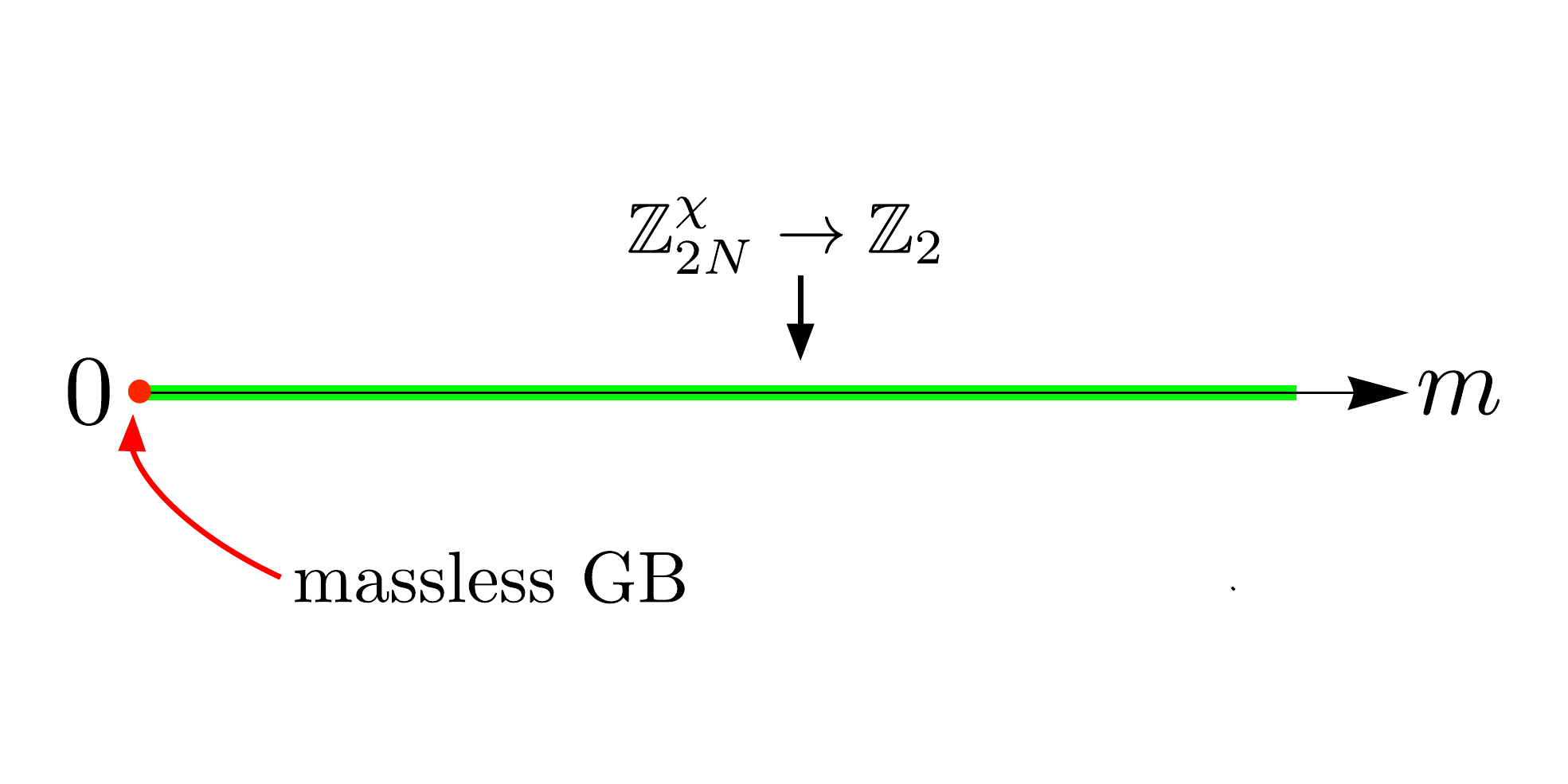}
  \caption{The phase diagram of the $\SU(N)$ QCD(f/adj) with one fundamental Dirac fermion and one massless adjoint Weyl fermion as we vary the mass $m$ of the Dirac fermion. For any non-zero $m$, the discrete chiral symmetry $\Z_{2N}^{\chi}$ is spontaneously broken to $\Z_2$, leaving us with $N$ vacua. At $m=0$, $\Z_{2N}^\chi$ is enhanced to $\U_{\chi}$, which also breaks spontaneously, giving rise to a massless Goldstone boson.}
  \label{fig:one-ferm-phase}
\end{figure}

One wonders if there could be a different phase opening in the intermediate regime. Symmetry-preserving TQFTs are excluded in 4d \cite{Cordova:2019bsd}. Spontaneous breaking of $\U_B$ neither saturates the nonperturbative $\mathbb Z_{2N}^\chi$ anomaly, nor the mixed $\mathbb Z_{2N}^\chi$--gravity anomaly. Further $\U_B$ cannot break by the Vafa-Witten theorem, which holds because of the positivity of the fermionic weight (see Appendix \ref{app:measure-positive-definite}). The massless composites are also excluded by the (Vafa-Witten-)Weingarten theorem \cite{Weingarten1983}. Therefore any intermediate bulk transition seems inconsistent with the anomalies and Vafa--Witten--Weingarten theorems. We conclude there is no bulk phase transition as we decrease $m$ all the way down to $m=0$, at which point $N$ vacua melt away and turn into a (pseudo-scalar) Goldstone boson. This behavior is summarized in Figure~\ref{fig:one-ferm-phase}.
\begin{figure}[tbp] %  figure placement: here, top, bottom, or page
   \centering
   \includegraphics[width=0.7\textwidth]{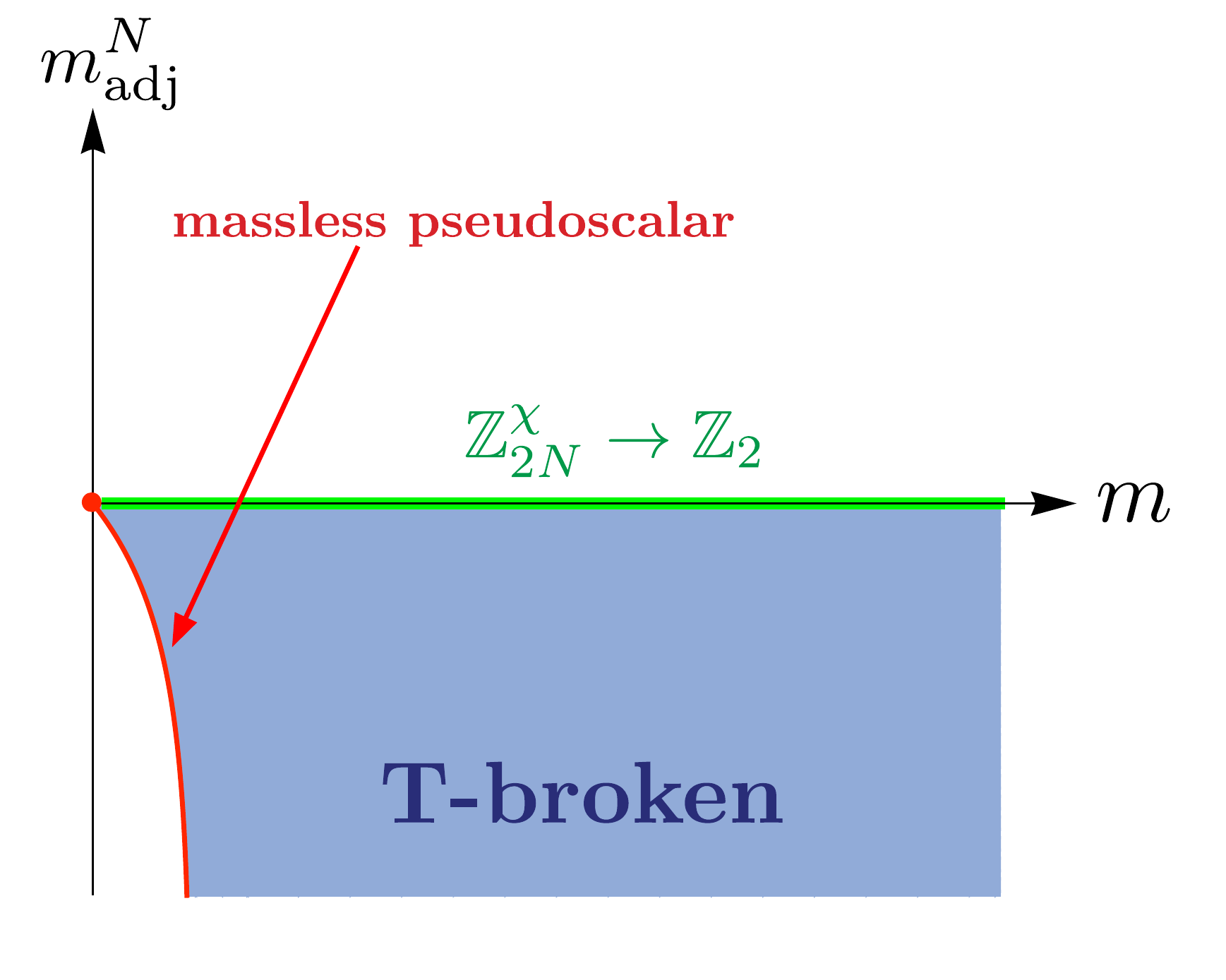} 
   \caption{ The $T$-invariant slice of the phase diagram of the $\SU(N)$ QCD(f/adj) with one Dirac-fermion fundamental flavor with mass $m$ which can be taken as real and positive, and one Weyl-fermion adjoint flavor with mass $m_{adj}$ which is taken as complex. When $m_{adj}=0$, the phase diagram reduces to that of Figure~\ref{fig:one-ferm-phase}, which we already discussed. The theory is still QCD(f/adj) at small values of m and m adj with a $\theta$ angle (given by the phase of $m_{adj}^N$). Then, in the $m\to \infty$ limit, it reduces to SYM with $m_{adj}$ and $\theta$. The same can be said for the other limit. When $\theta=\pi$, such a theory is expected to have $T$-broken vacuum. This phase is labeled by blue sheets. When $m$ becomes small enough, however, the vacuum undergoes a phase transition into the trivially gapped phase. The line on which this happens contains a massless pseudo-scalar. The massless pseudo-scalar phase connects continuously to the massless Goldstone boson phase of \eqref{eq:goldstone0}. }
   \label{fig:phase_adj_fund}
\end{figure}

Finally, we introduce the adjoint fermion mass $m_{adj}$. In this case, the theory has a physical $\theta$ parameter that can be removed by shifting the phases of $m_{adj}$ and $m$. Further, recall that there is a combined shift of the adjoint and fundamental phases, which does not affect the $\theta$ term. This means we can always make the fundamental mass $m$ real and positive, putting all the $\theta$ dependence into the $m_{adj}$ phase. In the following, we use a normalization where the bare $\theta$ parameter is zero and $m_{adj}$ is taken to be complex $m_{adj}=e^{i\frac{\theta}{N}}|m_{adj}|$. Since the theory is invariant under $\theta\rightarrow \theta+2\pi$, the phase diagram in the space of the real $m$ and complex $m_{adj}$ will be symmetric up to a $e^{i\frac{2\pi}{N}}$ phase shift of $m_{adj}$. When $|m_{adj}|$ is large, the theory is $1$-flavor QCD at $\theta$ angle given by the phase of $(m_{adj})^N$. At $\theta=\pi$, the theory was shown to have a massless $\eta'$ particle at some particular $m=m_0\sim \Lambda/N$  in the large-$N$ limit \cite{Gaiotto:2017tne}. The argument for this is as follows. The mass $m$ of a single fundamental flavor is complex, and its phase is associated with the $\theta$ term. Restricting to $T$-invariant theories, we set $\theta=0$ and take  $m$ real, but it can be both positive and negative. The negative mass theory is equivalent to the $\theta=\pi$ theory with a positive mass term.  If $|m|$ is large, the theory is pure Yang-Mills which is believed to break $T$-symmetry spontaneously at $\theta=\pi$ (large negative mass) \cite{Gaiotto:2017yup}, but is trivial at $\theta=0$. Hence, there should be a phase transition restoring the $T$ symmetry at some finite $m=m_0$, with a corresponding massless pseudo-scalar particle at that point. In \cite{Gaiotto:2017yup}, it was argued that this $m_0$ is negative, but this is not crucial. Since we have adjoint and fundamental fermion masses, we made $m$ real and positive by a non-anomalous chiral transformation and put all the phases in $m_{adj}$. Hence, when the modulus of $m_{adj}$ is large and the effective $\theta=\pi$, we should have a massless pseudo-scalar at $m=|m_0|$.

What happens as we reduce  $m_{adj}$ from infinity to small values? As $m_{adj}$ becomes comparable to $\Lambda$, it is natural to assume that the value of $m_0$ will change. However, the massless phase cannot just disappear as long as we pick $m_{adj}$ in such a way to preserve the $T$-symmetry\footnote{Naively, any choice of $m_{adj}$ which is not real breaks $T$ symmetry, mapping $m_{adj}\rightarrow m_{adj}^*$. However, recall that we can always perform a $\mathbb Z_{2N}^\chi$ rotation, which will not induce a $\theta$-term, to remove the $\mathbb Z_{N}$ phase from $m_{adj}$. The point is that the theory where the phase of $m_{adj}$ is the $(2N)$-th root of unity is $T$-symmetric.}, and instead, all that will happen is that $m_0$ -- the fundamental mass at which the massless pseudo-scalar exists -- will start moving as a function of $|m_{adj}|/\Lambda$ until at $m_{adj}=0$ it becomes $m_0=0$ where this massless pseudo-scalar phase fuses with that of the Goldstone pseudo-scalar of the spontaneous $\U_\chi$ symmetry breaking.  All of this is summarized in Figure ~\ref{fig:phase_adj_fund}.

This picture is analytically controlled for $|m_{adj}|\ll \Lambda$ at any\footnote{This differs from the opposite limit, $|m_{adj}|\gg \Lambda$, where one needs to invoke the large-$N$ limit to carry out the analysis, as was done in \cite{Gaiotto:2017yup}.} $N$. If a nonzero $m_{adj}$ is introduced, it will induce the following term in the Lagrangian
\begin{equation}
\mathcal L_{m_{adj}}\propto -\Lambda^3|m_{adj}|\cos(\varphi+\theta/N)\;.
\end{equation}
so that the potential for $\varphi$ is approximately
\begin{equation}
V=-a\cos(\varphi N)-b\cos(\varphi+\theta/N)\;.
\end{equation}
with some positive constants $a\propto m$ and $b\propto |m_{adj}|$ which scale as $N$ and $N^2$ respectively in the large $N$ limit. We can redefine $\varphi+\theta/N\rightarrow \varphi$ and obtain.
\begin{equation}
V=-a\cos(\varphi N-\theta)-b\cos\varphi\;.
\end{equation}
Now, restricting to $\theta=0,\pi$, we can capture both by setting $\theta=0$ and extending $a$ to be also negative. So positive $a$ corresponds to the $\theta=0$ regime, and negative $a$ corresponds to the $\theta=\pi$ regime. 

Now notice that if $a$ is positive, the global minimum of $\varphi$ is at $0\bmod 2\pi$. When $aN^2=-b$, the mass of $\varphi$ vanishes. Taking into account the large $N$-scaling, we have that $a\sim m\Lambda^3 N$ and $b\sim |m_{adj}|\Lambda^3 N^2$, so that at $m=m_0\sim -\frac{|m_{adj}|}{N}$ we have a massless pseudo-scalar.  If $a$ is dialed to be even smaller, then two vacua emerge, breaking $T$-symmetry (because $\varphi$ is a pseudo-scalar).

%%%%%%%%%%%%%%%%%%%
\subsubsection*{Domain walls}
%%%%%%%%%%%%%%%%%%

We have seen that the theory with massless adjoint Weyl fermion and one fundamental Dirac fermion with mass $m$ has no phase transition all the way to $m=0$. For every finite $m$, it supports $N$ discrete vacua and therefore has domain walls. Could there be a phase transition on the domain wall? In the limit of $m\rightarrow\infty$, the domain wall has a non-trivial inflow due to the mixed anomaly between the $\mathbb Z_{2N}^{\chi}$ chiral symmetry and $\mathbb Z_{N}^{[1]}$ 1-form symmetry, which is saturated by a TQFT \cite{Tachikawa:2016cha}. When a single fundamental fermion is introduced, the only remaining anomaly is the mixed $\U_B$--$\mathbb Z_{2N}^{\chi}$, and the inflow  on the $n$-domain wall we denote as $D_n$\footnote{Here integer $n$ signifies whether the domain wall connects neighboring vacua, with  $n=1$, or next-neighboring vacua $n=2$, or next-to-next neighboring vacua.} is
\begin{equation}
e^{\frac{i k}{4\pi N}\int_{4} F_B\wedge F_B}\;,
\end{equation}
where $F_{B}$ is the field strength of the $\U_B$ symmetry. Note that the 4d space over which the above integral runs is not the physical 4d space but the domain-wall world volume extension into a 5th-dimensional bulk.  The above anomaly polynomial has no inflow and corresponds to some anomalous conductivities on the domain-wall theory \cite{Closset:2012vp}. In other words, the domain wall theory has fractional Hall conductivity but can otherwise be gapped. 

Indeed, we know that in the limit $m\rightarrow \infty$, the theory is $\mathcal N=1$ super Yang-Mills, and we understand the domain wall theory well. The $\mathcal N=1$ SYM has $N$ vacua $v_i$, labeled by $i=1,2,\dots, N$. Domain walls $D_n$ that interpolate $v_i\rightarrow v_{i+n\bmod N}$ are all stable. The $D_n$ domain wall theory is conjectured to be the 3d  SYM theory with Chern-Simons level $n$, where we will take $n$ to have values $n=\pm1,\pm 2,\dots, \pm\lfloor N/2\rfloor$. Such a theory breaks the supersymmetry spontaneously, resulting in a Majorana Goldstino and a TQFT \cite{Acharya:2001dz,Gomis:2017ixy,Delmastro:2020dkz}     $\UU(n)_{N-n,N}$\,.

 Since a generic domain wall will not enjoy a time-reversal symmetry, the presence of fundamental matter is expected to induce a time-reversal non-invariant 3d mass term for a Goldstino, but a TQFT must remain robust  as long as the fundamental matter is heavy enough. The exception to this is a time-reversal-invariant domain wall which exists whenever $N$ is even and is given by $n=N/2$, where the TQFT $\UU(N/2)_{N/2,N}$ is non-trivially time-reversal invariant due to the level/rank duality \cite{Cordova:2017vab,Hsin:2016blu}. In this case, there is a non-vanishing $\bmod~16$ pure $T$-anomaly \cite{Witten:2015aba,Hsieh:2015xaa}, which descends from the $\mod 16$ anomaly of the $\Spin$-$\Z_4$ symmetry in the bulk \cite{Hason:2020yqf}. Both the Goldstino and the TQFT contribute nontrivially to this anomaly \cite{ Tachikawa:2016cha,Tachikawa:2016nmo,Gomis:2017ixy}, and hence a Goldstino is prohibited from acquiring a mass unless the TQFT is destroyed. Does this happen on the domain walls as the fundamental mass is reduced?

The effective theory is given by a pseudo-Goldstone boson $\varphi$ and the domain walls are just kinks of \eqref{eq:Sgoldstone}. Kinks carry tension $\sim\sqrt{m\Lambda}\Lambda^2$ which makes them light and within the realm of the effective theory. But one may worry that there are extra light degrees of freedom on the kinks from the anomaly-saturating massive fermions  \eqref{eq:LGoldstone_fermions}. That kinks do not have fermionic zero modes case is seen as follows. Consider the domain wall that takes $\varphi_0
\to \varphi_1 = \varphi_0+2\pi/N$, assuming a profile of $\varphi(x^3)$ in the 3-direction which interpolates between two vacua. We first ``diagonalize'' the symmetric mass matrix\footnote{This is an abuse of terminology, as the ``diagonalization'' is not performed by a matrix $P$ and its inverse $P^{-1}$ but by a unitary matrix $U$ and its transpose $U^T$ which is not necessarily its inverse. The ``diagonalization'' we refer to here is more properly called singular value decomposition.} $M_{ij}=m_{ij}e^{i(q^{\chi}_i+q^{\chi}_j)\varphi/2}$  to $M_D=U M U^T$, where $U$ is a unitary transformation\footnote{Note that the kinetic term is precisely invariant under $\chi\rightarrow U\chi$.}. Once diagonalized, the system becomes a set of $N+1$ 4d Majorana fermions, with masses given by the diagonal entries of $M_{D}$. We now want to check whether the Dirac operator
\begin{equation}\label{eq:Dirac_op}
i\slashed D_i = i \gamma_3\partial_3+ \begin{pmatrix}(M_D)_{ii} \mathbb{I}_2 & 0\\ 0& (M_D)^{*}_{ii}\mathbb{I}_2\end{pmatrix}\;,
\end{equation}
for each $i=0,1,\ldots, N$, has a zero mode. If it does, it will be the same as for
\begin{equation}
  \gamma_3\slashed D_i= \partial_3 + A_i \,,
\end{equation}
where
\begin{equation}
  A_i = -i\gamma_3  \begin{pmatrix}(M_D)_{ii} \mathbb{I}_2 & 0\\ 0& (M_D)^{*}_{ii}\mathbb{I}_2\end{pmatrix}\;,
\end{equation}
which is a Hermitian matrix by construction. It is well known that the zero modes of the above operators are in one-to-one correspondence with the spectral flow of the matrix $A$ (see Appendix \ref{app:spectral}). Now, $M_D$ cannot depend on the constant piece of $\varphi$, as a $\U_\chi$ rotation can remove the constant. Hence, the fermion mass spectrum does not change as a function of slowly varying $\varphi$. We can conclude that no fermion zero modes exist on the domain walls.

Further, in super Yang-Mills, the $n$-domain walls $D_n$ are stable and do not decay into $n\times D_1$ domain walls. This is not the case for the domain walls of \eqref{eq:Sgoldstone}, as the domain walls are the usual sine-Gordon kinks that repel each other. The massive fermions cannot induce attraction between the domain walls as they do not have any zero modes and can be integrated out. A sine-Gordon kink, e.g., varying in the $x^3$ direction, will have $\varphi$ vary slowly so that its derivative $\partial_3\varphi\sim \sqrt{m\Lambda}$. Since we assume $m\ll \Lambda$ we can integrate out the massive fermions and write an expansion in powers of the derivatives, i.e., in $\partial_\mu(...)/\Lambda$. Since the derivatives of the kink vary as $\sqrt{m\Lambda}$, and since no zero modes exist by the above reasoning, such a gradient expansion is valid, and the effective domain wall theory is empty.

\begin{figure}[tbp] %  figure placement: here, top, bottom, or page
   \centering
   \includegraphics[width=0.7\textwidth]{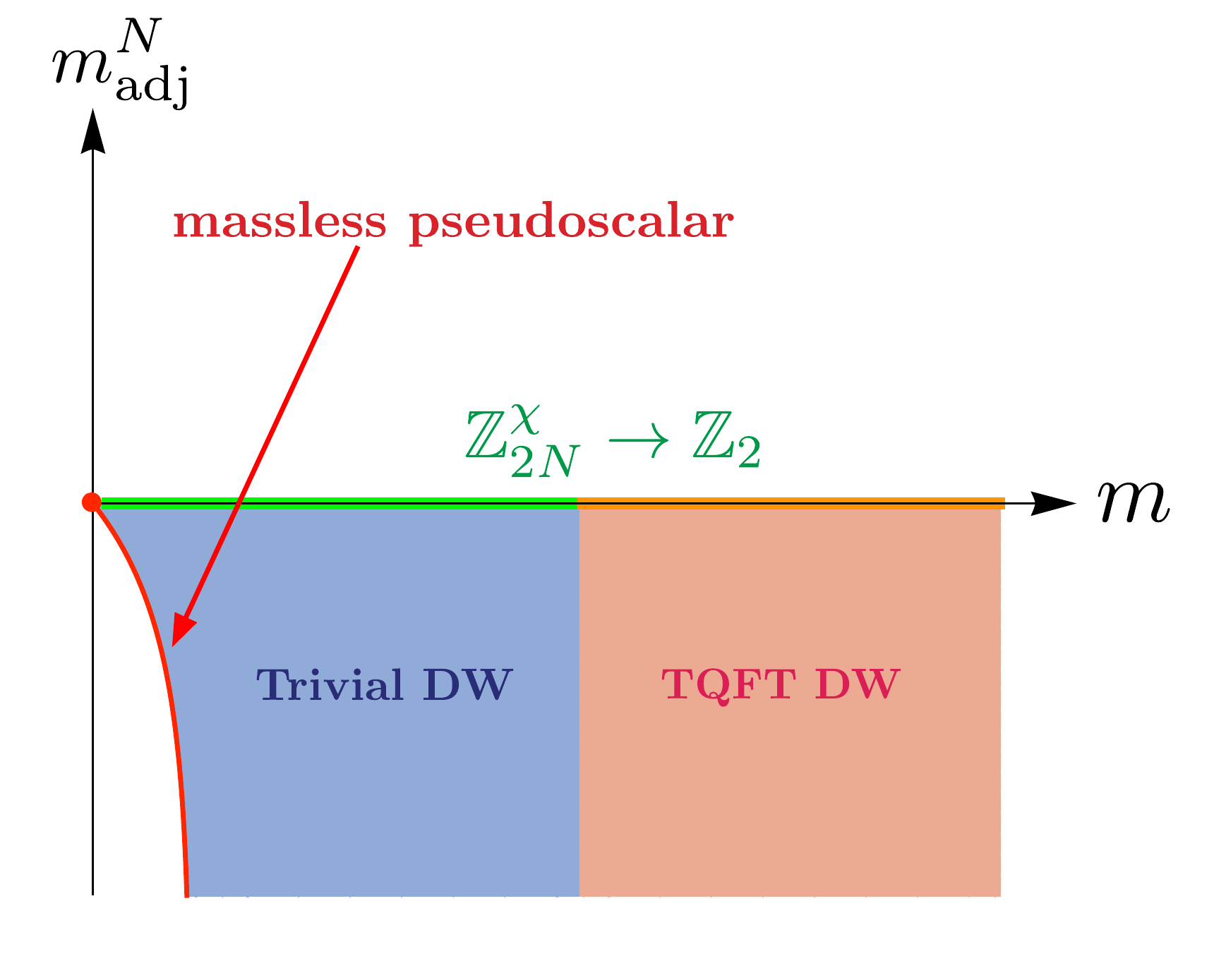} 
   \caption{The same as Figure~\ref{fig:phase_adj_fund} except that the phase transition on the domain wall is indicated. }
   \label{fig:phase_adj_fund_DW}
\end{figure}

The above makes it clear that while the domain walls support a TQFT for large fundamental mass $m$, no trace of a TQFT remains at small $m$, and $D_n$ domain walls disintegrate into $D_1$ domain walls. So while there is no phase transition in the bulk, a phase transition on the domain wall must occur at $m/\Lambda\sim 1$. The conservative scenario is that of a single transition, although we cannot exclude multiple transitions on the domain wall. Below, we elaborate on some possible scenarios of this sort. This discussion is continuously connected to that of a single-flavor QCD in \cite{Gaiotto:2017tne} (see also a related discussion in \cite{DiVecchia:2017xpu}), which is the limiting case $|m_{adj}|\gg \Lambda$ and the adjoint fermions decouple. We illustrate the phase diagram in Figure~\ref{fig:phase_adj_fund_DW}.

Note that this behavior is also expected from the general lore of the effective field theory (EFT). The potential of (\ref{eq:Sgoldstone}) is $V(\varphi)=m\Lambda^3 (\cos N\varphi-1)$. EFT description is robust as long as $\Delta V(\varphi)\ll \Lambda^4$. This is always true for small fluctuations in any of the vacua, which can be trivially checked by expanding about $\varphi=\frac{2\pi}{N}$ to find $\Delta V\sim m\Lambda^3(\Delta \varphi)^2$,  where $\Delta \varphi$ are the fluctuations near the vacuum.  Any fluctuations are small, i.e., $\Delta \varphi\ll 1$.  Thus, we trivially find $\Delta V\ll \Lambda^4$, even in the limit $m\sim \Lambda$. Now, consider large-field excursions as we traverse a domain wall and go from one vacuum to another. In this case $\Delta \varphi\sim 1$, and thus, we find $\Delta V\sim m\Lambda^3$. Then, for $m\ll \Lambda$, we still find $\Delta V\ll \Lambda^4$, and the effective field theory description is still robust; no rearrangement of degrees of freedom is needed on the wall to correct for anything. On the other hand, taking $m\sim \Lambda$, we find that a large-field excursion causes $\Delta V\sim\Lambda^4$. Now, the EFT description fails, and one expects some additional degrees of freedom on the domain walls to correct for the failure of the potential. Presumably, these degrees of freedom give rise to the TQFT or perhaps massless fermions in some intermediate phase (see below).

We can apply our general analysis above to the particular case of $N$ even, where there is a time-reversal preserving domain wall $D_{N/2}$, as $D_n\to D_{N-n}$ under time-reversal. In the limit of the large mass of the fundamental matter, the domain wall theory was conjectured to be the $T$-preserving $\mathcal N=1$ 3d Super Yang-Mills \cite{Delmastro:2020dkz}. When the mass of the fundamental matter is reduced in the bulk, it is natural to assume that the domain wall theory will be deformed by a massive fundamental multiplet.  Such theories were discussed in \cite{Lohitsiri:2022jyz}, where different scenarios of IR phases were considered. In particular, in the large-$N$ limit, it was argued that these 3d theories break the $T$-symmetry spontaneously. This is precisely what happens to the $T$-preserving domain wall $D_{N/2}$ of the effective Goldstone theory \eqref{eq:Sgoldstone}. Such a domain wall would correspond to a shift $\varphi\rightarrow \varphi+\pi$ as the domain wall is traversed. The domain wall $D_{N/2}$ breaks time-reversal invariance spontaneously, as $\varphi$ can wind forward or backwards\footnote{Recall that $\varphi$ is a pseudo-scalar so it reverses sign mod $2\pi$ under $T$.}. Now for $N>2$ and even, the $n=N/2$ domain wall is unstable for small $m$. Still, one could make it stable by explicitly breaking the $\mathbb Z_{2N}^\chi$ chiral symmetry down to $\mathbb Z_4$ by, for example, adding a term $\tr\lambda^{4}$ term\footnote{This operator is irrelevant. However, the fact that this deformation is relevant in the IR effective theory means that such a deformation is actually dangerously irrelevant in the UV. However, we cannot take the continuum limit with this term. Alternatively, we can consider a deformation by adding a real scalar $\phi$ and coupling it to the adjoint fermion as follows $i\phi(\tr \lambda\lambda+c.c.)$. Such a theory will retain the $\Z_4^\chi\subset \Z_{2N}^\chi$ chiral symmetry.}, and conclude that such a domain wall breaks time-reversal symmetry spontaneously.

\begin{figure}[t] %  figure placement: here, top, bottom, or page
   \centering
   \includegraphics[width=0.8\textwidth]{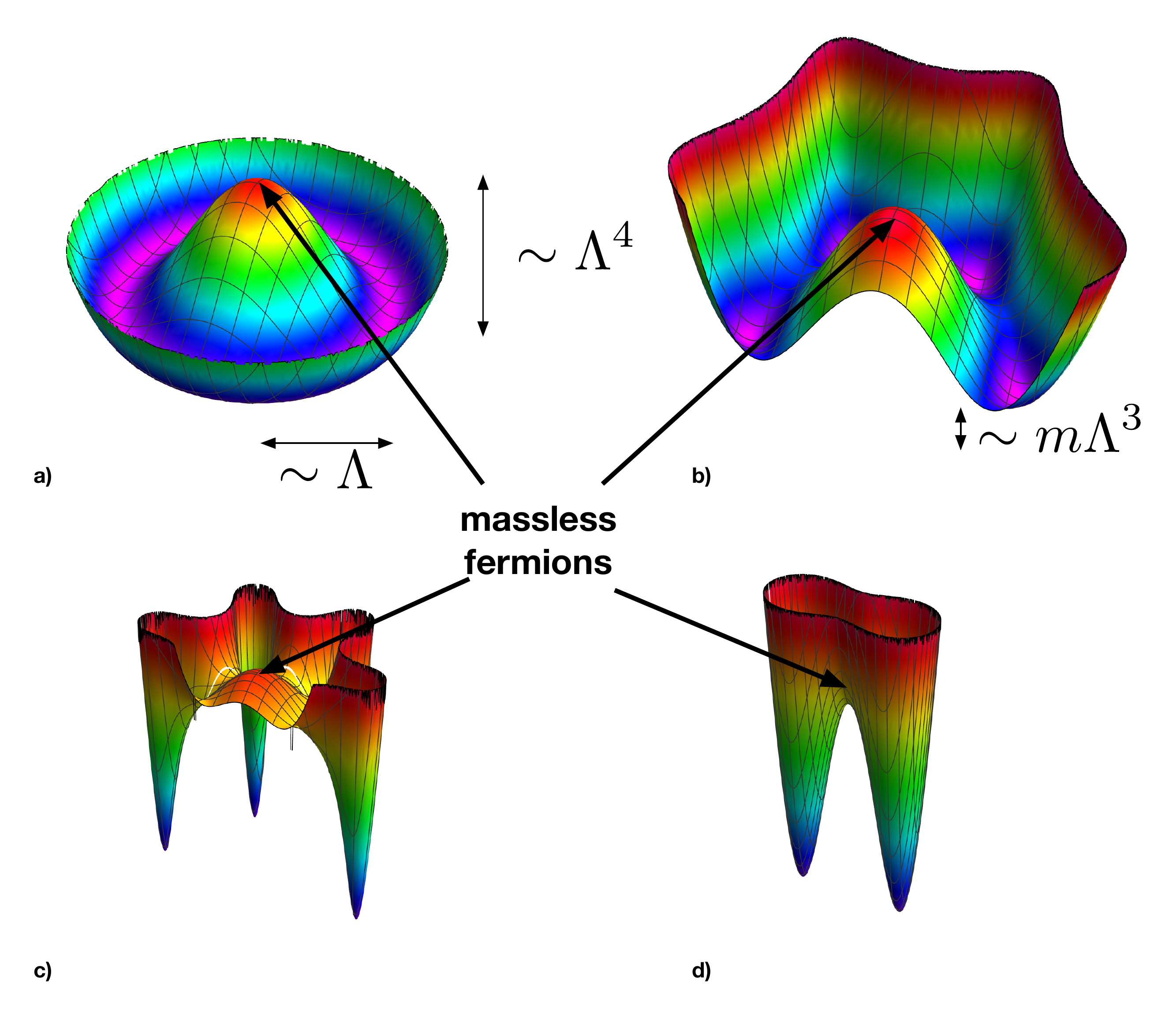} 
   \caption{Cartoons of vacua of the $\U_\chi$ order parameter $\phi$. (a) depicts the scenario of $m=0$, (b) depicts the scenario of $m\ll \Lambda$ with $N=6$, c) depicts a scenario with $m\sim \Lambda$ for $N=6$ and d) depicts a scenario with $m\sim \Lambda$ for $N=2$. Only (a) and (b) scenarios are under theoretical control, where (c) and (d) are speculative, intermediate phase descriptions of the theory where the hierarchy of scales is lost and reasoning is weak.  }
   \label{fig:vacua}
\end{figure}

As our analysis was done for small fundamental fermion masses, it does not a priori exclude an intermediate domain-wall phase. To gain insight, let us replace the model given by \eqref{eq:LGoldstone_fermions} and \eqref{eq:Sgoldstone} with a model that incorporates a full order parameter of the $\U_\chi$ breaking, namely a complex scalar $\phi$. The complex phase of $\phi$ can be identified with $\varphi$ in \eqref{eq:LGoldstone_fermions} and \eqref{eq:Sgoldstone}. To reproduce the anomaly, we must again couple fermions with terms like\footnote{We assume that $q_i+q_j>0$. If this is not the case,  the coupling can instead be written as $\sim\chi_i\chi_j (\phi)^{-q_i-q_j}$.} $\sim\chi_i\chi_j (\phi^*)^{q_i+q_j}$ where $q_{i}$ is the $\U_\chi$ charge of the fermion $\chi_i$, and with some sort of a Mexican hat potential $V(|\phi|)$, roughly of height $\Lambda^4$ and width $\Lambda$ . For the finite mass of the fundamental fermion $m$, we must deform the model with a term $\sim m \phi^N+c.c.$. The potential of the scalar for $m=0$ looks approximately as Figure~\ref{fig:vacua} (a), while a small mass theory $m\ll \Lambda$ looks like Figure~\ref{fig:vacua} (b).

At the origin $\phi=0$, there are massless fermions, which are gapped everywhere else. Importantly the $N$ vacua are separated by a very shallow barrier, controlled by the dimensionless parameter $m/\Lambda$. This parameter controls the tension of the elementary domain wall and can be made arbitrarily small. Further, the domain-wall $D_n$ with $n>1$ will not be stable, as the minimum energy configuration would prefer to go through elementary domain walls rather than through the high peak of order $\Lambda^4$ in the middle. In particular, the $T$-preserving domain wall of even $N$ theory will always prefer to surf through the shallow rim of the potential in this regime, thereby breaking $T$-symmetry. However, for $m\sim\Lambda$ one can imagine a potential like in Figure~\ref{fig:vacua}c) where the neighboring vacua are separated by the barrier of height $\Lambda^4$, which is of the same order as the barrier in the middle. Hence, it becomes a subtle issue whether the $T$-preserving domain wall prefers a direct, $T$-preserving route, where the mass of the fermions becomes zero, or the $T$-breaking route where the massless fermion point $\phi=0$ is avoided. The case $N=2$, however, is different as a natural domain wall would go through the $\phi=0$ point as depicted in Figure~\ref{fig:vacua} (d).

Nonetheless, it is important to note that when $m\sim \Lambda$, we have no reason to believe the effective description in terms of the would-be $\U_\chi$ order parameter, as this is reliable only for $m\ll\Lambda$. In other words, it could be that before the scenario of Figure~\ref{fig:vacua} (c) and (d), the composite fermions on the domain-wall restructure into quarks, as expected for $m\gg \Lambda$. Still, it is interesting to question whether there is an intermediate, massless composite phase for the domain wall theory.

\begin{figure}[htbp] %  figure placement: here, top, bottom, or page
   \centering
   \includegraphics[width=\textwidth]{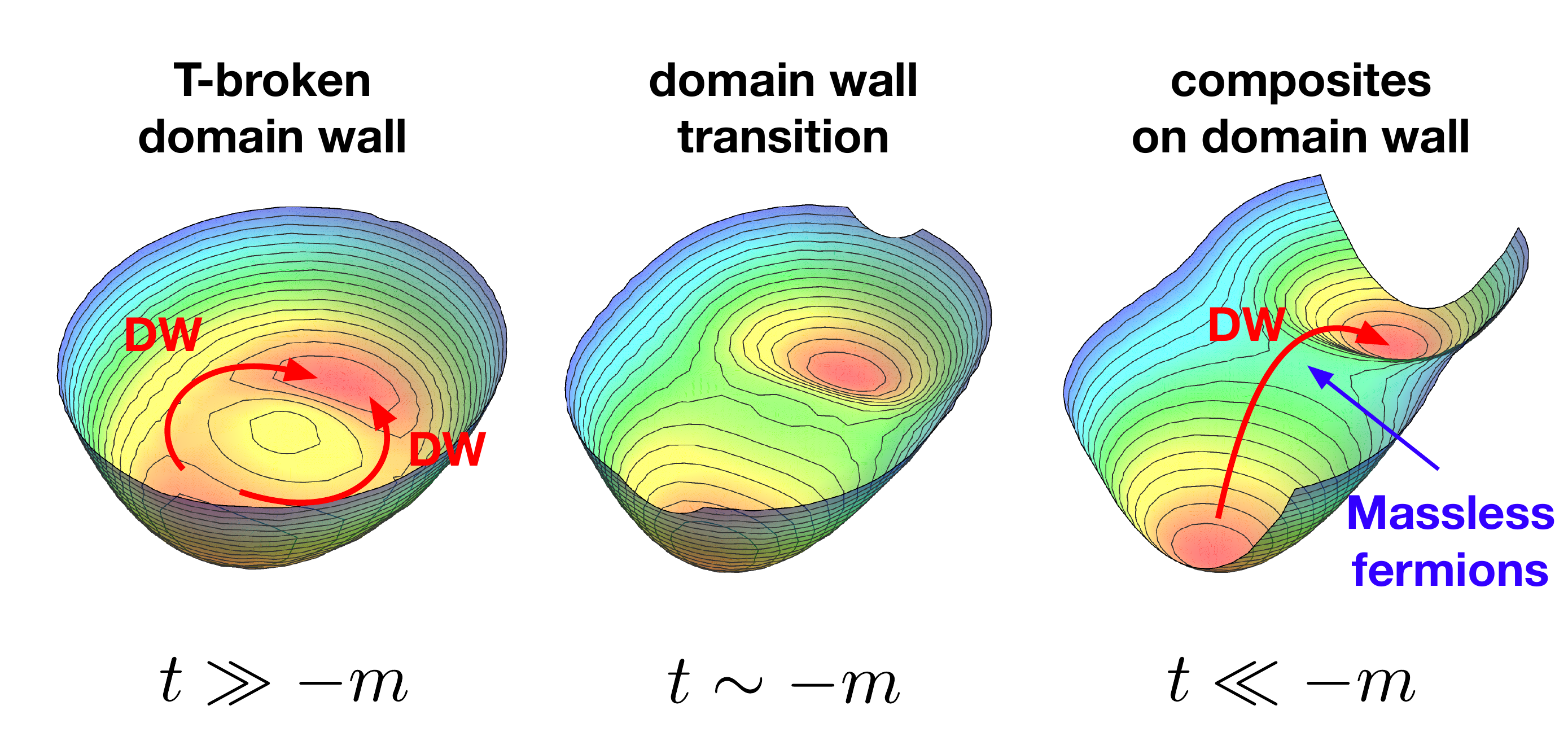} 
   \caption{The transition of the $\SU(2)$ fundamental Dirac/Higgs model. The plots show qualitatively the potential for the complex scalar  $\phi$ order parameter of the $\U_{\chi}$ symmetry. The potential is depicted for three different regimes as the mass of the fundamental fermions is increased from $m\ll -t$ to $m\gg-t$, where $t$ is a reduces mass-squared of the scalars. Increasing $m$ indicates a transition on the domain wall from a 2-vacuum, $T$-broken domain wall, to the unique domain wall with composite fermions.}
   \label{fig:SU2DWT}
\end{figure}
\begin{figure}[htbp] %  figure placement: here, top, bottom, or page
   \centering
   \includegraphics[width=\textwidth]{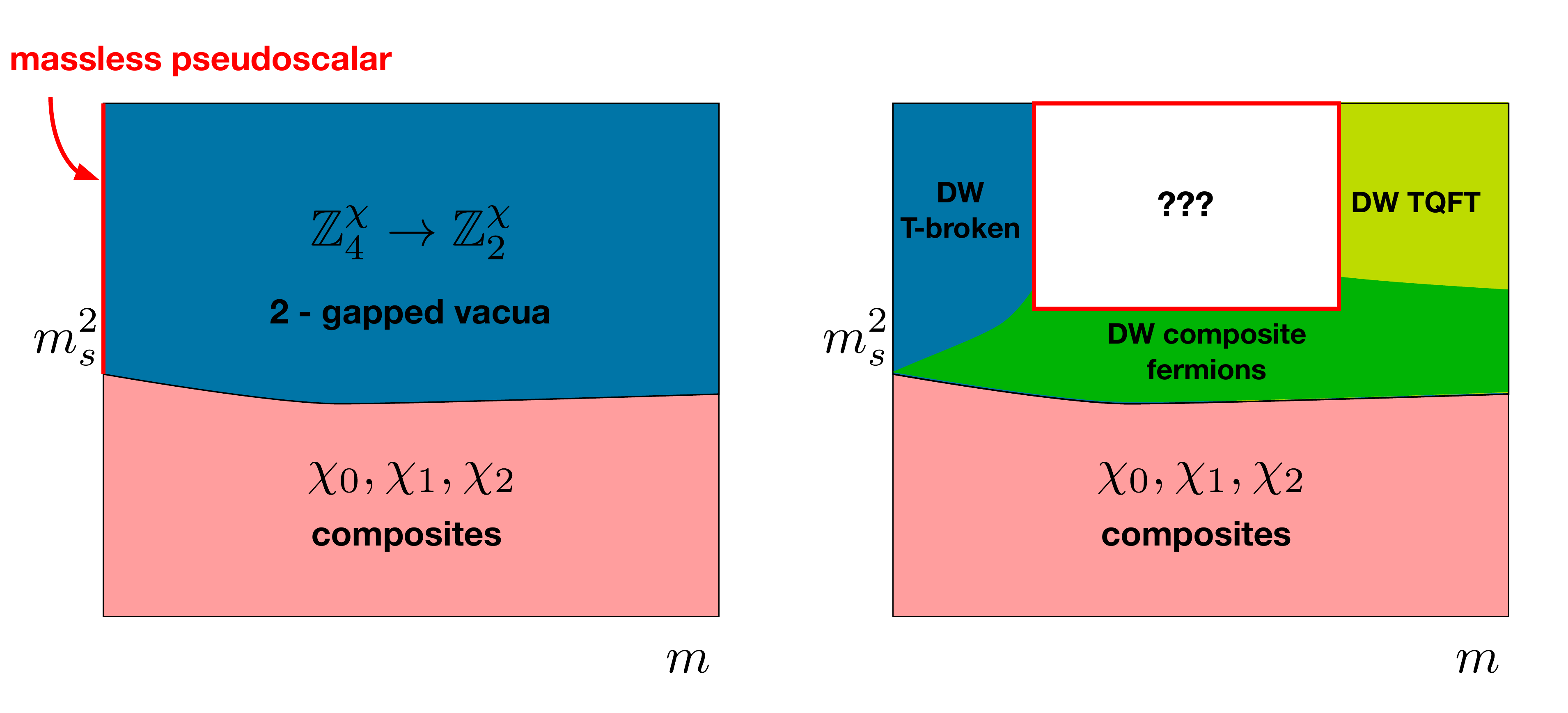} 
   \caption{The phase diagram of the one flavor of the fundamental Dirac fermion and one flavor of the fundamental scalar discussed at the end of \S~\ref{sec:warmup}. The horizontal axis depicts the mass $m$ of the fundamental Dirac fermion, while the vertical axis is the mass-squared $m_s^2$ of the fundamental scalar. The bulk transition is from a 2-vacuum chirally broken phase when $m_s^2$ is large and positive, to the composite fermion phase discussed in \S~\ref{sec:warmup}. Near this transition and near the point $m=0$, where the chiral symmetry enhances to $\U_{\chi}$ the theory can be studied using the effective theory of the complex scalar field order parameter $\phi$ of $\U_{\chi}$ breaking, coupled to the composite fermions $\chi_0,\chi_1,\chi_2$ as in \eqref{eq:eff_SU2_ferm_higgs}, which reveals two domain wall phases in the chiral symmetry broken bulk phase -- the T-broken phase and the composite fermion phase on the domain wall. An interesting question which we cannot answer is whether the fermionic domain wall phase persists as an intermediate phase as the scalars are decoupled.}
   \label{fig:fermion_higgs_phase_diag}
\end{figure}

Before we conclude this section, let us consider a model which indeed has a scenario described by Figure~\ref{fig:vacua}d) and is under full analytical control. To do this we will deform the $\SU(2)$ model with an adjoint Weyl and a fundamental Dirac fermion, by adding to it a scalar and introducing the Yukawa coupling exactly like we did in the second model described in \S~\ref{sec:warmup}. 

Let us label the bare mass of the scalar as $m_s$ in this model. Then as scalars are condensed we saw in \S~\ref{sec:warmup} that three free fermion phase develops $\chi_0,\chi_1$ and $\chi_2$, with charges exactly consistent with our proposed all-$N$ formulas in the $m\rightarrow 0$ limit (see discussion below \eqref{eq:U1_anom_condition}). When $m=0$ the transition from $m_s^2\gg\Lambda^2$ to $m_s^2\ll -\Lambda^2$ is changing from a $\U_\chi$ broken to a composite fermion phase. We can hence study the theory around $m_s^2=M_{c}^2$ -- the critical mass-squared. Introducing the parameter of mass dimension one $t=\frac{M_c^2-m_s^2}{\Lambda}$, the effective theory for $t,m\ll\Lambda$ is described by a complex scalar $\phi$ and three fermions $\chi_0,\chi_1,\chi_2$ with the following interactions
\begin{equation}\label{eq:eff_SU2_ferm_higgs}
\Delta\mathcal L_{eff}\sim (g_1\phi\chi_0\chi_1+g_2\frac{\phi^3}{\Lambda^2} \chi_2\chi_2+c.c.)+t \Lambda|\phi|^2+\lambda |\phi|^4+(m\Lambda \phi^2+c.c.)\;.
\end{equation}
with some couplings $g_1$ and $g_2$. The above Lagrangian is valid for small $|t|$ and $m$. As $t$ is increased from negative to positive values a restoration of the $\U_{\chi}$ symmetry ensues. However if $t$ is negative, depending on whether the $|t|\gg m$ or $|t|\ll m$  the effective potential for the complex scalar $m$ can look as in Figure~\ref{fig:SU2DWT}. So as $m$ is increased for fixed $t$, the domain wall undergoes a transition from a twice degenerate, $T$-broken phase to the composite fermion phase. Moreover, the effective theory still remains valid when $m$ is large\footnote{This reduces the model to the first model of \S~\ref{sec:warmup}, and the transition in question is the Higgsing transition.} in the sense that the order parameter becomes a real scalar $\phi$ with the $\Z_2$ symmetry  $\phi\rightarrow -\phi$, which is a subgroup of $\Z_4^\chi$. Since there are still many residual 't Hooft anomalies involving this chiral symmetry, one expects that the qualitative effective model remains, allowing us to study the transition in the bulk by demoting the field $\phi$ to a real scalar and dropping the last term of \eqref{eq:eff_SU2_ferm_higgs}. However here it is easy to see that the chiral-broken phase supports domain-wall fermions\footnote{This follows immediately from the reality of the order parameter $\phi$. In the phase where $\phi$ develops a vev, there are two vacua. The fermions therefore must become massless on the domain wall. }. This is nicely consistent with the study of an analogous 3d theory in \cite{Lohitsiri:2022jyz}. We summarise this model's bulk phase diagram and the domain wall phase diagram in Figure~\ref{fig:fermion_higgs_phase_diag}. We are unable to determine whether, in the scalar decoupling limit, the composite fermion phase on the domain wall persists. This is indicated by the question marks in the figure.

%%%%%%%%%%%%%%%%%%%%%%%%%%%%%
\subsection{Comments on the large-$N$ scaling}
\label{sec:large_N}
%%%%%%%%%%%%%%%%%%%%%%%%%%%%%%%

We will here briefly comment on the large $N$ scaling of the effective Goldstone theory \eqref{eq:Sgoldstone}. To take the large $N$ limit, we normalize the UV Lagrangian in a standard way
\begin{equation}
\mathcal{L}= \frac{N}{\lambda_{t}}\left(\frac{1}{2}\tr{|F|^2}+i\tr\bar\lambda D\lambda+i\bar\psi D\psi+i\bar{\tilde\psi} D\tilde\psi\right)
\end{equation}
where $\lambda_t$ is the 't Hooft coupling.
The current associated with the chiral symmetry is given by
\begin{equation}
j^\mu_\chi \propto N\tr\bar\lambda \bar\sigma^\mu\lambda-N^2\bar\psi\bar\sigma^\mu\psi-N^2\bar{\tilde\psi}\sigma^\mu \tilde\psi
\end{equation}
where the extra factor of $N$ in the last two terms are because the fields $\psi$ and $\tilde\psi$ are charged with charge $N$ under the $\U_\chi$. The current current correlator then has three pieces, which are schematically given by
\begin{equation}
\avg{j^\mu j^\nu}\propto A N^4\avg{J^\mu_\psi J^\nu_\psi}+BN^3 \avg{J^\mu_\psi J^\nu_\lambda}+CN^2\avg{J^\mu_\lambda J^\nu_\lambda}\;.
\end{equation}
where $J^\mu_{\lambda}=\tr \bar\lambda \bar\sigma^\mu\lambda$, $J^\mu_{\psi}=\bar\psi\bar\sigma^\mu\psi+\bar{\tilde \psi}\bar\sigma^\mu\tilde\psi$ and $A,B,C$ are some order one coefficients. We have that $\avg{J^\mu_\psi J^\nu_\psi},\avg{J^\mu_\psi J^\nu_\lambda}\sim 1/N$, $\avg{J^\mu_\lambda J^\nu_\lambda}\sim 1$. So despite the correlator in the first term above being suppressed by $1/N$, the enhancing factor of $N^4$ causes the first term to dominate, so
\begin{equation}
\avg{j^\mu j^\nu}\sim N^3\;.
\end{equation}
To reproduce this, we have to take that in the large $N$ limit the coefficient of the kinetic term $(\partial_\mu\varphi)^2$ of the effective Goldstone theory needs to scale as $N^3$.

Let us give another, more heuristic, way to understand this result. Namely if we note that there are two fermion-bilinear order parameters of $\U_\chi$ symmetry breaking $\tr\lambda\lambda \sim e^{i\varphi}= U_\lambda$ and $\psi\tilde\psi\sim e^{iN\varphi}=U_\psi$, we then expect that the effective Lagrangian is given by
\begin{equation}
N^2|\partial_\mu U_\lambda|^2+N|\partial_\mu U_\psi|^2\;.
\end{equation} 
In other words, the order parameter $U_\lambda$ will have an a priori dominant kinetic terms, while the $U_\psi$ is expected to be $1/N$ suppressed because it couples to fundamental matter. The second term however is not sub-leading and is of order $N^3$ because the order parameter $U_\psi$ carries charge $N$, i.e. $|\partial_\mu U_\psi|^2=N^2(\partial_\mu\varphi)^2$. So the second term above is dominant and is given by $\sim N^3(\partial_\mu\varphi)^2$ in the large $N$ limit.

It is perhaps not surprising that the fundamental contribution is important even at leading order, because the $\U_\chi$ symmetry crucially depends on the presence of massless fundamentals. Indeed the way that importance shows up in the large $N$ limit is through the dominance of the charge of the fundamentals under the $\U_\chi$.

Now if one inserts the mass terms for the adjoint fields and the fundamental fields, we expect them to scale as $N^2$ and $N$ respectively, i.e. the effective theory schematically becomes
\begin{equation}
\mathcal{L} \propto N^3(\partial_\mu\varphi)^2-m_{adj}\Lambda N^2\cos\varphi-m\Lambda N\cos(N\varphi)\;.
\end{equation}
Notice that if we set the fundamental mass to zero, i.e. $m=0$, then the pseudo-Goldstone has a mass $1/N$ allowing an analysis to be carried out for large $N$ at any $m_{adj}$. This is exactly what one expects when the adjoint fermion decouples. On the other hand setting $m_{adj}=0$, the pseudo-Goldstone mode gets an order one mass in the large $N$ limit, and the analysis breaks down when $m$ is of order $\Lambda$--the strong scale.

\section{Theory with one fundamental scalar}
\label{sec:one-fund-scal}

Now we replace the fundamental fermion with a fundamental scalar 
$\phi$. The action for the matter content is given by
\begin{equation}
S_{\text{matter}} = \int \dd^4 x\, \left(  i \bar{\lambda} \left( \slashed{\partial} - i \slashed{a}_{adj} \right)\lambda + \abs{(\partial- i a) \phi}^2 + V(\phi) \right)\;,
  \label{eq:S-1-scalar}
\end{equation}
where $V(\phi)$ is the potential for the scalar field, which we can take to be of the form
\begin{equation}
V(\phi) = m^2 \abs{\phi}^2 + \mathcal{O} (\abs{\phi}^4)\;.
\end{equation}
where we will not be concerned with interaction terms much, as long as they are there to stabilize the potential when $m^2<0$. 

\subsection{Symmetry and anomalies}
\label{sec:symmetry-anomalies-one-scalar}

Similar to the massive fundamental fermion case, the faithful global symmetry group is
\begin{equation}
  \begin{split}
    G^{\text{Global}} &= \frac{\U_q}{\Z_N}\times  \Z_{2N}^{\chi}\;,\\
    &= \U_B \times \Z_{2N}^{\chi}\;.
  \end{split}
\end{equation}
where the representation of each field under $G^{\text{Global}}$ is given in Table
\ref{tab:scalar-adj-reps}. Here, we again denote the global baryon symmetry as
$\U_B\cong\U_q/\Z_N $, in terms of the quark symmetry $\U_q$. The lowest charge of a gauge invariant operator under $\U_B$ is unity, while under $\U_q$ it is necessarily a multiple of $N$ so that $\U_B$ is the faithful global symmetry. The (gauge non-invariant) field $\phi$, however, is fractionally charged under $\U_B$.
\begin{table}[h]
  \centering
  \begin{tabular}{c||c|cc}
    & $\SU(N)$ & $\U_q$ & $\Z_{2N}^{\chi}$\\
    \hline
    $\phi$ & ${\tiny\yng(1)}$ & $+1$ & $0$ \\
    $\lambda$ & $\textbf{adj}$ & $0$ & $+1$
  \end{tabular}
  \caption{Representations of the scalar field and the adjoint fermion
    under the gauge and global symmetry groups}
  \label{tab:scalar-adj-reps}
\end{table}

Just like in the fermionic case, there is a mixed anomaly between
$\Z_{2N}^{\chi}$ and $\U_B$  such that the $\Z_2$ subgroup of $\Z_{2N}^{\chi}$
remains anomaly-free. This anomaly can also be seen as follows. We turn on a fractional instanton color flux $Q_c\in \mathbb Z/N$. One also must turn on the fractional baryon-number flux to render the theory well-defined in the presence of scalars, which see both the color and baryon-number fluxes.  The adjoint fermions, however, are uncharged under $U(1)_B$. Thus, in the background of the fractional color flux, the partition function transforms as 
\begin{eqnarray}
{\cal Z}\xrightarrow{\mathbb Z_{2N}^{d\chi}}{\cal Z}\exp\left[\frac{i2\pi}{2N}(2N Q_c)\right]={\cal Z}\exp\left[i\frac{2\pi }{N}\right]
\label{the Z2N mixed anomaly with scalars}
\end{eqnarray}
under a discrete chiral rotation. The phase is the above-mentioned mixed anomaly.

 The theory also exhibits a mixed $\mathbb Z_{2N}^{\chi}$-gravitational anomaly.
Both $\mathbb Z_{2N}^{\chi}$-$\U_B$ and $\mathbb Z_{2N}^{\chi}$-gravitational anomalies are identical to the anomalies of the theory with a massive fundamental fermion. This is expected since massive fermions and scalars carry the same global charges under $\U_B$.

%%%%%%%%%%%%%%%%%%%%
\subsection{The IR phases}
\label{sec:one-scalar-IR-phase}
%%%%%%%%%%%%%%%%%%%%

We already discussed the phase structure of the $N=2$ case in \S~\ref{sec:warmup}, so here we restrict ourselves to $N>2$.

{\bf (I)} When $m^2$ is positive and much larger than $\Lambda^2$, the scalar decouples, and the theory is pure super Yang-Mills with $N$ degenerate vacua.

{\bf (II)} When $m^2$ is negative and large, $\phi$ will condense and acquire a vacuum
expectation value\footnote{This is an abuse of notation since this consideration
  is not gauge invariant. Instead, one should talk about fixing a gauge 
  to be more precise.}. Assuming that this happens at a scale above the
strong scale of $\SU(N)$, the condensation will higgs the gauge group $\SU(N)$
down to $\SU(N-1)$,
under which the adjoint fermion of $\SU(N)$ decomposes as
\begin{equation}
    \textbf{adj}_{\SU(N)} \qquad = \qquad \textbf{adj}_{\SU(N-1)} \oplus \textbf{N-1} \oplus \overline{\textbf{N-1}} \oplus \textbf{1}\,.
\end{equation}
In other words, the Higgs regime is effectively described by an $\SU(N-1)$ gauge theory with one Weyl fermion $\tilde\lambda$ in the adjoint of $\SU(N-1)$,  two Weyl fermions $\psi,\tilde\psi$ in the fundamental and anti-fundamental of $\SU(N-1)$, and one neutral Weyl fermion $\nu$.
%\footnote{The operators $\psi,\tilde\psi$  and $\nu$ in the effective $\SU(N-1)$ gauge theory can be constructed as follows $\psi\propto \lambda\phi, \tilde\psi\propto \phi^\dagger\lambda$ and $\nu\propto \phi^\dagger \lambda \phi$ where we view $\lambda$ as a matrix with a fundamental and an anti-fundamental index of $\SU(N)$. }. 

It will be beneficial to discuss the deep Higgs regime $m^2\rightarrow -\infty$ separately from the Higgs regime where $m^2$ is merely large and negative. We will see that finite Higgs VEV will induce dangerously irrelevant terms, which will change the IR physics.

 \underline {In the deep Higgs regime} with $m^2\rightarrow -\infty$, the Higgs field
fluctuation is suppressed together with the massive gauge
bosons which now have infinite masses. Concretely, by gauge-fixing so that the VEV of $\phi$ takes the form
\begin{equation}
\expval{\phi} =
\begin{pmatrix}
 v \\ 0 \\ \vdots \\ 0 
\end{pmatrix}\,,
\end{equation}
the decomposition of the adjoint fermion $\lambda$ is given by
\begin{equation}
  \label{eq:adj-decomp}
\lambda =
\begin{pmatrix}
  \nu & \tilde{\psi} \\ \psi & \hspace{1em}\tilde{\lambda}-\frac{\nu}{N-1}\mathbb I_{N-1}
\end{pmatrix}\,,
\end{equation}
where $\mathbb I_{N-1}$ is an $(N-1)\times (N-1)$ identity matrix.
Note that mass terms  $\tilde{\lambda} \tilde{\lambda}$, $\tilde{\psi}\psi$, and $\nu\nu$ are forbidden as $\tilde{\lambda}$, $\psi,\tilde\psi$ and $\nu$ are charged under the original global $\mathbb Z_{2N}^\chi$ symmetry. Thus, the IR theory is fully equivalent to the $\SU(N-1)$ gauge theory with a massless fundamental
Dirac fermion and a massless adjoint Weyl fermion discussed in
\S~\ref{sec:one-Dirac-massless}, together with a decoupled Weyl
fermion $\nu$. The global symmetry enhances in the IR to
\begin{equation}\label{eq:Gglobal_higgsed}
  \begin{split}
    G^{\text{Global}}_{\text{IR}} &= \frac{\U_q^{\prime}}{\Z_{N-1}} \times \U_\chi  \times \U_{\nu}\\
    &= \U_B^\prime\times \U_\chi \times \U_{\nu}
  \end{split}
\end{equation}
where $\U_B^{\prime}$ is also identified with a combination of $\U_B$ and
$\SU(N)$ that leaves the Higgs VEV invariant.  The charge assignment is given in Table \ref{tab:higgs_charge_assigment}. There are many anomalies as discussed
in \S~\ref{sec:one-Dirac-massless}.  Thus, the effective theory of the
deep Higgs regime is just a Goldstone theory of a spontaneously broken
$\U_\chi$, along with a decoupled $\nu$, i.e.
\begin{equation}
\mathcal L_{eff}= N^3\Lambda^2 (\partial\varphi)^2+N\bar\nu i\bar\sigma^\mu\partial_\mu \nu\;.
\end{equation}
Now, let us discuss the theory when $m^2$ is large and negative but finite.

% Requires the booktabs if the memoir class is not being used
\begin{table}[htbp]
   \centering
   %\topcaption{Table captions are better up top} % requires the topcapt package
   \begin{tabular}{@{} lccc @{}} % Column formatting, @{} suppresses leading/trailing space
      \toprule
      \cmidrule(r){1-2} % Partial rule. (r) trims the line a little bit on the right; (l) & (lr) also possible
		      		& $\U'_q$ & $\U_{\chi}$ & $\U_\nu$\\
      \midrule
      $\psi$      		& $1$ 		& $1-N$   & $0$ 		\\
       $\tilde\psi$         	& $-1$     	&  $1-N$  &	$0$	\\
      $\tilde\lambda$      	& $0$  		& $1$ 	    &	$0$	\\
      $\nu$       		& $0$		& $0$ 	    &	$1$	\\
      \bottomrule
   \end{tabular}
   \caption{The assignment of charges to the degrees of freedom between the Higgs $v$ scale and the strong scale $\Lambda$.}
   \label{tab:higgs_charge_assigment}
\end{table}
\begin{figure}[htbp] %  figure placement: here, top, bottom, or page
   \centering
   \includegraphics[width=4in]{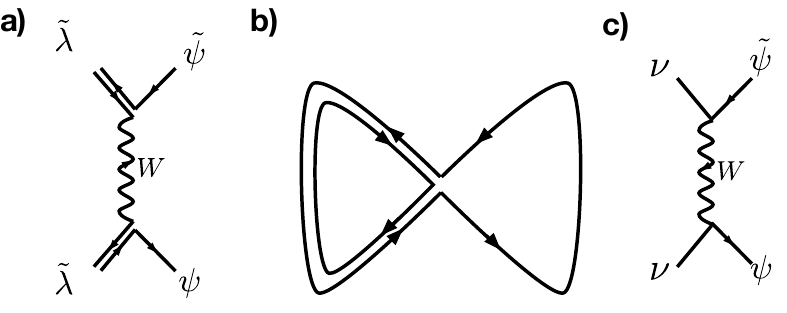} 
   \caption{a) The leading tree-level diagram generating the $N\tilde\psi(\tilde\lambda\tilde\lambda)^\dagger\psi$ vertex. b) The large $N$ scaling of the term $N\tilde\psi(\tilde\lambda\tilde\lambda)^\dagger\psi\sim Ne^{iN\varphi}$. The 4-fermi vertex scales as $N$, while the adjoint and fundamental propagators are $1/N$ suppressed, whose contribution is cancelled by the 2 color loops. c) The leading tree-level diagram contributing to the mass of $\nu$.}
   \label{fig:4-fermi-vertex}
\end{figure}

\begin{figure}[h!]
  \centering
  \includegraphics[width=\textwidth]{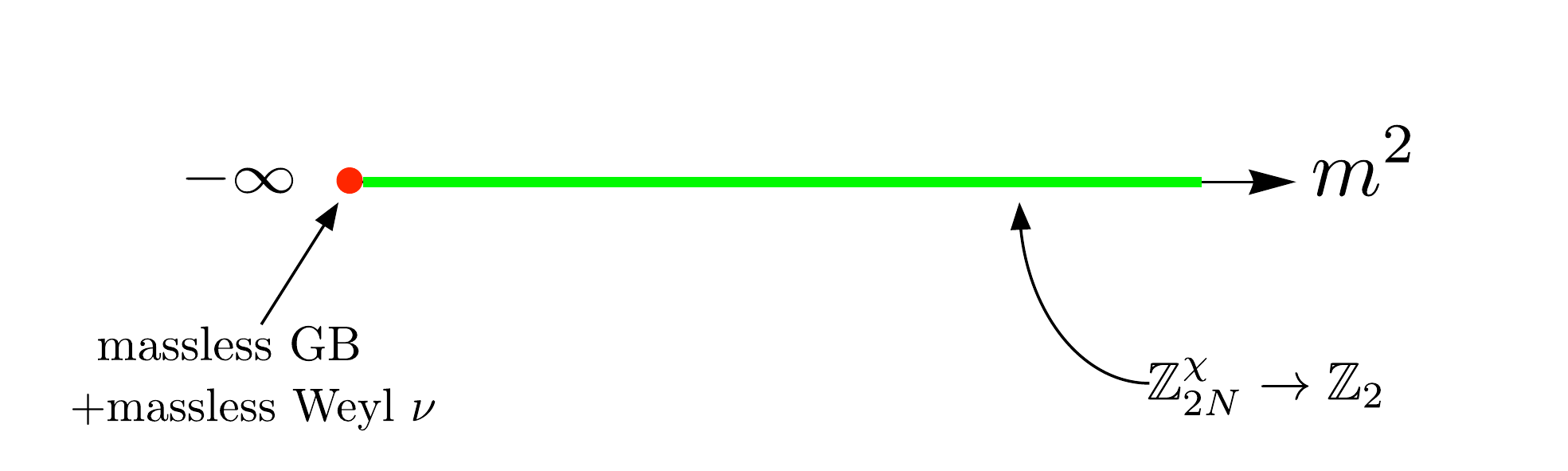}
  \caption{The phase diagram of $\SU(N)$ QCD (f/adj) with one fundamental scalar flavor as $m^2$ is varied from $-\infty$ to $+\infty$.% {\bf(a)} 
  The IR phase consists of $N$ degenerate vacua at all $m^2$ except in the deep Higgs regime where there is a massless Goldstone boson and a massless Weyl fermion $\nu$.  }
  \label{fig:phase_1scalar}
\end{figure}

 \underline{When $-\infty < m^2 <-\Lambda^2$} so that the scalar VEV is $v\gg \Lambda$. Naively not much changes, as the masses for $\psi,\tilde\psi,\nu$ and $\tilde\lambda$ are still forbidden because mass terms are not invariant under the $\mathbb Z_{N}$ subgroup of the global $\mathbb Z_{2N}^\chi$. However, higher fermi interactions will generally be induced, although suppressed by powers of $1/v$. Since higher fermion terms are naively irrelevant, one may erroneously conclude that they can be neglected. But such terms are symmetry breaking terms reducing $\U_\chi\times\U_\nu\rightarrow \mathbb Z_{2N}$, they will be relevant in the IR regime of the Goldstone boson. Such perturbations are called \emph{dangerously irrelevant}. In particular the lowest fermion operator which does this is $\tilde\psi\tilde\lambda^\dagger \tilde\lambda^\dagger\psi \sim e^{iN\varphi}$. Such a term is invariant under the $\mathbb Z_{2N}^\chi$ symmetry and is therefore allowed. The term is generated with a factor of $N$ in the large-$N$ limit. This scaling comes from considering the single $W$-boson exchange shown in Fig.~\ref{fig:4-fermi-vertex}a), where the two vertices contribute $N^2$ and the $W$-boson propagator contributes $1/N$. Further a vertex $N\nu\nu (\tilde\psi\psi)^\dagger$ is generated through the diagram of Fig.~\ref{fig:4-fermi-vertex}c). Since $(\tilde\psi \psi)^\dagger\sim e^{i(N-1)\varphi}$ is also generated, and it induces a mass term for $
 \nu$:\footnote{The large $N$ scaling can be deduced from the Fig.~\ref{fig:4-fermi-vertex}b) which, before contracting $\tilde\psi$ and $\psi$, is  of order $\sim N$. Contracting $\tilde\psi$ and $\psi$ is order $\sim 1$ because the $1/N$ from the propagator will cancel $N$ from the color loop. } $N\nu\nu e^{i(N-1)\varphi}$, so the effective theory becomes%\footnote{Note that this Lagrangian breaks down in the presence of the $\varphi$-vortex, because the vortex has an anomaly inflow. }
 \begin{equation}\label{eq:1scalar-DW-theory}
\mathcal L_{eff}=N^3\Lambda^2(\partial\varphi)^2-AN\Lambda^4\left(\frac{\Lambda}{v}\right)^2\cos(N\varphi)+N\bar\nu i\slashed\partial\nu+BN\Lambda\left(\frac{\Lambda}{v}\right)^2(\nu\nu e^{i(N-1)\varphi}+c.c.)\,.
 \end{equation}
The constants $A$ and $B$ are dimensionless numbers of order $1$. The suppression by $1/v^2$ of the $\cos(N\varphi)$ term and the $\nu\nu$ term comes from the fact that these descend from the 4-fermi vertices generated at the Higgs scale, which must be suppressed by $1/v^2$ on dimensional grounds. We plot the phase diagram in Figure~\ref{fig:phase_1scalar}.

Let's now give a non-zero mass $m_{adj}$ to the adjoint fermion $\lambda$. The now-physical $\theta$-angle can be absorbed as a phase of the Majorana mass: $m_{adj}= \abs{m_{adj}}e^{i\theta/N}$. For $\abs{m_{adj}}\gg \Lambda$ at a generic $\theta$, there is no $T$-symmetry and after we integrate out the massive adjoint fermion the theory is an $\SU(N)$ gauge theory with the fundamental Higgs field and a $\theta$-angle. When $m^2 \gg \Lambda^2$, we can also integrate out the Higgs, landing on a pure $\SU(N)$ Yang--Mills with the $\theta$-angle. Instead, if we tune $m^2\ll -\Lambda^2$ so that the Higgs condenses, the theory flows to a pure $\SU(N-1)$ gauge theory with the $\theta$-angle. Since both $\SU(N)$ and $\SU(N-1)$ gauge theories are believed to be in the same phase for any $\theta$-angle\footnote{There is a proposal, however that $\SU(2)$ gauge theory may be massless at $\theta=\pi$. See \cite{Gaiotto:2017yup} and \cite{Kitano:2020mfk}. }, it is natural to assume that there is no bulk phase transition as the Higgs condenses.

In the opposite limit when $\abs{m_{adj}}\ll \Lambda$, we can analyse the theory around the massless adjoint regime \eqref{eq:1scalar-DW-theory}. The adjoint mass term $m_{adj} \tr \lambda\lambda +c.c.$ descends to $m_{adj} \tr \tilde{\lambda} \tilde{\lambda} + m_{adj} \tilde{\psi}\psi +m_{adj}\nu\nu+ c.c.$, which induce the term proportional to 
\begin{multline}
m_{adj}\lambda\lambda+c.c.\sim -N^2\abs{m_{adj}} \Lambda^3 (\cos \left( \varphi+\theta/N \right)+C/N\cos(\varphi(1-N)+\theta/N))\\+ND|m_{adj}|e^{i\frac{\theta}{N}}\nu\nu+c.c.
\end{multline}
in the effective action \eqref{eq:1scalar-DW-theory}, where $C,D$ are order $1$ dimensionless numbers. Shifting $\varphi \to \varphi-\theta/N$, the potential for $\varphi$ takes the form
\begin{equation}
V(\varphi) \sim \Lambda^4N^2\left[-\frac{\abs{m_{adj}}}{\Lambda} \left(\cos \varphi+\frac{C}{N}\cos((1-N)\varphi+\theta)\right) - \frac{D}{N} \left( \frac{\Lambda}{v} \right)^{2} \cos \left(N \varphi - \theta\right)\right]  
\end{equation}
while mass term for $\nu$ takes on the form
\begin{equation}
\mathcal L_{m_{\nu}}=N\nu\nu\left(E\abs{m_{adj}}+\Lambda\left(\frac{\Lambda}{v}\right)^2e^{i(N-1)\varphi-i\theta}\right)+c.c.
\end{equation}
for some positive numerical constants $C$ and $D$. The relative signs between the three terms in $V(\varphi)$ is fixed by the expectation that the decoupling limit of the adjoint fermions results in a trivially gapped phase if $\theta=0$. 

At $\theta=\pi$, however, the local minima of $V(\varphi)$ for ${m_{adj}}=0$ are instead at $\varphi=\pm(2k+1)\pi/N$ for $k=0,1,2,\ldots,N-1 $. As we turn on a non-vanishing $\abs{m_{adj}}\ll \Lambda$, most of the degeneracy is lifted except two minima near $\varphi\pm \pi/N$, which generically break the $T$ symmetry. This is what we expect in the decoupling limit of the adjoint fermion at $\theta=\pi$, as the Higgs regime in that case is a pure $\SU(N-1)$ gauge theory at $\theta=\pi$.

\begin{figure}[h!]
   \centering
  \includegraphics[width=0.7\textwidth]{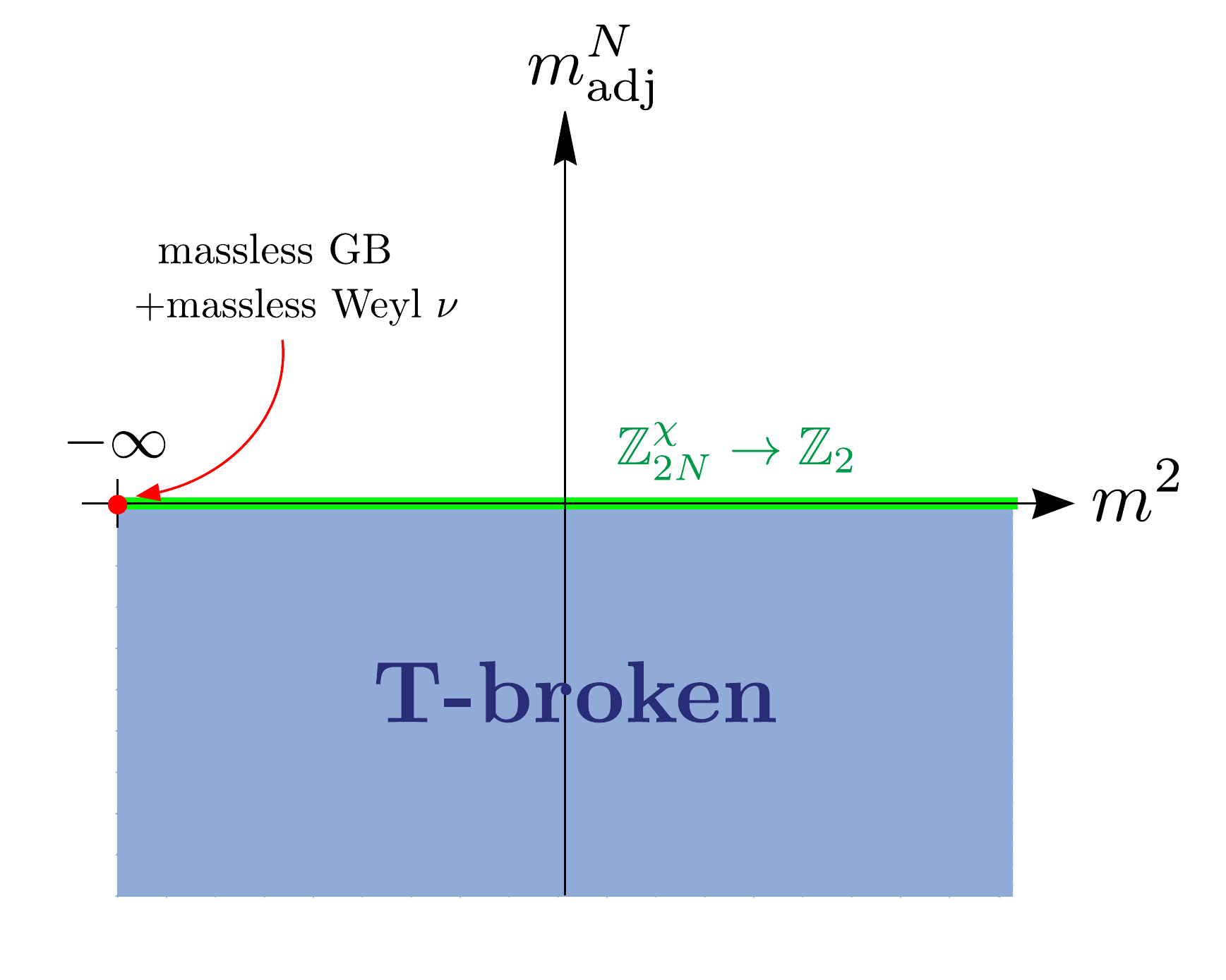}
  \caption{The $T$-symmetric slice of the phase diagram of $\SU(N)$ QCD (f/adj) with one fundamental scalar as both the scalar mass squared $m^2$ and the adjoint fermion mass $m_{adj}$ are varied. When $m_{adj}=0$, the diagram reduces to Figure~\ref{fig:phase_1scalar}. When $m_{adj} \neq 0$, the theory is mostly in the trivially gapped phase, except at  $ \theta= \arg m_{adj}^N = \pi$, where it can be in a $T$-broken phase shown as a blue sheet in the diagram.\label{fig:1scalar-madj-diag}}
\end{figure}

Assembling the IR phases from different corners of the parameter space results in a phase diagram that likely looks like the one shown in Figure~\ref{fig:1scalar-madj-diag}. Observe the similarities and differences to the phase diagram of $\SU(N)$ QCD (f/adj) with one fundamental Dirac fermion in Figure~\ref{fig:phase_adj_fund}: here, unlike in the fundamental Dirac version, the IR is always in the $T$-broken phase as $\abs{m_{adj}} \to \infty$ at $\theta=\pi$ regardless of $m^2$. Moreover, the massless Goldstone boson phase in the deep Higgs regime is always accompanied by a massless Weyl fermion $\nu$, unlike in Figure~\ref{fig:phase_adj_fund} at $m=0$ where there is no additional massless particle.

Like what we found in the fundamental fermion case, even though there
is no phase transition in the bulk over a range of parameters where
domain walls exist, there can be a phase transition on the domain
walls themselves. When the adjoint mass vanishes, there is a phase
transition on the domain wall joining adjacent $\Z_{2N}\to \Z_2$
vacua. As $m^2\to \infty$, we again recover the $\mathcal{N}=1$ SYM, and the
domain wall must also be decorated with a TQFT. On the other hand, as
$m^2\ll -\Lambda^2$, we see that the domain wall can be effectively described by
a chiral Lagrangian, indicating that the domain wall theory is
trivial. Thus, there must be a phase transition somewhere in the
middle as we vary the parameter $m^2$ between the two extremes.

\begin{figure}[h]
  \centering
  \includegraphics[width=0.7\textwidth]{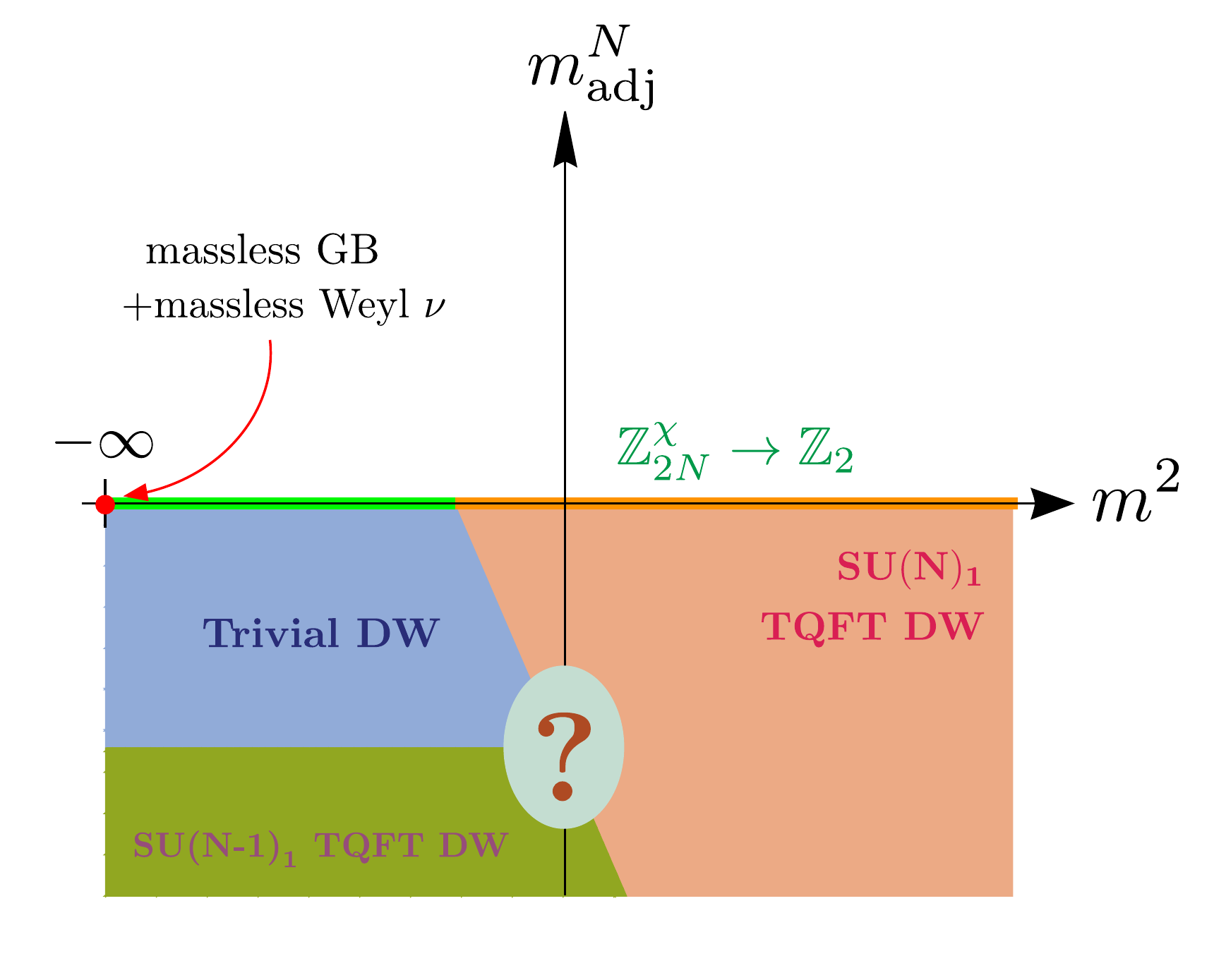}
  \caption{The same as Figure~\ref{fig:1scalar-madj-diag}, but also showing tentative phase transitions on the domain walls. Intermediate regions cannot be reliably analysed and we indicate this by a question mark.}
  \label{fig:1scalar-madj-diag-DW}
\end{figure}

Now, take the adjoint fermion to be so massive it decouples from the
theory at $\theta=\pi$, where $T$-symmetry is spontaneously broken. As
$m^2\to \infty$, the theory becomes pure $\SU(N)$ Yang--Mills at
$\theta=\pi$, whose $T$-breaking domain wall is equipped with the
$\SU(N)_1$ Chern-Simons TQFT \cite{Gaiotto:2017yup}. In the opposite
limit, the scalar field condenses and higgses the gauge group down to
the pure $\SU(N-1)$ Yang--Mills at $\theta=\pi$, whose domain wall theory is
now the $\SU(N-1)_1$ Chern-Simons TQFT, which is different from the
$\SU(N)_1$ theory. Again, we have an indication that there must be a
phase transition between these two limits. Together with the trivial
domain wall phase close to the $m_{adj}=0$ axis, these 3
possible phases are shown tentatively in
Figure~\ref{fig:1scalar-madj-diag-DW}.

%%%%%%%%%%%%%%%%%%%%%%%%%%%%%%%
\section{Theory with multiple fundamental fermions}
\label{sec:multi-flavor-fermion}
%%%%%%%%%%%%%%%%%%%%%%%%%%%%%%%%%

We consider $\SU(N)$ Yang-Mills theory endowed with a single adjoint
Weyl $\lambda$ and $N_f$ fundamental Dirac fermions $\Psi_i$,
$i=1,\ldots, N_f$, all with the same mass $m$.  The Lagrangian for the
matter sector is given by
\begin{equation}
  S = \int \dd^4x\, \left( i \bar{\lambda}(\slashed{\partial} - i \slashed{a}_{adj})\lambda +  i \sum_{i=1}^{N_f}\bar{\Psi}_i (\slashed{\partial}- i \slashed{a} -m)\Psi_i\right)
\end{equation}

\subsection{Symmetry and anomalies}
\label{sec:symmetry-anomalies-mult-ferm}

\subsubsection*{The massive case}
\label{sec:mult-flav-ferm-massive}

When $m\neq 0$, the faithfully acting symmetry group is
\begin{equation}
  \begin{split}
    G^{\text{Global}} &= \frac{\SU(N_f)\times \U_q}{\Z_{N_f}\times \Z_N} \times \Z_{2N}^{\chi} \\
    &\cong \frac{\UU(N_f)}{\Z_N} \times \Z_{2N}^{\chi}\;,
  \end{split}
\end{equation}
The only addition here compared to the symmetry group of
\S~\ref{sec:one-Dirac-massive} is the $\SU(N_f)$ factor. The matter
fields transform under $G^{\text{Global}}$ in the representations
given in Table \ref{tab:multi massive-fermion-reps} , where we
decompose the Dirac fermion into $2$ left-handed Weyl fermions $\psi$ and
$\tilde \psi$ in the fundamental and anti-fundamental representations of
$\SU(N)$, respectively. The extra quotient by $\Z_{N_f}$ arises
because $U = \ee^{2\pi i k/N_f}\textbf{1}_{N_f}$,
$k=0,1,\ldots, N_f-1$, in the centre of $\SU(N_f)$ is the same as
transforming $\Psi_i$ by $\ee^{2\pi i k/N_f}\in \U_B$. It is also worth
pointing out that, just like in the one-flavor case, the $\Z_2$
subgroup of $\Z_{2N}^{\chi}$ acts identically as $\Z_2^F$, allowing us to
define the theory on a non-spin manifold that admits a
$\Spin$-$\Z_{2N}^{\chi}$ structure.
\begin{table}[h]
  \centering
  \begin{tabular}{c||c|c|c|c}
    & $\SU(N)$ & $\SU(N_f)$  & $\U_q$ & $\Z_{2N}^{\chi}$\\
    \hline
    $\psi$ & ${\tiny\yng(1)}$ & ${\tiny\yng(1)}$ &  $+1$ & $0$ \\
    $\tilde{\psi}$ & ${\tiny \overline{\yng(1)}}$  & ${\tiny \overline{\yng(1)}}$ & $-1$ & $0$ \\
    $\lambda$ & $\textbf{adj}$ & $\textbf{1}$ &  $0$ & $1$
  \end{tabular}
  \caption{Action of gauge and global symmetries in the multi-flavor
    massive $\SU(N)$ QCD(f/adj) theory.}
  \label{tab:multi massive-fermion-reps}
\end{table}

To analyse the anomalies, we first turn on background gauge fields
$A_f$ for $\SU(N_f)$ and $\mathcal{A}_q$ for $\U_q$. However, since the
background field must be in a $U(N_f)/\Z_N$ bundle and not
$\SU(N_f)\times \U$, these fields do not have their proper
normalization. To specify a
$U(N_f)/\Z_N = (\SU(N_f) \times \U_q)/(\Z_{N_f} \times \Z_N)$ bundle, we specify
$P\SU(N_f)$ bundle and a $\U_q/\Z_{D}$ bundle (with
$D= \text{lcm}(N,N_f)$) on the underlying manifold $M$ in terms of the
obstructions to lifting them to $\SU(N_f)$ and $\U_q$ bundles which
are related in a certain way by the quotient structure. The
obstruction to lifting a $P\SU(N_f)$ bundle to an $\SU(N_f)$ bundle is
the mod $N_f$ second Stiefel--Whitney class
$w_2^{(N_f)} \in H^2(M;\Z_{N_f})$, while that of $\U_q/\Z_{D}$ is
$ D \mathcal{F}_q/2\pi ~\mod D$. They are related by
\begin{equation}
D \frac{\mathcal{F}_q}{2\pi} = \frac{D}{N} w_2^{(N)} + \frac{D}{N_f} w_2^{(N_f)} ~\mod D\;,
\end{equation}
where $w_2^{(N)}\in H^2(M;\Z_N)$, the obstruction to lifting a
$P\SU(N)$ gauge bundle to an $\SU(N)$ gauge bundle is involved because
the quotient $\Z_N$ is between $\U_q$ and $\SU(N)$ gauge group. Then,
the fractional parts of the $\SU(N)$ and $\SU(N_f)$ instanton numbers
are given in terms of $w_2^{(N)}$ and $w_2^{(N_f)}$ by \cite{Witten:2000nv}
\begin{equation}
  \label{eq:fractional-instanton-Pontryagin}
  \begin{aligned}
    \int_M \frac{\Tr f\wedge f}{8\pi^2} &= -\frac{1}{N} \int_M \frac{\mathcal{P}(w_2^{(N)})}{2} ~ \mod 1 \,,\\
    \int_M \frac{\Tr F_f \wedge F_f}{8\pi^2} &= -\frac{1}{N_f} \int_M \frac{\mathcal{P}(w_2^{(N_f)})}{2} ~ \mod 1\,.
  \end{aligned}
\end{equation}
where $f$ is the $\SU(N)$ gauge field strength and $\mathcal{P}$ the Pontryagin
square operation. For the mod $N$ cohomology classes, when $N$ is
even, it is defined to be the cohomology operation \cite{Browder:1962}
$\mathcal{P}: H^2(M;\Z_{N}) \to H^4(M;\Z_{2N})$. The image is even when
$M$ is spin, so $\mathcal{P}(w_2^{(N)})/2$ is a well-defined cohomology class in
$H^4(M;\Z_N)$. When $N$ is odd, we define $\mathcal{P}$ to be simply the cup
product, following Ref. \cite{Aharony:2013hda}. Division by 2 makes
sense because it is invertible in $\Z_N$ when $N$ is odd. The same
definition goes for the mod $N_f$ classes. We can then clearly see
that the $\SU(N)$ and $\SU(N_f)$ instanton numbers are fractions in
the units of $1/N$ and $1/N_f$, respectively.

It turns out that the anomaly structure is very similar to that
described in \S \ref{sec:one-Dirac-SUN-symmetry}: there is again a
mixed anomaly between $\Z_{2N}^{\chi}$ and the flavor symmetry
$\UU(N_f)/\Z_N$ (the CFU anomaly).  Under a $\Z_{2N}^{\chi}$
transformation $\lambda \to e^{2\pi i k/2N} \lambda$, in the non-trivial background
field for $\UU(N_f)/\Z_N$, the action effectively shifts by
\begin{equation}
  \begin{split}
    \Delta S &= 2\pi i k \frac{1}{8\pi^2} \int \Tr f \wedge f\\
    &= -\frac{2\pi i k}{N} \int \frac{\mathcal{P} (w_2^{(N)})}{2}\;,
    \end{split}
\end{equation}
Since $\int \mathcal{P}(w_2^{(N)})/2 $ is an integer modulo $N$ on a spin manifold,
such a $\Z_{2N}^{\chi}$ transformation is anomalous unless
$k = N ~\mod 2N$, signifying that the $\Z_2$ subgroup is
non-anomalous.

\subsubsection*{The massless case}
\label{sec:mult-flav-ferm-massless}

Without the mass term, the theory now possesses
$\SU(N_f)_L\times \SU(N_f)_R$ flavor as well as $\U_q$ quark and
$\U_\chi$ axial symmetries (with appropriately modded common discrete centers). The action of the gauge and global symmetries
on the fermion content is displayed in the following table. We choose
the minimal charge assignments for $\U_q$ and $\U_\chi$.
\begin{table}[h]
  \centering
  \begin{tabular}{c||c|c|c|c|c}
    & $\SU(N)$ & $\SU(N_f)_L$ & $\SU(N_f)_R$ & $\U_q$ & $\U_\chi$\\
    \hline
    $\psi$ & ${\tiny\yng(1)}$ & ${\tiny\yng(1)}$ & $\textbf{1}$ &  $+1$ & $-q:=-\frac{N}{\text{gcd}(N,N_f)}$ \\
    $\tilde{\psi}$ & ${\tiny \overline{\yng(1)}}$ & $\textbf{1}$ & ${\tiny \overline{\yng(1)}}$ & $-1$ & $-q$ \\
    $\lambda$ & $\textbf{adj}$ & $\textbf{1}$ & $\textbf{1}$ & $0$ & $Q:= \frac{N_f}{\text{gcd}(N,N_f)}$
  \end{tabular}
  \caption{Action of gauge and global symmetries in the multi-flavor massless $\SU(N)$ QCD(f/adj) theory.}
  \label{tab:multi massless-fermion-reps}
\end{table}

This section explains the origin behind the mixed anomalies discussed in the previous sections. In doing so, we give the details that link the above derivations to the CFU anomaly computations of \cite{Anber:2019nze}.

The theory possesses the following traditional 't Hooft anomalies:
\begin{equation}
  \begin{split}
  \left[\SU(N_f)_L\right]^3=-   \left[\SU(N_f)_R\right]^3:& \quad \mathcal{C}_{L^3}=-\quad \mathcal{C}_{R^3}=N\,, \\
\left[\U_\chi\right]^3:&\quad \mathcal{C}_{A^3} = \frac{-2N^4N_f+N_f^3(N^2-1)}{\text{gcd}(N,N_f)^3}\,,\\
\left[\U_\chi\right]\left[\SU(N_f)_L\right]^2 =\left[\U_\chi\right]\left[\SU(N_f)_R\right]^2:& \quad \mathcal{C}_{AL^2} = \mathcal{C}_{AR^2}=-\frac{N^2}{\text{gcd} (N,N_f)}\,,\\
\left[\U_\chi\right]\left[\U_q\right]^2:& \quad \mathcal{C}_{AB^2}=-\frac{2N^2N_f}{\text{gcd}(N,N_f)}\,,\\
\left[\U_q\right]\left[\SU(N_f)_L\right]^2 =-\left[\U_q\right]\left[\SU(N_f)_R\right]^2 :& \quad \mathcal{C}_{BL^2}= -\mathcal{C}_{BR^2} = N\,,\\
\left[\U_\chi\right]\left[\mbox{gravity}\right]:& \quad \mathcal{C}_{A\text{-grav}} = -\frac{N_f\left(1+N^2\right)}{\text{gcd}(N,N_f)}\,.
\end{split}
\label{all set of anomalies}
\end{equation}

More constraining anomalies can be found by utilizing the faithful global group. To find the latter, one first needs to determine a subgroup in the center of
\begin{equation}
G = \SU(N) \times \SU(N_f)_L \times \SU(N_f)_R \times \U_q \times \U_\chi
\end{equation}
that acts trivially on all the fields. Consider the transformation
\begin{equation}
\left( e^{\frac{2\pi i n_c}{N}} \textbf{1}_N, e^{\frac{2\pi i n_L}{N_f}} \textbf{1}_{N_f}, e^{\frac{2\pi i n_R}{N_f}} \textbf{1}_{N_f}, e^{2\pi i \beta}, e^{2\pi i \alpha} \right) \in G \nn
\end{equation} 
The condition that it acts trivially on all the fields $\psi$,
$\tilde \psi$, and $\lambda$ are :
\begin{equation}
  \begin{split}
\psi\;:&\quad \underbrace{e^{\frac{i 2\pi n_c}{N}}}_{\SU(N)}\underbrace{e^{\frac{i 2\pi n_L}{N_f}}}_{\SU(N_f)_L}\underbrace{e^{i 2\pi \beta}}_{\U_B}\underbrace{e^{-i 2\pi \alpha N}}_{\U_\chi}=1\,,\\
\tilde \psi\;:&\quad \underbrace{e^{\frac{-i 2\pi n_c}{N}}}_{\SU(N)}\underbrace{e^{\frac{-i 2\pi n_R}{N_f}}}_{\SU(N_f)_R}\underbrace{e^{-i 2\pi \beta}}_{\U_B}\underbrace{e^{-i 2\pi \alpha N}}_{\U_\chi}=1\,,\\
\lambda\;: &\quad \underbrace{e^{i2\pi \alpha N_f}}_{\U_\chi}=1\,,
\end{split}
\label{consistency conditions}
\end{equation}
where $n_c$ are integers $\mbox{mod}\, N$, $n_{L,R}$ are integers
$\mbox{mod}\, N_f$, and $\alpha, \beta$ are $\U$ phases. The conditions (\ref{consistency conditions}) ensure that the transition functions of the gauge and global symmetry bundles satisfy the cocycle conditions. The set of solutions
to these conditions forms a subgroup of $G$ that must be quotiented
out from $G$ to determine the faithful global symmetry group. We find
that this subgroup is $\Z_N\times \Z_{N_f} \times \Z_Q$, where $Q=\frac{N_f}{\gcd(N,N_f)}$ is the charge of $\lambda$ under the global $\U_\chi$. This subgroup is generated by
\begin{equation}
  \begin{split}
   &\Z_{N}\;\;: \left( e^{\frac{2\pi i}{N}}\textbf{1}_N, \textbf{1}_{N_f},\textbf{1}_{N_f}, e^{-\frac{2\pi i}{N}}, 1 \right),\\
   &\Z_{N_f}\;:  \left( \textbf{1}_N, e^{\frac{2\pi i}{N_f}}\textbf{1}_{N_f}, e^{\frac{2\pi i}{N_f}}\textbf{1}_{N_f}, e^{-\frac{2\pi i}{N_f}},1 \right),\\
   &\Z_Q\;\;:  \left( \textbf{1}_N, e^{\frac{2\pi i N}{N_f}}\textbf{1}_{N_f}, e^{-\frac{2\pi i N}{N_f}} \textbf{1}_{N_f}, 1, e^{\frac{2\pi i}{Q}} \right), \qquad Q=\frac{N_f}{\gcd(N,N_f)}
 \end{split}
 \label{eq:quotient-generators}
\end{equation}
Thus, the faithful global symmetry of the theory is
\begin{eqnarray}
G^{\scriptsize\mbox{Global}}=\frac{\SU(N_f)_L\times \SU(N_f)_R\times \U_q\times \U_\chi}{\Z_N\times \Z_{N_f} \times \Z_Q}\,.
\label{the global symmetry of multi flavor}
\end{eqnarray}
To fully study the anomalies we note that the symmetries allow us to define a
$\Spin$-$G^{\text{Global}} := \left( \Spin \times G^{\text{Global}}
\right)/\Z_2^F$ structure on our manifold and not just a $\Spin$
structure. This is possible when there is a $\Z_2$ subgroup of
$G^{\text{Global}}$ that acts identically to
$\Z_2^F \subset \Spin(4)$. Indeed, there is such a subgroup.  An element
\begin{equation}
\left( e^{\frac{2\pi i n_L}{N_f}}\textbf{1}_{N_f}, e^{\frac{2\pi i n_R}{N_f}}\textbf{1}_{N_f}, e^{ i \beta}, e^{ i \alpha}  \right) \in \SU(N_f)_L\times \SU(N_f)_R\times \U_B\times \U_\chi
\end{equation}
acts identically to $(-1)^F$ if the following conditions are
satisfied:
\begin{equation}
  \begin{split}
\psi\;:&\quad e^{\frac{ 2\pi  i n_L}{N_f}}e^{i \beta}e^{- i  q \alpha }=-1\,,\\
\tilde \psi\;:&\quad e^{\frac{- 2\pi i n_R}{N_f}} e^{-i  \beta}e^{- i q \alpha }= -1\,,\\
\lambda\;: &\quad e^{ i Q \alpha} =-1\,.
\end{split}
\label{(-1)F conditions}
\end{equation}
When both $q$ and $Q$ are odd, we can take
$\alpha=\pi, \beta=n_L = n_R =0$ as a solution. When $Q$ is odd and
$q$ is even, we can take $\alpha=\beta=\pi$ and $n_L=n_R =0$ as a
solution. Lastly, when $Q$ is even and $q$ is odd, we can take
$n_L=0$, $\alpha = \pi/Q$, $\beta=\pi+q\pi/Q$, and $n_R = N$ as a solution. Thus, it
is possible to put the theory on a non-spin manifold that admits a
$\Spin$-$G^{\text{Global}}$ structure.  Since the quotient $\Z_2^F$
only involves $\U$ symmetries (either $\U_q\times \U_\chi$ or with a Cartan of
$\SU(N_f)_R$), we can always turn on only the $\Spinc$
structure. Since all orientable manifolds admit a $\Spinc$ structure,
we can define our theory on all orientable manifolds, including
e.g., $\CP^2$ (unlike in the massive case). 

Let us now analyze the anomalies in more detail. For this purpose, we
turn on the background field strengths $F_L$, $F_R$, $\mathcal{F}_q$,
$\mathcal{F}_\chi$, for $\SU(N_f)_L$, $\SU(N_f)_R$, $\U_q$, and $\U_\chi$,
respectively. Then, the 6d anomaly polynomial for our theory is given
by
\begin{multline}
  \Phi = \frac{1}{3!} \frac{1}{(2\pi)^3} \bigg[N \left( \Tr F_L^3 -\Tr
    F_R^3 \right) + 3N \left( \Tr F_L^3- \Tr F_R^3\right) \mathcal{F}_q \\-\frac{3N^2}{\text{gcd}(N,N_f)} \left( \Tr F_L^2+\Tr F_R^2 \right)\mathcal{F}_\chi- \frac{6N^2N_f}{\text{gcd}(N,N_f)}\mathcal{F}_q^2\mathcal{F}_\chi \\
  + \frac{1}{\text{gcd}(N,N_f)^3}\left( N_f^3(N^2-1)-2N^4N_f \right)\mathcal{F}_\chi^3 \bigg] + \frac{N_f(N^2+1)}{\text{gcd}(N,N_f)} \frac{p_1}{24} \frac{\mathcal{F}_\chi}{2\pi} \;,
\end{multline}
and $p_1\equiv -\frac{1}{8\pi}\mbox{tr}R\wedge R$  is the first Pontryagin class and $R$ is the curvature $2$-form. The
terms inside the square bracket capture 't Hooft anomalies of
$G^{\text{Global}}$ while the last term is the mixed
$\left[\U_\chi\right]$-gravitational anomaly. Through the
anomaly descent equations, we can see that, under a transformation
$e^{ i \alpha}\in \U_\chi$, the partition function changes as
\begin{multline}
  {\cal Z} \mapsto {\cal Z} \exp\bigg[  i \alpha\bigg(-\frac{ N^2}{\text{gcd}(N,N_f)} \int  \frac{\Tr F_L^2+ \Tr F_R^2}{8\pi^2}  -\frac{2 N^2N_f}{\text{gcd}(N,N_f)} \int \frac{\mathcal{F}_q^2}{8\pi^2} \\+\frac{1}{3}\frac{N_f^3(N^2-1)-2N^4N_f}{\text{gcd}(N,N_f)^3} \int \frac{\mathcal{F}_\chi^2}{8\pi^2} + \frac{N_f(N^2+1)}{\text{gcd}(N,N_f)}\int \frac{p_1}{24}  \bigg)\bigg]\,.
  \label{eq:mult-ferm-axial-anom}
\end{multline}

Note that, because of the discrete quotient in $G^{\text{global}}$,
various instanton numbers that appear above can be fractional. These
are dubbed the {\it color-flavor-$U(1)$} (CFU) fluxes in \cite{Anber:2019nze, Anber:2021iip}. More
precisely,  one can give the fractional parts of these instantons in terms of the obstruction to lifting a $G^{\text{Global }}$
bundle to a $\SU(N_f)_L\times \SU(N_f)_R \times \U_B\times \U_\chi$ bundle, which we
will call the product bundle. One specifies $G^{\text{Global}}$
bundle by specifying $P\SU(N_f)_L$, $P\SU(N_f)_R$,
$\U_B/\Z_{qQ}$, and $\U_\chi/\Z_Q$ bundles. The obstructions to
lifting these bundles to the product bundle are given by the second
``Stiefel--Whitney classes''
\begin{equation}
w_2^{(L)} , w_2^{(R)} \in H^2(M;\Z_{N_f}), \quad w_2^{(N)}\in H^2(M;\Z_N), \quad w_2^{(Q)}\in H^2(M;\Z_Q)\;.
\end{equation}
$w_2^{(N)}$, $w_2^{(N_f)}$, and $w_2^{(Q)}$ are directly related to the $\U$ fluxes by
\begin{equation}
  Q \frac{\mathcal{F}_\chi}{2\pi} = w_2^{(Q)} ~\mod Q\;,\quad qQ \frac{\mathcal{F}_q}{2\pi} = Q w_2^{(N)} + q w_2^{(N_f)} ~\mod qQ\;.
\end{equation}
Again, since the quotient $\Z_N$ involves the dynamical $\SU(N)$, the
dynamical gauge field is in a $P\SU(N)$ bundle whose obstruction to
lifting to an $\SU(N)$ bundle is given precisely by $w_2^{(N)}$. Just
like in \eqref{eq:fractional-instanton-Pontryagin}, the fractional
$\SU(N)$ instanton number is given in terms of $w_2^{(N)}$ by
\begin{equation}
N \int \frac{\Tr f\wedge f}{8\pi^2} = -\int \frac{\mathcal{P} (w_2^{(N)})}{2} ~\mod N\;.
\end{equation}
Similarly, $w_2^{(L)}$ and $w_2^{(R)}$, which obstruct lifting
$P\SU(N_f)_{L,R}$ bundles to $\SU(N_f)_{L,R}$ bundles are related to
the fractional instanton numbers by 
\begin{equation}
  \label{eq:fractional-instantons-Pw2}
N_f\int \frac{\Tr F_{L,R}^2}{8\pi^2} = -\int \frac{\mathcal{P}\left( w_2^{(L,R)} \right)}{2} ~\mod N_f\;.
\end{equation}
 The structure of the
quotient given by the generators in \eqref{eq:quotient-generators}
relates $w_2^{(L,R)}$ to the other Stiefel--Whitney classes by
\begin{equation}
w_2^{(L)} = w_2^{(N_f)} + N w_2^{(Q)} ~\mod N_f, \quad w_2^{(R)} = w_2^{(N_f)} - N w_2^{(Q)} ~\mod N_f \;.
\end{equation}
These are well-defined modulo $N_f$: as
$w_2^{(Q)}\sim w_2^{(Q)} + N_f/\text{gcd}(N,N_f)$, we have
$Nw_2^{(Q)}\sim Nw_2^{(Q)} + qN_f = Nw_2^{(Q)}~\mod N_f$.

To understand how these fractional instanton numbers could alter our
anomalies, it is instructive to consider concrete examples. Let us
define the topological charges associated with $\SU(N)$,
$\SU(N_f)_{L,R}$, $\U_B$, and $\U_\chi$, respectively, by
\begin{equation}
  \begin{split}
    Q_c := \int\frac{\Tr f \wedge f}{8\pi^2},& \quad Q_{L,R} := \int \frac{\Tr F_{L,R} \wedge F_{L,R}}{8\pi^2},\\
    Q_B := \int \frac{\mathcal{F}_q \wedge \mathcal{F}_q}{8\pi^2},& \quad Q_\chi := \int \frac{\mathcal{F}_\chi \wedge \mathcal{F}_\chi}{8\pi^2},\\
    Q_{AB} := \int \frac{\mathcal{F}_\chi \wedge \mathcal{F}_q}{4\pi^2}\,.&
  \end{split}
\end{equation}
One can also calculate the Dirac indices in these center fluxes:
\begin{equation}
  \begin{split}
{\cal I}_\psi &= N_f Q_c+ N Q_L+N N_f \left( Q_B + q^2Q_\chi \right) - q NN_f Q_{AB}\,,\\
{\cal I}_{\tilde \psi} &= N_f Q_c+ N Q_R+ NN_f \left( Q_B+q^2Q_\chi \right) + qNN_fQ_{AB}  \,,\\
{\cal I}_{\lambda} &= 2NQ_c+(N^2-1) Q^2 Q_\chi\,,
\end{split}
\end{equation}
which are always integers in a consistent background: there is a
one-to-one correspondence between the solutions of (\ref{consistency
  conditions}) and the integrality of the Dirac indices.  The finest
fractional charges are reached when we put on the background fields
with lowest, non-trivial Stiefel--Whitney classes $w_2^{(N)}$, $w_2^{(N_f)}$, and $w_2^{(Q)}$. To
achieve this, let's consider the theory on the product manifold
$S^2\times S^2$. Then, we can take $w_2^{(N)}$, $w_2^{(N_f)}$, and $w_2^{(Q)}$ to have the form
$\alpha+\beta$ modulo $N$, $N_f$, and $Q$, respectively, where $\alpha,\beta$ are the two generators of
$H^2(S^2\times S^2;\Z)$. In this configuration, the various topological
charges (CFU fluxes) are given by
\begin{equation}
  \begin{split}
    & Q_\chi = \left(\frac{1}{Q}+n_1\right)^2, \quad Q_B = \left( \frac{1}{N}+\frac{1}{N_f} +n_2\right)^2, \quad Q_c = k_c - \frac{1}{N},\\
    & Q_L = k_L - \frac{1}{N_f}\left( 1+N \right)^2, \quad Q_R = k_R -\frac{1}{N_f}\left( 1-N \right)^2,\\
    & Q_{AB} = 2\left(\frac{1}{Q}+n_1\right)\left( \frac{1}{N}+ \frac{1}{N_f}+n_2 \right)\,,
\end{split}
\label{most general topological charges}
\end{equation}
where $n_{1}, n_{2}$, $k_c$, $k_{L,R}$ are integers. 

Let ${\cal Z}[\hat A_f]$ be the partition function in the background of the vector-like flavor symmetry $\UU(N_f)/\mathbb Z_N$, which in general activates the CFU fluxes. Then, under a $\U_\chi$ rotation we have that
\begin{equation}
  \begin{split}
     {\cal Z}[\hat A_f]\xrightarrow{\U_\chi}& \;{\cal Z}[\hat A_f]\exp\left[ i \alpha\left(-q({\cal I}_{\psi}+{\cal I}_{\tilde \psi})+Q{\cal I}_\lambda\right)\right] \,,
  \end{split}
\label{most general CFU anomaly}
\end{equation}
and it is easy to see that the part that multiplies $Q_c$ cancels out. This should be expected since the theory is not endowed with a genuine $\mathbb Z_N^{[1]}$ $1$-form symmetry, thanks to the fundamentals. Thus, what we find is an anomaly of mixed type between $\U_\chi$, $\U_B$, and $\SU(N_f)$.  The anomaly is exactly the same one we obtain from (\ref{eq:mult-ferm-axial-anom}). 
As a special case, we can consider fractional fluxes of $F_\chi$ turned off and set $n_1=n_2=0$:
\begin{eqnarray}
Q_c=k_c-\frac{1}{N}\,, \quad Q_L =k_L-\frac{1}{N_f}, \quad Q_R = k_R-\frac{1}{N_f}, \quad Q_B = \left( \frac{1}{N}+\frac{1}{N_f} \right)^2.
\end{eqnarray}
The corresponding Dirac indices are
\begin{eqnarray}
{\cal I}_\psi= 2+N k_L +N_f k_c,\quad  {\cal I}_{\tilde \psi}=2+N k_R + N_f k_c, \quad {\cal I}_\lambda=2(N k_c-1)\,.
\end{eqnarray}
 Then, under a $\U_\chi$ rotation we find
\begin{equation}
  \begin{split}
     {\cal Z}[\hat A_f]\xrightarrow{\U_\chi}& \;{\cal Z}[\hat A_f]\exp\left[ i \alpha\left(-q ({\cal I}_\psi+ {\cal I}_{\tilde{\psi}})+Q {\cal I}_\lambda\right)\right] \\
    = &\;{\cal Z}[\hat A_f]\exp\left[ -i  \alpha(2 Q +q (4+(k_L+k_R)N))\right ] \,.
  \end{split}
\label{special CFU anomaly}
\end{equation}
If we further  turn off the $\SU(N_f)$ fluxes, with only $Q_c=k_c-1/N$ and
$Q_B= 1/N^2$ non-trivial, the Dirac indices become
\begin{equation}
{\cal I}_\psi= {\cal I}_{\tilde \psi}= N_f k_c, \quad {\cal I}_\lambda=2(N k_c-1)\,,
\end{equation}
and CFU anomaly is 
\begin{equation}
    {\cal Z}[\hat A_f]\xrightarrow{\U_\chi}\;{\cal Z}[\hat A_f]\exp[ -2 i  \alpha \frac{N_f}{\gcd (N,N_f)} ] \,.
\end{equation}
This is exactly  the $A\wedge {\cal F}_B\wedge {\cal F}_B$ (the mixed $\U_\chi$-$\U_B$) anomaly in (\ref{mixed BCF anomaly}) when we set $N_f=1$.

Finally, we briefly discuss the anomalies when we put the theory on a non-spin manifold. For concreteness, we consider the theory on $ {\CP}^2$ and assume both $q$ and $Q$ are odd. In this case, the topological charges are given by \cite{Anber:2020gig}
\begin{eqnarray} 
Q_\chi&=&\frac{1}{2}\left(\frac{1}{2}+n\right)^2\,, n\in \mathbb Z,\quad Q_G=\int_{\CP^2}\frac{p_1}{24}=-\frac{1}{8}\,.
\end{eqnarray}
While the Dirac indices are
\begin{equation}
\begin{split}
{\cal I}_\lambda&=(N^2-1)\left[Q^2Q_\chi+Q_G \right]\,,\\
{\cal I}_{\psi}&={\cal I}_{\tilde \psi}=N_f N\left[ q^2 Q_\chi+Q_G \right]\,,
\end{split}
\end{equation}
which are always integers when both $q$ and $Q$ are odd.
Then, the anomaly on $\mathbb {CP}^2$ reads
\begin{eqnarray}
{\cal Z}[\mathbb {CP}^2]\xrightarrow[]{\U_\chi} {\cal Z}[\mathbb {CP}^2]
\exp\left[{i  \alpha\left(Q{\cal I}_\lambda-2q{\cal I}_\psi\right)}\right]\,.
\end{eqnarray}
Similarly, one can work out the anomalies when $q$ and $Q$ are even or mixed even/odd.

%%%%%%%%%%%%%%%%%
\subsection{The IR phases}
\label{sec:mult-ferm-IR-phase}
%%%%%%%%%%%%%%%%%%

In the infinite mass limit, we can again integrate out the Dirac
fermions, leaving us with  $\mathcal{N}=1$ SYM with
gauge group $\SU(N)$. As discussed in \S~\ref{sec:one-ferm-IR-phase}, there are $N$ degenerate
vacua and domain walls connecting them.

At the massless point, the $\Z_{2N}^{\chi}$ discrete chiral symmetry enhances
to $\U_\chi$. All anomalies are now given in terms of the anomaly
polynomial. Saturating the anomalies can be achieved in the IR in one
of three ways:
\begin{enumerate}
\item Composite massless fermions charged under the global symmetry.
\item Spontaneous symmetry breaking (SSB).
\item Interacting conformal field theory.
\end{enumerate}
The first choice can be ruled out by Weingarten's theorem
\cite{Weingarten1983} because the measure of the theory is positive
definite (see Appendix \ref{app:measure-positive-definite}).  Let's
now discuss the second option, where anomalies are saturated by
breaking the global symmetries
$\SU(N_f)_L\times \SU(N_f)_R\times \U_\chi\times \U_B$ down to
$\SU(N_f)_V\times \U_B$, giving rise to $N_f^2-1$ Goldstones and a
$\U$ vortex.  The condensate $\tilde \psi\psi$ is charged under the full
$\SU(N_f)_L\times \SU(N_f)_R\times \U_\chi$, and breaks the group down to
$\SU(N_f)_V\times \mathbb Z_{2q}$. The $\mathbb Z_{2q}$ phase is the
unbroken subgroup of $\U_\chi$ under $\tilde \psi\psi$. However, the unbroken
phase is anomalous, as can be checked using
$\left[\U_\chi\right][\mbox{gravity}]$ and the CFU anomalies given by
\eqref{special CFU anomaly} by setting $\alpha=\frac{2\pi}{2q}$. To avoid
this problem, another condensate has to form in order to break $\U_\chi$
to a non-anomalous group. The minimal choice is $\lambda\lambda$, which transforms
under $\U_\chi$ as:
$\lambda\lambda\xrightarrow[]{\U_\chi}e^{-i 2Q\alpha}\lambda\lambda$, and thus, the formation of
$\lambda\lambda$ breaks $\U_\chi$ down to $\mathbb Z_{2Q}$.  The combined condensates
$\lambda\lambda$ and $\tilde\psi\psi$ break $\U_\chi$ down to
$\mathbb Z_{2\text{gcd}(q,Q)} = \Z_2$. it is easy to see that this
$\Z_2$ has no mixed anomaly with $\SU(N_f)_V$ nor $\U_B$ by setting
$\alpha=\pi$ and $k_L = k_R$ in \eqref{special CFU anomaly}. Moreover, since
$\Omega^{\Spin}_5(B\Z_2)= 0$ \cite{Kapustin:2014dxa}, there are no global
anomalies in $\Z_2$ itself. So for sufficiently small number or flavors $N_f<N_f^*$ we expect that the theory flows to a Goldstone phase corresponding to $N_f$ Goldstone bosons.

\begin{figure}[h]
  \centering
  \includegraphics[scale=0.5]{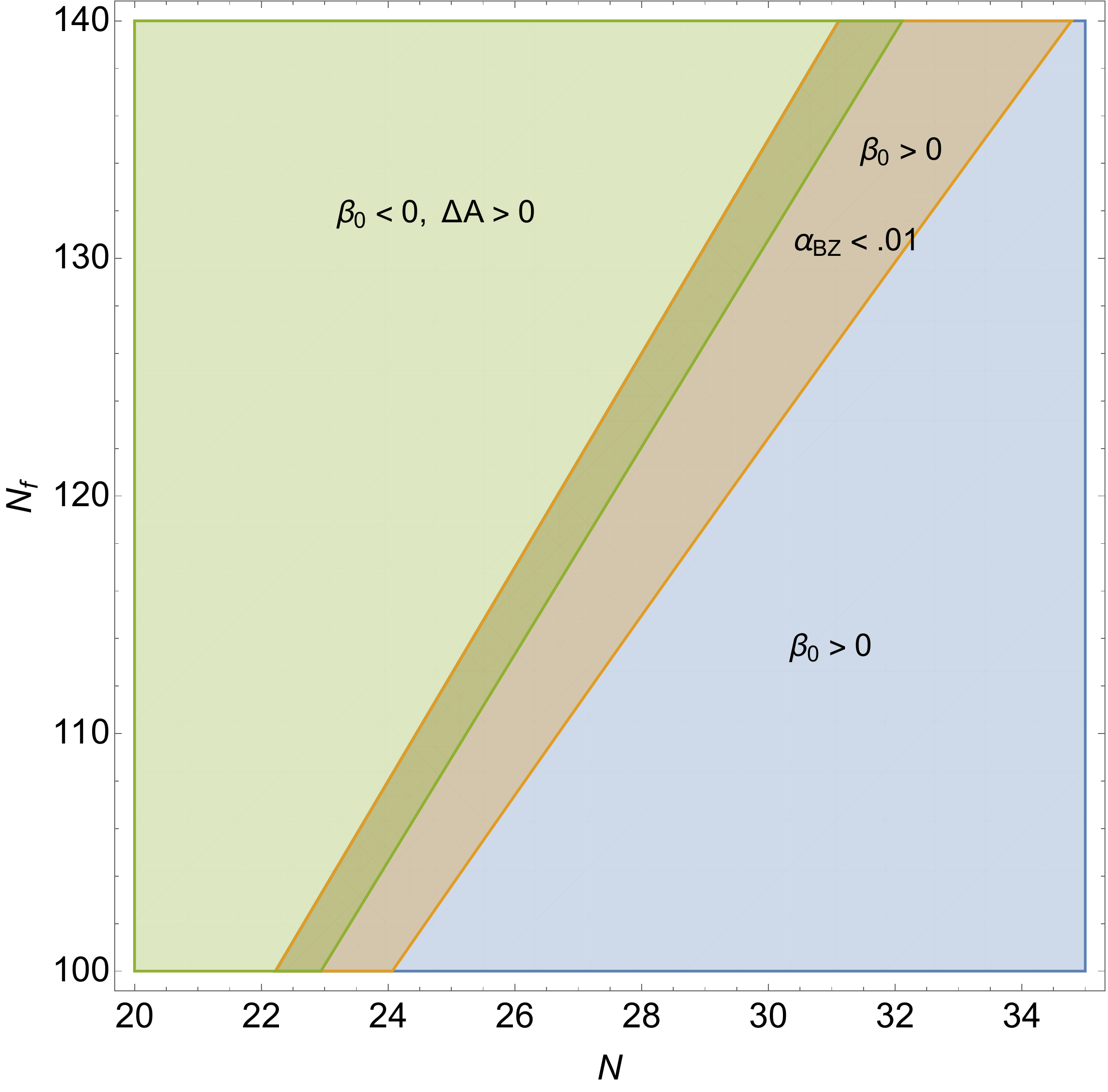}
  \caption{The horizontal axis is the number of colors $N$, while the
    vertical axis is the number of flavors $N_f$. Here, we prefer not to use the 't Hooft coupling $\lambda=Ng^2$ since we are not in the regime of strict large-$N$ limit.  We
   display the asymptotically free region, $\beta_0>0$ in blue, while the region that supports a Banks-Zack fixed point, with $\alpha_*\equiv\frac{g_*^2}{4\pi}<0.01$ at the fixed point, is displayed in orange. This coupling constant value is taken for convenience. Increasing the value of $\alpha_*$ will increase the Banks-Zack region. The phase with
    $\Delta {\cal A}>0$ is displayed in green.  Most of this region lies outside the asymptotically free region. There is a small window where the regions with $\Delta {\cal A}>0$ and $\alpha_*<0.01$ intersect. As can be easily seen, the intersection region happens very close to the boundary of the asymptotically free region. Thus, the $\Delta A$ analysis aligns with the $\beta$-function calculations of the fixed point, provided the latter remains perturbative within the $\epsilon$-expansion framework. See the bulk of the paper for more details.} 
  \label{composite fermions vs goldstones}
\end{figure}

Another scenario is that the theory flows to a phase that
preserves all the global symmetries, e.g., a conformal window. One may wonder which scenario is
preferred. The answer to this question comes from comparing the number of
effective massless degrees of freedom (DOF) between the UV theory and
the SSB scenario \cite{Appelquist:1999hr}. The, effective degrees of
freedom $\mathcal{A}$ of $n_B$ massless real scalars and $n_f$
massless Weyl fermions are given in terms of the free energy density
$F$ as (we turn on a small temperature $T\ll \Lambda$, where $\Lambda$ is the strong
scale)
\begin{eqnarray}
\mathcal{A} = \frac{90}{\pi^2 T^4} F= n_B+\frac{7 n_f}{4} \,.
\end{eqnarray}
The effective degrees of freedom defined via the free energy is not guaranteed to decrease
along RG flows. Yet, as we shall see, the outcomes of this method are consistent with the findings via renormalization group analysis. It is expected that the phase with lower free energy, i.e., a smaller
number of DOF, is preferred.  Define the difference between the number
of DOF in the UV and the SSB scenario by
\begin{equation}
  \begin{split}
    \Delta {\cal A}:={\cal A}^{\scriptsize\mbox{Goldstones}}-{\cal
      A}^{\scriptsize\mbox{UV}}&=(N_f^2-1)-\left[\underbrace{2(N^2-1)}_{\text{gluons DOF}}+\frac{7}{4}\left(N^2-1+2NN_f\right)\right]\\
    &= -\frac{7 N N_f}{2}-\frac{15}{4} N^2+N_f^2+\frac{11}{4}\,.
\end{split}
\end{equation}
If $\Delta {\cal A}>0$, the SSB phase is disfavored, and the theory should flow to the conformal phase in the IR, provided that it is asymptotically free. In Figure~\ref{composite fermions vs goldstones}, we plot both $\Delta {\cal A}$ and the $\beta$-function versus $(N, N_f)$. The phase with $\Delta {\cal A}>0$   is displayed in green, while the asymptotically free region is in blue. There is only a small intersection window between the two regions. The intersection window, however, lies completely inside the conformal window, as is evident from computing the $2$-loop Bank-Zacks fixed point. The latter region is displayed in orange.

In order to make the last point more quantitative, we study the theory
in the Veneziano limit. Thus, we take both $N$ and $N_f$ infinite,
keeping the ratio ${\cal R}=N_f/N$ finite.  The $2$-loop $\beta$-function is given
by (see Appendix \ref{beta function section})
\begin{eqnarray}
g^{-1}\beta(g)=-\beta_0\frac{g^2}{(4\pi)^2}-\beta_1\frac{g^4}{(4\pi)^4}+\mbox{3-loop correction}\,,
\label{the 2 loop expansion of the beta function}
\end{eqnarray}
and in the Veneziano limit, we have 
\begin{eqnarray}
\beta_0=N\left(3-\frac{2}{3}{\cal R}\right)\,,\quad \beta_1=N^2\left(6-\frac{13 }{3}{\cal R}\right)\,.
\end{eqnarray}
The theory is asymptotically free for ${\cal R}<4.5$, while it
develops a Banks-Zaks fixed point at (writing the RG equation using the 't Hooft coupling $\lambda_t\equiv g^2N$)
\begin{eqnarray}
\lambda_t=\frac{8\pi^2(9-2{\cal R})}{(13{\cal R}-18)}\,,
\label{the BZ FP}
\end{eqnarray}
provided that ${\cal R}>1.38$. Notice that the fixed point is well
under control in the limit $N\rightarrow \infty$, provided that we stay close to the boundary of the asymptotically free region so that higher-order loops are parametrically small compared to the first two terms in the loop expansion. We lose asymptotic freedom when $N_f\geq 4.5 N$. Thus, by taking $N_f/N=4.5(1-\epsilon)$ and $\epsilon\ll 1$, the first $2$ terms in (\ref{the 2 loop expansion of the beta function}) are of the same order $\sim N \epsilon$, while the third term is $\sim N\epsilon^2$ and can be safely neglected.  The $\beta$-function analysis of (\ref{the 2 loop expansion of the beta function}, \ref{the BZ FP}) predicts that the conformal window lies in the range $1.38 \lesssim {\cal R}\lesssim 4.5$, with less control on the lower value of the window as we depart from the well-controlled $\epsilon$-expansion. At finite but large $N$, we should
expect the conformal window to be in the range
\begin{equation}
\mathcal{R}_{*} \leq \mathcal{R} \leq 4.5\,, 
\end{equation}
where the lower bound $\mathcal{R}_{*}$ is harder to compute.

Let us compare this result with what we get from the constraint
$\Delta{\cal A}>0$. In the Veneziano limit, the conformal behavior is favored when
\begin{eqnarray}
\Delta{\cal A}=N^2\left({\cal R}^2-\frac{7}{2}{\cal R}-\frac{15}{4}\right)>0\,,
\end{eqnarray}
which is solved by requiring $\mathcal{R}\gtrsim 4.36$. This implies that the theory is in the conformal window when
${\cal R}$ is in the range $ 4.36 \lesssim {\cal R}\lesssim 4.5$.  This result is consistent with the $\beta$-function analysis in the $\epsilon$-expansion. On the other hand, when $\Delta \mathcal{A} <0$, the inequality only
tells us that the SSB phase is favored compared to the weakly coupled
conformal phase (whose DOF are the same as the UV theory), but it does
not exclude the strongly interacting conformal phase, whose DOF
are not as easily computed. When $N$ is large, $\mathcal{R}_{*}$
should still remain close to the value computed using the $\beta$-function in the Veneziano
limit, and so should be lower than $4.36$. Thus, when $\mathcal{R}$ is
in the range $\mathcal{R}_{*} \leq \mathcal{R} \leq 4.36$, the IR phase
could be an interacting CFT.

Also, our investigation included identifying theories with IR fixed points at finite $N$ and $N_f$ using $\Delta{\cal A}>0$, but otherwise lacking such fixed point from the $\beta$-function analysis.  We found no evidence of such fixed points in the $N$-$N_f$ plane.

To summarize, the $\Delta \cal A$ calculations put a stringent constraint on the conformal window, consistent with the $\beta$-function analysis calculations in the $\epsilon$ expansion. This is summarized in Figure \ref{composite fermions vs goldstones}.

\subsection{Fermion masses and the phase diagrams}

When $N_f$ is low enough that chiral symmetry breaking occurs, we can
broaden the scope of our analysis by turning on the mass of the
adjoint fermion in addition to the fundamental fermions' mass.  When
$m_{adj} \gg \Lambda$, it can be integrated out so that we are left
with $\SU(N)$ QCD with $N_f$ fundamental Dirac fermions. The IR phase
structure of this theory has been analyzed in Ref.
\cite{Gaiotto:2017tne}, which we briefly recount here. At non-zero
$m$, the theory has no time-reversal symmetry at all theta-angle
except at $\theta=0,\pi$ and flows to the trivially gapped phase in the
IR. At $\theta=0$, the $T$-symmetry is unbroken and the theory is still in
the trivially gapped phase. On the other hand, the $T$-symmetry is
spontaneously broken at $\theta=\pi$, resulting in a phase with $2$
inequivalent vacua. This persists for all value of $\abs{m}$ down to
$m=0$ where the flavor symmetry $\SU(N_f)_V$ enhances to
$\SU(N_f)_L\times \SU(N_f)_R$. This enhanced chiral symmetry breaks
spontaneously down to $\SU(N_f)$, leaving us with non-Abelian
Nambu--Goldstone bosons as the theory flows to the IR.

Nothing much can be said quantitatively in the intermediate regime,
except when the masses are small, $m, m_{adj} \ll \Lambda$, where we can
analyze the IR theory in more detail through the chiral Lagrangian.
We start by writing down the chiral Lagrangian at the massless
point. As previously discussed, both $\tr \lambda\lambda$ and
$\psi\tilde{\psi}$ condense, inducing the spontaneous symmetry breaking
pattern \footnote{ Recall that $Q:= N_f/\gcd(N,N_f)$ and
  $q:=N/\gcd(N,N_f)$.}
\begin{equation}
\frac{\SU(N_f)_L\times \SU(N_f)_R\times \U_q\times \U_{\chi}}{\Z_N\times \Z_{N_f}\times \Z_Q} \to \frac{\SU(N_f)_V\times \U_{q}}{\Z_N\times \Z_{N_f}} \times \Z_2^F\,. \nn
\end{equation}
The target space of the chiral Lagrangian is then the coset space
\begin{equation}
\mathcal{M}_0 =  \frac{ \U_{\chi}/\Z_2 \times \SU(N_f)}{\Z_Q}
\end{equation}
which we parametrize by the pair
\begin{equation}
\left( e^{i \varphi}, U \right) \in \U_{\chi}/\Z_2 \times \SU(N_f)
\end{equation}
with the identification
\begin{equation}
  \label{eq:ZQ-equivalence}
\left( e^{i \varphi}, U \right)\sim \left( e^{2\pi i/Q}\ee^{i \varphi}, e^{2\pi i N/N_f} U \right)
\end{equation}
enforcing the $\Z_Q$ discrete quotient. Another way to see that we
need this $\Z_Q$ identification is by noting that such a $\Z_Q$
transformation leaves the condensates
\begin{equation}
\expval{\tr \lambda\lambda} \sim e^{i Q \varphi}, \quad \expval{\psi \tilde{\psi}} \sim e^{-iq\varphi} U
\end{equation}
invariant. Note also that our parametrization implies that $\varphi$ has
charge $2$ under the original chiral symmetry $\U_{\chi}$ like the
one-flavor case. The lowest derivative terms are given by 
\begin{equation}
  \begin{split}
    \mathcal{L}_{0} &= \frac{\tilde{f}_{\varphi}^2}{2} \left( \partial\varphi \right)^2 +  \frac{f_{\pi}^2}{2} \tr \left( \partial_{\mu}(e^{-iq\varphi}U)\partial^{\mu}(e^{-iq\varphi}U)^{\dagger} \right)\\
    &= \frac{f_{\varphi}^2}{2} \left( \partial\varphi \right)^2 + \frac{f_{\pi}^2}{2} \tr \left( \partial_{\mu}U\partial^{\mu}U^{\dag} \right) + \frac{ i q f_{\pi}^2}{2}\partial_{\mu} \varphi \tr \left( U^{\dag}\partial^{\mu}U-U\partial^{\mu}U^{\dag} \right)
  \end{split}
\end{equation}
where $f_{\pi}^2$ and
$f_{\varphi}^2 := \tilde{f}_{\varphi}^2+N_f q^2f_{\pi}^2$ are two different `pion
decay constants', scaling with $N$ as $f_{\pi}^2 \sim N^2$ and $f_{\varphi}^2 \sim N^3$. Additional terms (including the WZW term) are needed
to match the 't Hooft anomalies of the UV theory.

Turning on the positive masses $m_{adj}$ for the adjoint
fermion $\lambda$ and $m$ for the fundamental fermions $\psi$,
$\tilde{\psi}$, as well as the $\SU(N)$ theta-angle $\theta$, induces a
potential on $\mathcal{M}_0$:
\begin{equation}
V = -\Lambda^3 \left[ (N^2-1)m_{adj} \cos \left( Q \varphi \right) + \frac{m}{2} \tr\left( \ee^{- i(q\varphi-\theta/N_f)} U +  \ee^{ i(q\varphi-\theta/N_f)} U^{\dag} \right) \right]
\end{equation}
For the purpose of finding the vacua, we can focus on $U$ of the form
$U= \ee^{2\pi i k/N_f} \textbf{1}_{N_f}$ so that the symmetry
$\SU(N_f)_V$ is preserved. The potential now reads
\begin{equation}
  \label{eq:full-V-phi-k}
V(\varphi,k) = -\Lambda^3 \left[ (N^2-1)m_{adj} \cos (Q\varphi) + mN N_f \cos \left( q\varphi - \frac{2\pi k+\theta}{N_f} \right) \right].
\end{equation}
To complete the chiral Lagrangian, additional terms (including the WZW
term) are needed to match the 't Hooft anomalies of the UV theory, but
these will not be necessary for what we want to discuss next.

\paragraph{\underline{$m_{adj}=0$}}

\begin{figure}[h]
  \centering
  \includegraphics[width=0.7\textwidth]{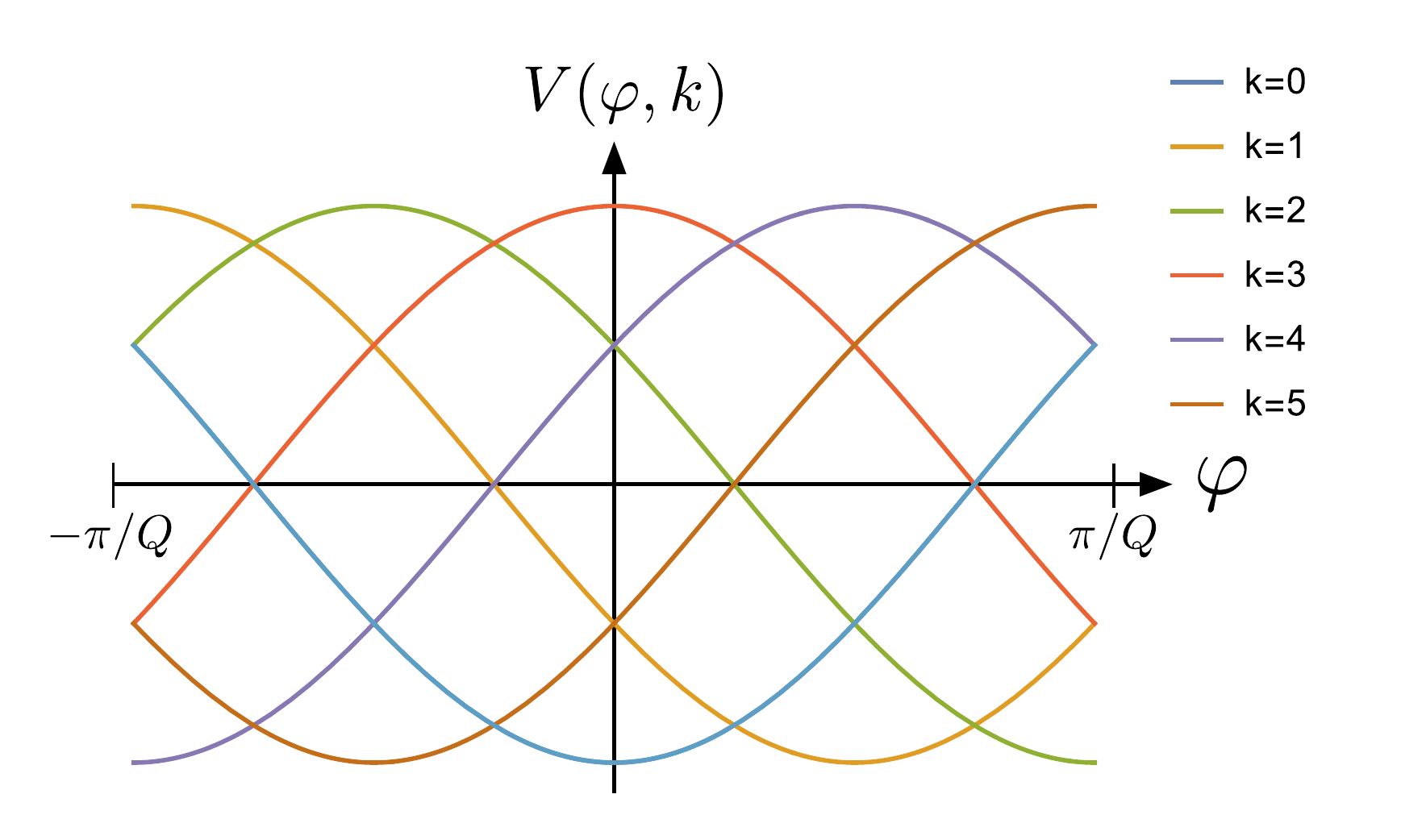}
  \caption{$V(\varphi,k)$ for the case $N=4$, $N_f=6$. The $\Z_Q$ quotient means we should identify $\pi/Q$ and $-\pi/Q$. It is clear there are $N=4$ vacua here.\label{fig:potential-madj0}}
\end{figure}

When we tune the adjoint mass to zero, the theta angle becomes
unphysical as it can be rotated away by an anomalous chiral
rotation. The potential for $\varphi$ thus reduces to
\begin{equation}
V(\varphi,k) = -mN N_f \Lambda^3 \cos \left( q\varphi - \frac{2\pi k}{N_f} \right),
\end{equation}
with obvious minima at $\varphi = 2\pi k/QN$. However, the number of distinct
vacua are smaller than $QN$ due to the $\Z_Q$ quotient. Denote the
$k$\textsuperscript{th} vacuum by
\begin{equation}
  \ket{k} := \left( \ee^{\frac{2\pi  i k}{QN}}, \ee^{\frac{2\pi  i k}{N_f}}\textbf{1}_{N_f} \right).
\end{equation}
Then the equivalence \eqref{eq:ZQ-equivalence} implies that
\begin{equation}
  \ket{N} = \left( \ee^{2\pi  i /Q}, \ee^{2\pi  i N/N_f}\textbf{1}_{N_f} \right) \sim \left( 1,\textbf{1}_{N_f} \right) = \ket{0}.
\end{equation}
Thus, there are only $N$ distinct vacua. See Figure~\ref{fig:potential-madj0} for the
visualization of a specific case.

There are domain walls connecting neighbouring vacua $\ket{k}$ and
$\ket{k+1}$. The domain wall configurations cannot preserve the full
global symmetry
\begin{equation}
\frac{\SU(N_f)_V \times \U_q \times \Z_2}{\Z_{N_f}\times \Z_N} \nn
\end{equation}
of each vacuum. Following \cite{Gaiotto:2017tne}, one can show that
the non-Abelian symmetry $\SU(N_f)_V$ is necessarily broken down to
$S[\U\times \UU(N_f-1)]$ by domain wall configurations, resulting in a
non-linear $\sigma$-model on the domain wall with the target space
\begin{equation}
\frac{\SU(N_f)_V}{S[\U\times \UU(N_f-1)]} \cong \mathbb{CP}^{N_f-1}
\end{equation}
coupled to a topological term induced by the WZW term in the
bulk.  We can also conclude that there must be a phase transition on
the domain wall as we crank up the fundamental mass $m$, just like in
the one-flavor case, because in the large mass limit, the domain-wall
theory is a TQFT with no massless degrees of freedom.

\paragraph{\underline{$m=0$}}

When we set $m=0$ instead of $m_{adj}$, the potential becomes
\begin{equation}
  V(\varphi) = -\Lambda^3 m_{adj} (N^2-1) \cos \left( Q \varphi \right),
\end{equation}
independent of $U$. The potential has its minima at $\varphi = 2\pi k/Q$ for any integer
$k$, and any $U\in \SU(N_f)$. The vacuum manifold
$\mathcal{M}_0 = (\frac{\U_\chi}{\mathbb Z_2}\times \SU(N_f))/\Z_Q$ is reduced to to
\begin{equation}
\frac{\Z_Q\times \SU(N_f)}{\Z_Q} \cong \SU(N_f).\nn
\end{equation}
Therefore, when $m=0$ and $0<m_{adj}\ll\Lambda$, only the non-Abelian
NGBs remain massless. The Abelian NGB $\varphi$ becomes massive.

\begin{figure}[h]
  \centering
  \begin{subfigure}[b]{0.45\textwidth}
    \includegraphics[width=\textwidth]{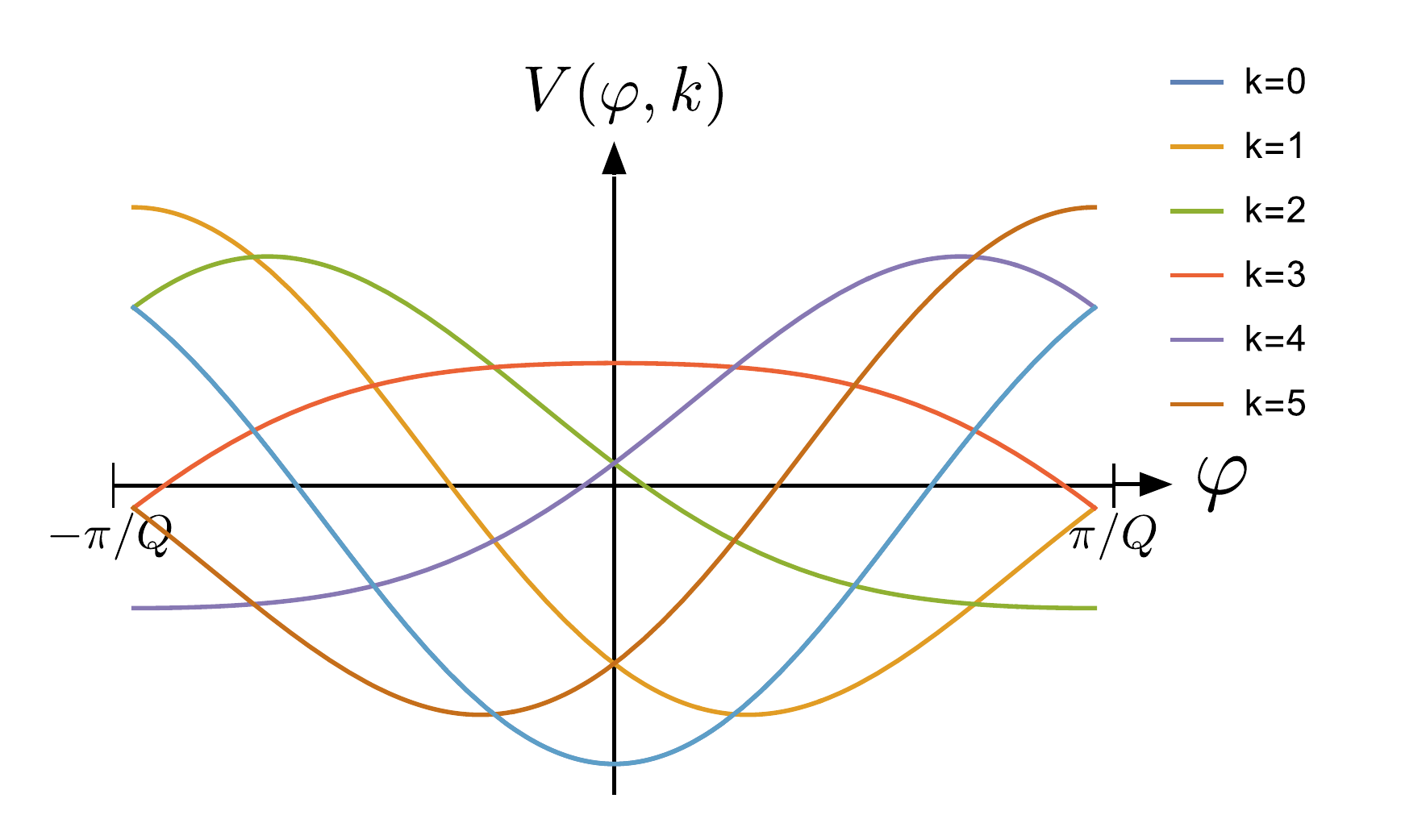}
    \caption{}
  \end{subfigure}
  \begin{subfigure}[b]{0.45\textwidth}
    \centering
    \includegraphics[width=\textwidth]{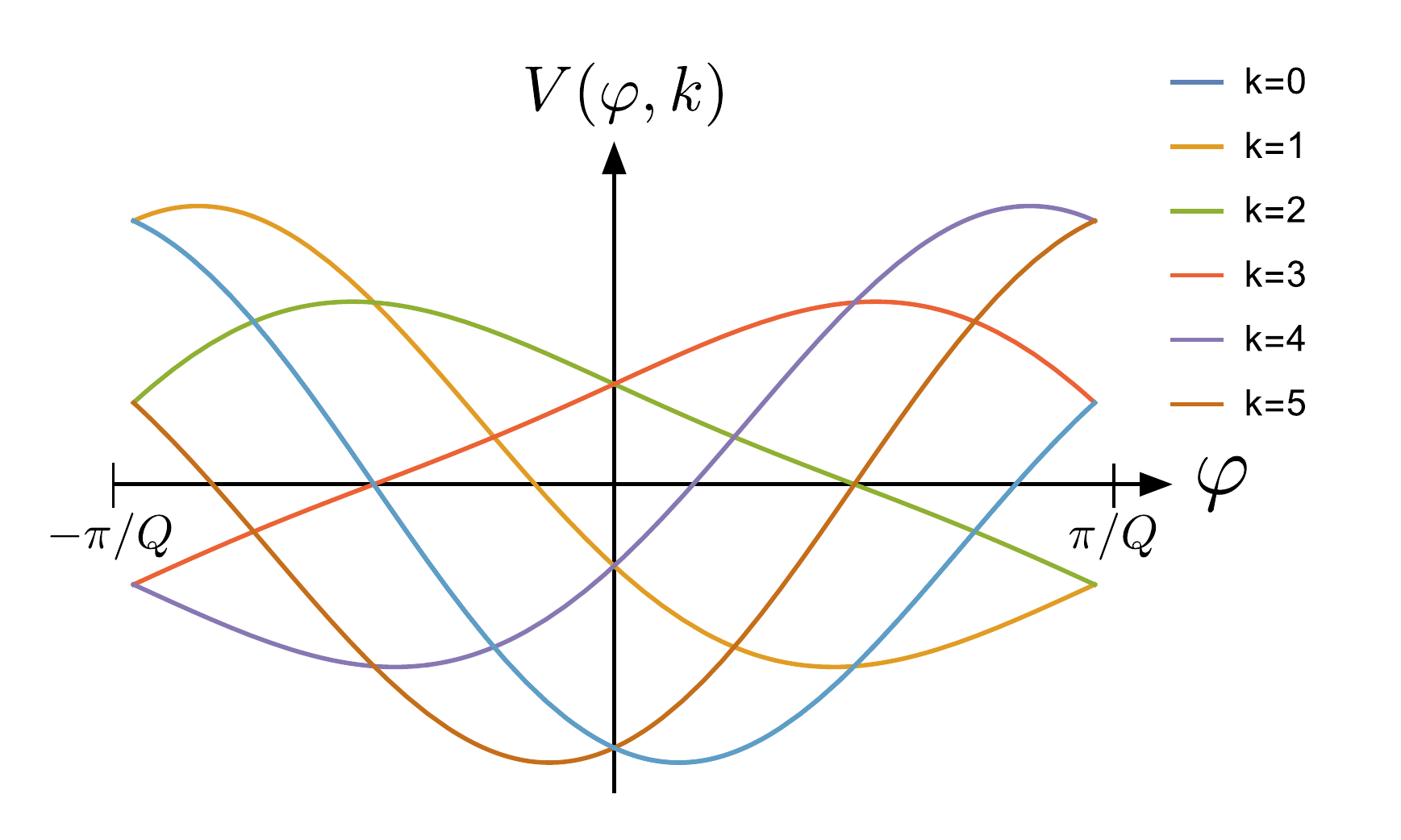}
    \caption{}
    \end{subfigure}
  \caption{The potential $V(\varphi,k)$ for $N=4$, $N_f=6$ when $m,m_{adj} \neq 0$ at \textbf{(a)} $\theta= 0$ and \textbf{(b)} $\theta=\pi$.\label{fig:potential-theta-0-Pi}}
\end{figure}

\paragraph{\underline{$m,m_{adj}>0$}}
When both masses are non-zero, we need to look at the full potential \eqref{eq:full-V-phi-k}:
\begin{equation}
V(\varphi,k) = -\Lambda^3 \left[ (N^2-1)m_{adj} \cos (Q\varphi) + mN N_f \cos \left( q\varphi - \frac{2\pi k+\theta}{N_f} \right) \right]. \nn
\end{equation}
There is a $\Z_2$ time-reversal symmetry at $\theta= 0, \pi$, which is broken
explicitly at other values of $\theta$. The symmetry transformation is
given by
\begin{equation}
\varphi \mapsto -\varphi, \quad k \mapsto -k,\nn
\end{equation}
at $\theta=0$, and by
\begin{equation}
\varphi \mapsto -\varphi, \quad k \mapsto -k-1,\nn
\end{equation}
at $\theta=\pi$.  Most of the vacuum degeneracy is lifted by the non-zero
$M$ term in the potential, due to the fact that $Q$ and $q$ are
coprime. The new vacua are those closest to $\varphi=0$. At
$\theta =0$, there is a single vacuum at $\varphi=0, k=0$, invariant under the
$\Z_2$ time-reversal symmetry. At $\theta=\pi$, however, there are two
degenerate vacua related to each other by the time-reversal symmetry
(see Figure \ref{fig:potential-theta-0-Pi}). It is clear that there can be no
second order phase transition to the trivially gapped phase, unlike
what we saw earlier in the one-flavor cases for both the Dirac fermion
and the scalar fields. There is a $\mathbb{CP}^{N_f-1}$ non-linear
  sigma-model on the domain wall connecting the two vacua just like in
  the case with $m_{adj}=0$ consistent with the proposal \cite{Gaiotto:2017tne} for QCD.

\begin{figure}[h]
  \centering
  \includegraphics[scale=0.6]{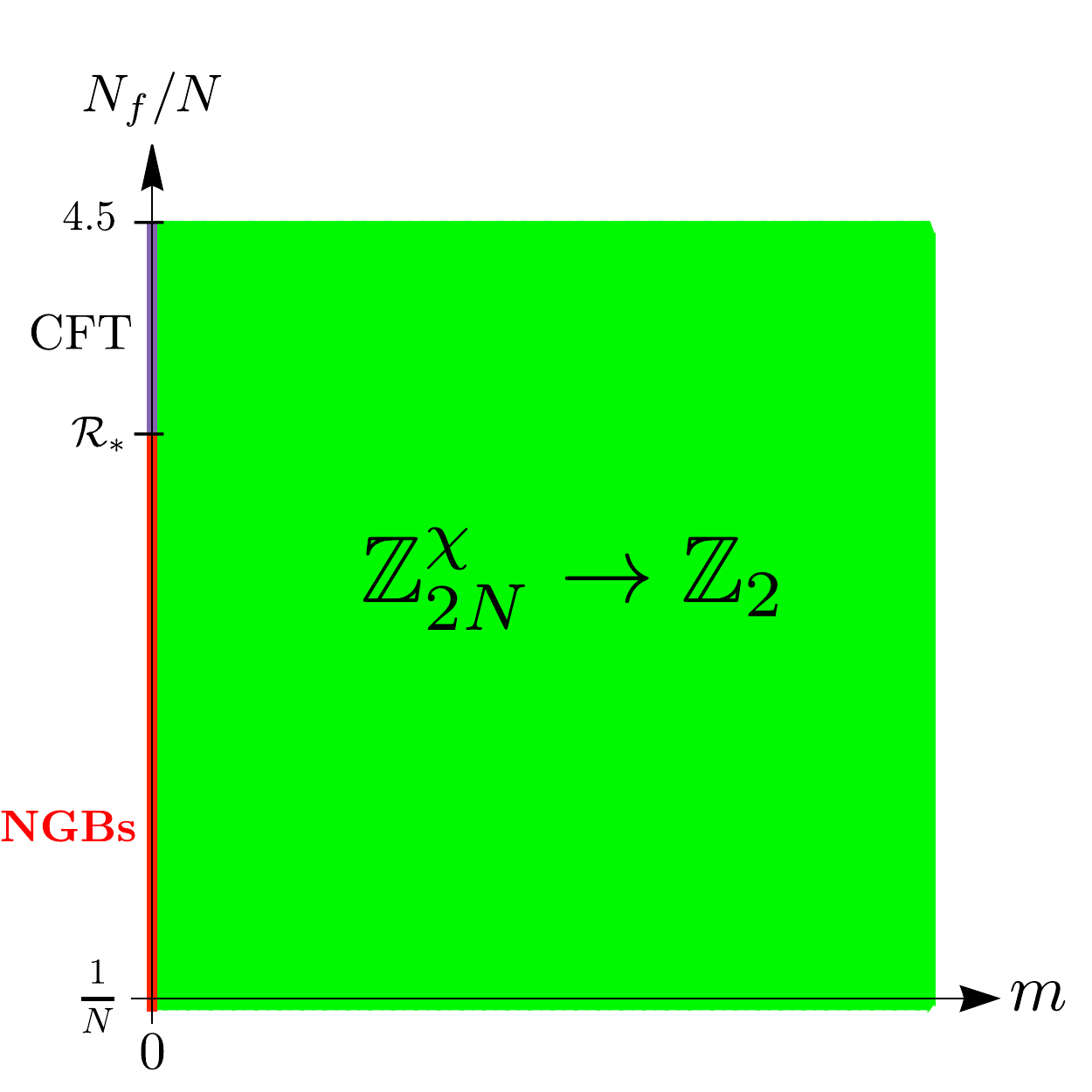}
  \caption{Phase diagram for the $\SU(N)$ QCD (f/adj) with fundamental
    Dirac fermions. It is color-coded as follows.  Purple means it is
    conformal; red means it is in a Goldstone phase; green means
    there is a broken discrete symmetry with domain walls joining
    different vacua, called the domain wall (DW) phase. The theory
    loses asymptotic freedom and is not well-defined in the UV when
    $N_f/N >4.5$}
  \label{fig:mult-ferm-phases}
\end{figure}
To summarise, when the adjoint fermion is massless, we can combine our
results from \S\S ~\ref{sec:one-Dirac-SUN-symmetry},
\ref{sec:multi-flavor-fermion} together to obtain a phase diagram in
terms of the fundamental fermions' mass $m$ and
, the ratio of the number of flavors to the number of
colors $\mathcal{R}=N_f/N$. The phase diagram is shown in Figure~\ref{fig:mult-ferm-phases}.

\begin{figure}[h]
  \centering
  \includegraphics[width=0.7\textwidth]{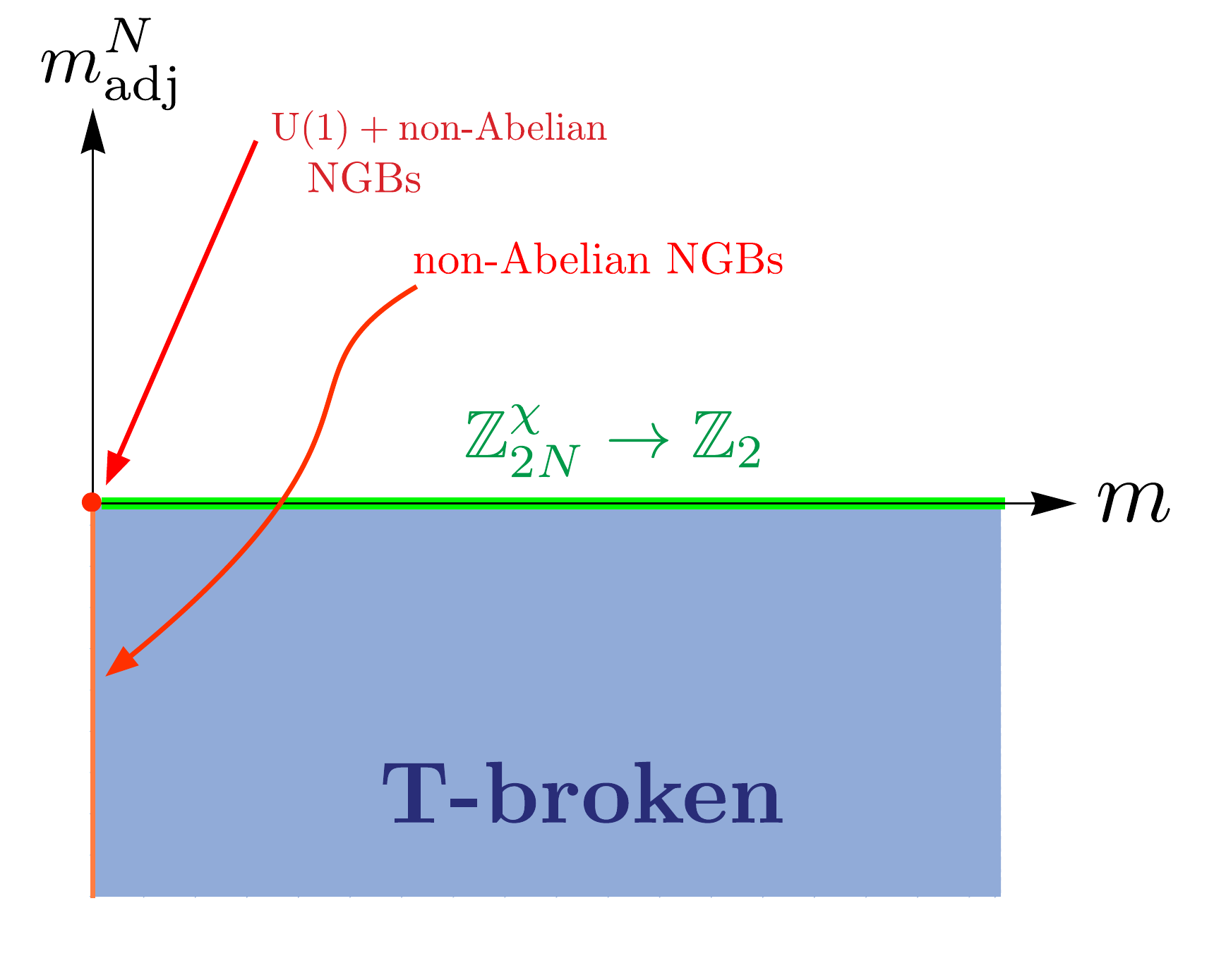}
  \caption{The $T$-invariant slice of the phase diagram for QCD
    (f/adj) with multiple fundamental fermion flavors. Note that when
    the fundamental fermions' mass vanishes, the theta-angle is not
    physical. We choose to represent the adjoint mass along the
    negative mass axis to emphasize that it is the end point of the
    $\Z_2$ broken phase. \label{fig:mult-ferm-phase-2d}}
\end{figure}

\begin{figure}[h]
  \centering
  \includegraphics[width=0.7\textwidth]{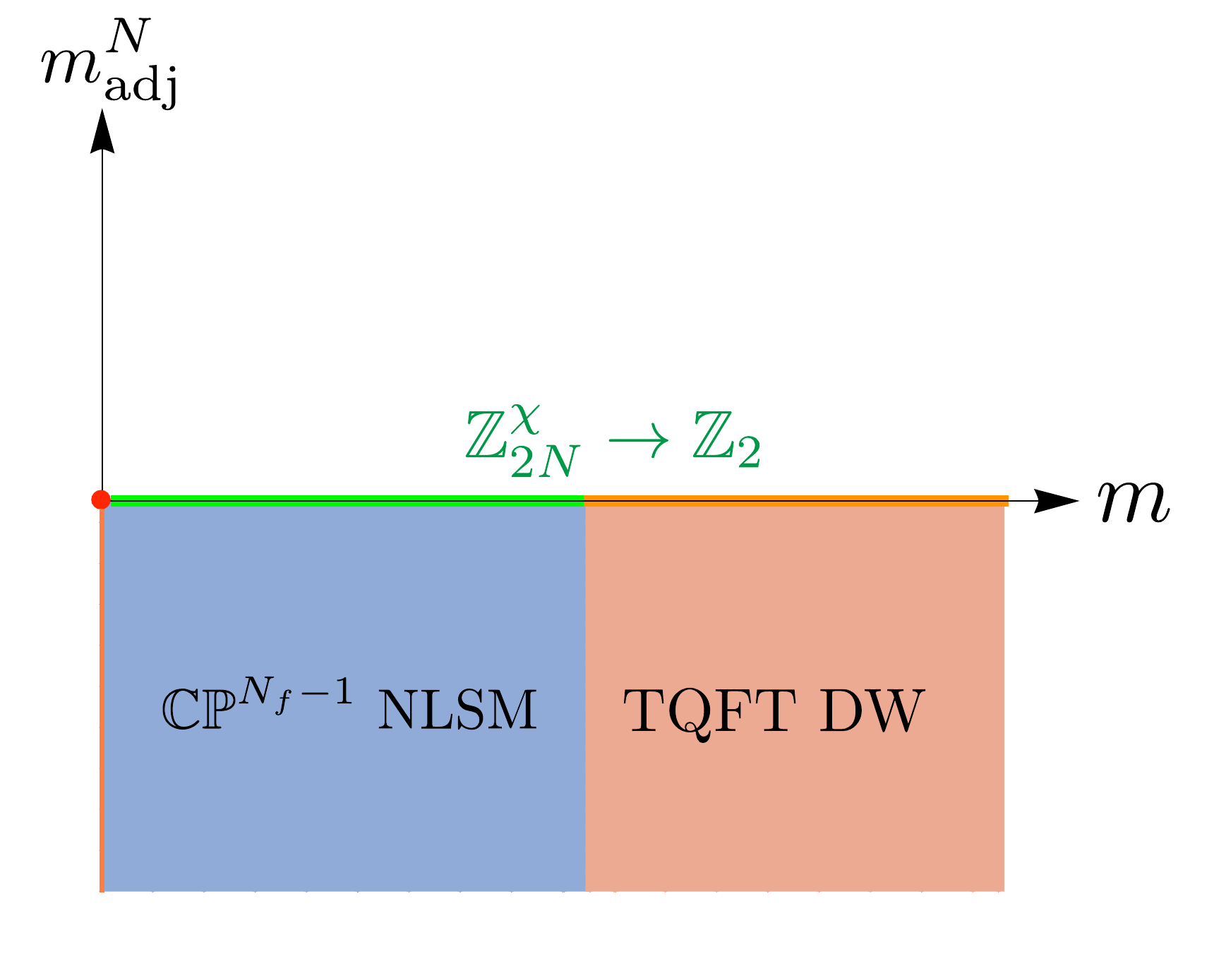}
  \caption{The same as Figure~\ref{fig:mult-ferm-phase-2d}, but showing the phases on the domain walls instead of the bulk phase. \label{fig:mult-ferm-phase-DW-2d}}
\end{figure}

On the other hand, when $N_f$ is fixed such that there is chiral
symmetry breaking, we can vary the adjoint mass $m_{adj}$ and obtain
the phase diagram shown in as shown in
Figure~\ref{fig:mult-ferm-phase-2d} by piecing together various limits
explored earlier. When all masses vanish, there are both a massless
Abelian Nambu--Goldstone boson as well as non-Abelian Nambu--Goldstone
bosons; only the non-Abelian ones remain when we turn on the adjoint
mass. Contrast this with the single fundamental fermion case in
Figure~\ref{fig:phase_adj_fund}. There, the existence of an Abelian
Nambu--Goldstone boson persists for all value of the adjoint fermion
mass, whereas in the multi-flavor case, it only appears when all
fermions are massless. There are no phase transition in the bulk
  as we dial the mass $m$ down to zero, but there are phase
  transitions on the domain walls, just like in the one-flavor case,
  from a TQFT to a $\mathbb{CP}^{N_f-1}$ non-linear sigma-model (NLSM)
  as shown in Figure~\ref{fig:mult-ferm-phase-DW-2d}.

\section{Theory with multiple fundamental scalars}
\label{sec:mult-scal-flav}

To cap off our analysis, let us turn to multiple scalar flavors
scenario in this section. Even though the anomaly story goes much the
same way as before, we will see that the IR behaves qualitatively
differently as we increase the number of the scalar fields.

The action for the matter fields reads
\begin{equation}
S_{\text{matter}} = \int \dd^4 x\, \left(  i \bar{\lambda} \left( \slashed{\partial} - i \slashed{a}_{adj} \right)\lambda + \sum_{i=1}^{N_b}\abs{(\partial- i a) \phi_i}^2 - V(\phi_i) \right)\;,
  \label{eq:S-mult-scalar}
\end{equation}
with the potential
\begin{equation}
V(\phi_i) = \sum_{i=1}^{N_b}m^2 \abs{\phi_i}^2+ \mathcal{O} (\abs{\phi}^4)\;.
\end{equation}
We give the same mass to all $N_b$ scalar fields, just like what we
did in \S~\ref{sec:multi-flavor-fermion}, to preserve as much global
symmetry as possible. Terms with larger power in $\phi$ are also assumed
to preserve the maximal symmetry.

\subsection{Symmetry and anomalies}
\label{sec:symmetry-anomalies-mult-scalar-flav}

The global symmetry group is
\begin{equation}
  \begin{split}
    G^{\text{Global}} &= \frac{\SU(N_b)\times \U_B}{\Z_N\times \Z_{N_b}} \times \Z^{\chi}_{2N}\\
    &= \frac{\UU(N_b)}{\Z_N} \times \Z^{\chi}_{2N}\;,
  \end{split}
\end{equation}
under which the matter content transforms in the representations given
by Table \ref{tab:rep-mult-scalars}. There is a mixed anomaly between
$\UU(N_f)/\Z_N$ and $\Z^{\chi}_{2N}$, exact as explained in \S
\ref{sec:mult-flav-ferm-massive}.
\begin{table}[h]
  \centering
  \begin{tabular}{c||c|ccc}
    & $\SU(N)$& $\SU(N_b)$ & $\U_B$ & $\Z^{\chi}_{2N}$\\
    \hline
    $\phi$& ${\tiny\yng(1)}$ & ${\tiny\yng(1)}$ & $+1$ & $0$\\
    $\lambda$ & $\textbf{adj}$ & $\textbf{1}$ & $0$ &$+1$
  \end{tabular}
  \caption{Representations of the matter content under the gauge and the global symmetry groups in the multiple scalar flavors case}
  \label{tab:rep-mult-scalars}
\end{table}

\subsection{The IR phases}
\label{sec:ir-phases-mult-scalar-flav}

When the mass parameter $m^2>0$, the scalar fields do not condense. At
the scale below the mass scale, we can integrate them out and again
obtain the same phase as the $\mathcal{N}=1$ $\SU(N)$ SYM: there are
$N$ distinct vacua and there are domain walls connecting
them. 

When $m^2<0$, we need to
include the quartic terms in the potential for stability. Let us
combine the scalar fields into an $N\times N_b$ matrix $\Phi$, which transform
under the $\SU(N)$ gauge group and the $\SU(N_b)$ global symmetry
group as
\begin{equation}
  \Phi \mapsto U \Phi V, \quad U \in \SU(N),\; V \in \SU(N_b)\;.\nn
\end{equation}
Then the most general potential invariant under the gauge and the
global symmetry group is
\begin{equation}
  V(\Phi) = -\abs{m}^2 \Tr \Phi^{\dag}\Phi + \kappa_1 \Tr \left( \Phi^{\dag}\Phi \Phi^{\dag} \Phi \right) + \kappa_2 \left( \Tr \Phi^{\dag}\Phi \right)^2 + \ldots  
\end{equation}
and we require $\kappa_1+N_b\kappa_2>0$ for stability. By adding appropriate
constant terms to this potential, we can complete the square and write
the potential as
\begin{equation}
  V(\Phi) = \kappa_1 \Tr \left( \Phi^{\dag}\Phi- v^2 \textbf{1}_{N_b} \right)^2 + \kappa_2 \left[ \Tr \left( \Phi^{\dag}\Phi - v^2 \textbf{1}_{N_b} \right) \right]^2
\end{equation}
with $v^2=\abs{m}^2/(\kappa_1+N_b\kappa_2)$.  $\Phi$ must now acquire vacuum
expectation value to minimize the potential, which is achieved by the
configuration $\Phi^{\dag}\Phi = v^2\textbf{1}_{N_b}$. In this Higgs regime, the
IR phases are sensitive to the number of scalar flavors $N_b$.  We
will now consider each different scenario in turn.

\subsubsection*{\underline{$N_b<N$}}

When $N_b<N$, $\expval{\Phi}\neq 0$ partially higgses the gauge group down
to $\SU(N-N_b)$ without spontaneous breaking of the global symmetry,
similar to the one flavor case. To see this, let us first rotate
$\expval{\Phi}$ by $\SU(N)$ and $\SU(N_b)$ transformations to be of the
diagonal form \footnote{This is none other than the {\it singular
    value decomposition} (SVD). We can always write any $N\times N_b$
  complex matrix $\Phi$ as $ \Phi = U \Phi_D V^{\dagger}$ where
  $\Phi_D$ is diagonal, and $U, V$ are $\SU(N)$ and $\SU(N_b)$ matrices.}
\begin{equation}
\expval{\Phi} = \left(
\begin{array}{ccccc}
  v_1 & 0 & \cdots & 0 & 0\\
  0 &  v_2 & \cdots  &0 &0\\
  \vdots & \vdots & \ddots & \vdots & \vdots\\
  0 & 0 &\cdots &0 & v_{N_b}\\
  \hline
  0 & 0 & \cdots & 0 & 0\\
  \vdots & \vdots & \ddots & \vdots & \vdots\\
  0 & 0 & \cdots & 0 & 0
\end{array}
\right)\,.
\end{equation}
This configuration minimizes the potential if and only if $v_i = v$
for all $i$. To see what symmetry is left unbroken by such a VEV, 
let's act on $\expval{\Phi}$ by a combined $\SU(N)\times \UU(N_b)$
transformation of the form
\begin{equation}
\expval{\Phi} \to U \expval{\Phi} V , \quad U =
\begin{pmatrix}
  U_1 & \\
   & U_2
\end{pmatrix} \in \SU(N), \quad V \in \UU(N_b),
\end{equation}
where $U_1$ and $U_2$ are $N_b \times N_b$ and $(N-N_b)\times (N-N_b)$ unitary
matrices, respectively. The transformed VEV is
\begin{equation}
  \label{eq:color-flavor-locking}
U \expval{\Phi} V =
\begin{pmatrix}
  v U_1 V \\ \textbf{0}^T\\ \vdots \\ \textbf{0}^T
\end{pmatrix}\,.
\end{equation}
We see that the transformation leaves the VEV invariant if and only if
we take $U_1 = V^{\dag}$, and $U_2 = e^{ i \theta} \tilde{U}$, where the
phase $\theta$ is determined by $V$ (because we need
$\det U_1 \det U_2 =1$), but are free to choose $\tilde{U}$ to be any
$\SU(N-N_b)$ matrix. We can clearly see that the gauge group $\SU(N)$
is higgsed down to $\SU(N-N_b)$, while the rest of the gauge group is
locked with the flavor symmetry $\UU(N_b)$, resulting in the
color-flavor locked $\UU(N_b)_{cf}$ global symmetry that survives in
the Higgs regime. The discrete chiral symmetry $\Z^{\chi}_{2N}$ does not
act on the scalar fields, so it remains intact in this phase.

The $\SU(N)$ adjoint fermion $\lambda$ decomposes into an $\SU(N-N_b)$
adjoint fermion $\tilde{\lambda}$, $N_b$ Weyl fermions in the fundamental
representation of $\SU(N-N_b)$, $\psi$, $N_b$ Weyl fermions in the
anti-fundamental representation, $\tilde{\psi}$, a neutral fermions in
the adjoint representation of $\UU(N_b)$, $\eta$, and one neutral Weyl
fermion $\nu$. The matter fields transform under the IR gauge group
$\SU(N-N_b)$ and the IR global symmetry
\begin{equation}
  G^{\text{Global}}_{\text{IR}}=  \frac{\SU(N_b)\times \U^{\prime}_B}{\Z_{N_b}\times \Z_{N-N_b}} \times \Z^{\chi}_{2N}
  \label{eq:GglobalIR-mutlscalars}
\end{equation}
in the representations given in Table \ref{tab:NblessNrep}. Again,
even though the free massless fermion matter content has an enhanced
global symmetry, as discussed in \S
\ref{sec:symmetry-anomalies-one-scalar}, there are irrelevant terms in
the Lagrangian that reduce the symmetry down to the one we have in the
UV. The enhancement that lifts $\Z^{\chi}_{2N}$ to a continuous chiral
symmetry happens only at $v=\infty$. The difference in the discrete
quotient between $G^{\text{Global}}_{\text{IR}}$ in
Eq. \eqref{eq:GglobalIR-mutlscalars} and $G^{\text{Global}}$ in the UV
is not a cause for concern. It simply reflects the fact that we assign
the $\U^{\prime}_B$ charge $\pm 1$ to the fundamental and anti-fundamental
fermions $\psi$, $\tilde{\psi}$. Note that the flavor symmetry that acts on
the scalar fields in the UV now acts on the fermions in the IR through
color-flavor locking, as we have already seen in the one-flavor scalar
case.
\begin{table}[h!]
  \centering
  \begin{tabular}{c||c|ccc}
    &$\SU(N-N_b)$ & $\SU(N_b)$ & $\U^{\prime}_B$ & $\Z^{\chi}_{2N}$\\
    \hline
    $\tilde{\lambda}$ & $\textbf{adj}$ & $\textbf{1}$ & $0$ & $1$\\
    $\psi$ & ${\tiny \yng(1)}$ & ${\tiny \yng(1)}$ & $1$ & $1$\\
    $\tilde{\psi}$ & $\overline{{\tiny \yng(1)}}$ & $\overline{{\tiny \yng(1)}}$ & $-1$ & $1$\\
    $\eta$ & $\textbf{1}$ & $\textbf{adj}$ & $0$ & $1$\\
    $\nu$ & $\textbf{1}$ & $\textbf{1}$ & $0$ & $1$
  \end{tabular}
  \caption{Representations of the fermions in the Higgs regime when $N_b<N$ \label{tab:NblessNrep}}
\end{table}

When $N_b/N>9/11$, the resulting gauge theory is IR-free, leading to
the free fermion CFT phase. When $N_b/N <9/11$, this $\SU(N-N_b)$
gauge theory is asymptotically free. We learn about its IR dynamics by
looking back at the dynamics of the fermionic theory that we studied
in \S~\ref{sec:multi-flavor-fermion}, because it is the theory that
emerges in the intermediate region, barring a few extra Weyl fermions
neutral under the gauge group that only couple to the rest via
higher-order terms. Thus, for any finite $v$, we expect to have domain
walls connecting $N$ vacua in the IR. The fermions that transform
non-trivially under the $\SU(N-N_b)$ gauge group are gapped out by the
gauge dynamics, while the neutral fermions are gapped out by the
interaction with the would-be Goldstone boson for the $\U$ chiral
symmetry.

In the deep Higgs regime where $v \gg \Lambda$, still assuming asymptotic
freedom, there are two options for the IR phases, depending on the
ratio between the number of flavors and the number of colors. There
exists a critical point $N_b/N = \mathcal{R}_{b*} $, below which we have chiral
symmetry breaking, and above which the phase enters a conformal
window. The value of $R_{b*}$ cannot be ascertained in the generic
case due to the strong dynamics involved. In the Veneziano limit,
however, the Banks-Zaks computation can be trusted, and we can
estimate $\mathcal{R}_{b*}$ to be
\begin{equation}
\mathcal{R}_{b*} = \frac{18}{31}  \approx 0.58\;,
\end{equation}
up to corrections in $1/N$.

\subsubsection*{\underline{$N_b=N-1$}}

When $N_b=N-1$, the gauge group is completely higgsed. The global
symmetry is still not spontaneously broken. The IR dynamics is that of
free fermions in multiplets of the global symmetry
$G^{\text{global}}$. The massless composite fermions $\eta$, $\psi$, and
$\tilde{\psi}$, now transform under the global symmetry group
\begin{equation}
  G^{\text{Global}}_{\text{IR}} = \frac{\U^{\prime}_B\times \SU(N_b)}{\Z_{N_b}} \times \Z^{\chi}_{2N} \cong \UU(N_b) \times \Z^{\chi}_{2N}
\end{equation}
in the adjoint representation, the fundamental representation, and the
anti-fundamental representation of $\UU(N_b)$, respectively. All of
them have unit charge under the discrete chiral symmetry
$\Z^{\chi}_{2N}$. We can dress these fermions with scalars to form gauge-invariant composites, as we did in (\ref{eq:xi123}). The special case $N=2$ was considered as a warmup exercise in Section \ref{A warmup: Higgs phase of SU(2) gauge theory with a single scalar}. Here, the global symmetry $\U_B$ is enhanced to $SO(3)$ custodial symmetry; see Footnote \ref{enhanced symmetry}.

\subsubsection*{\underline{$N_b=N$}}

When the number of flavors $N_b$ is equal to the number of colors $N$,
the gauge group is completely higgsed just like in the case when $N_b$
is one less than $N$. However, there is not enough room in the gauge
group $\SU(N)$ to fully preserve $\UU(N_b)= \UU(N)$ global symmetry
through color-flavor locking when $\expval{\Phi}\neq 0$. The $\U$ baryon
symmetry must now be broken spontaneously by the VEV scalar fields,
while the non-abelian flavor symmetry $\SU(N_b)$ remains unbroken. The
IR phase consists of one $\U$ Goldstone boson, as well as a massless
composite Weyl fermion in the adjoint representation of the global symmetry
$\SU(N_b)$.

\subsubsection*{\underline{$N_b>N$}}

Things get more complicated when $N_b>N$. We see from the SVD that the
equation $\Phi^{\dag}\Phi = v^2\textbf{1}_{N_b}$ has no solutions. Nonetheless,
the diagonal form
\begin{equation}
\expval{\Phi} =
\begin{pmatrix}
  v\textbf{1}_N & \textbf{0} & \ldots & \textbf{0}
\end{pmatrix}
\end{equation}
still minimizes the potential. The gauge group is still fully
higssed. Moreover, the global symmetry now breaks spontaneously to
$\SU(N)_{cf}\times \UU(N_b-N)$, which can be shown by the same argument
around Eq. \eqref{eq:color-flavor-locking} but in the opposite
direction. In the IR, there are Goldstone bosons described by the
coset space
\begin{equation}
\mathcal{M} = \frac{\UU(N_b)}{\SU(N)\times \UU(N_b-N)}
\end{equation}
apart from the composite fermions that we have previously. The 't
Hooft anomalies in the UV can be matched by the composite fermions and
the WZW term in the effective Lagrangian.

To summarise, there are many IR phases of the theory as the number
$N_b$ of the fundamental scalars and the mass squared $m^2$ of the
scalars varies. The ultraviolet (UV) theory has
$\mathbb Z_{2N}^{\chi}$ symmetry that acts on the massless adjoint. In
addition, baryon-number $\U_B$ and flavor $\SU(N_b)$
symmetries\footnote{The global symmetry is actually
  $\SU(N_b)/\mathbb Z_{\text{GCD}(N_b,N)}$, because the would-be
  global $\SU(N_b)$ and the $\SU(N)$ may have a common center.} act on
the scalars. The theory admits mixed anomalies between
$\mathbb Z_{2N}^{\chi}$ and $\U_B$ as well as between
$\mathbb Z_{2N}^{\chi}$ and gravity.  For $m^2>0$, we can integrate out
the scalars ending up with ${\cal N}=1$ super Yang-Mills theory. When
$m^2<0$, we need to distinguish between different scenarios depending
on $N$ and $N_b$. {\bf (A)} $N_b<9N/11$, the gauge group is higgsed
down to $\SU(N-N_b)$ that is still strongly-coupled. In the IR,
the theory enjoys emergent continuous symmetries. Assuming that the
Higgs vev is not much larger than $\Lambda$, the continuous symmetries are
explicitly broken down to $\mathbb Z_{2N}^{\chi}$ by dangerously
irrelevant operators.  The IR theory breaks the
$\mathbb Z_{2N}^{\chi}$ chiral symmetry spontaneously, leading to $N$
vacua and domain walls. {\bf (B)} $9N/11 < N_b < N-1$, the gauge
  group is again higgsed down to $\SU(N-N_b)$. However, the new gauge
  coupling becomes weaker under the RG flow. The composite fermions,
  i.e., the UV fermions dressed by scalars that transform covariantly
  under $\SU(N-N_b)$, decouples from the gauge fields in the IR. The
  theory then thus flows to the composite free fermion phase. {\bf
  (C)} $N_b=N-1$, the gauge group is fully higgsed, in which case
composite free fermions match the anomalies. These are the UV fermions
dressed by scalars in a fashion that preserves gauge invariance. {\bf
  (D)} $N_b=N$, the gauge group is fully higgsed, the flavor
$\SU(N_b)$ is intact, and $\U_B$ is spontaneously broken. The IR phase
contains one massless adjoint fermion in the global flavor group and
one Goldstone boson. {\bf (E)} $N_b>N$, the gauge group is fully
higgsed, and the continuous symmetry is broken, leading to many
Goldstones. These various cases are neatly summarised in a phase
diagram, shown in Fig.  \ref{fig:mult-scalars-diag} below.
\begin{figure}[h]
  \centering
  \includegraphics[scale=0.6]{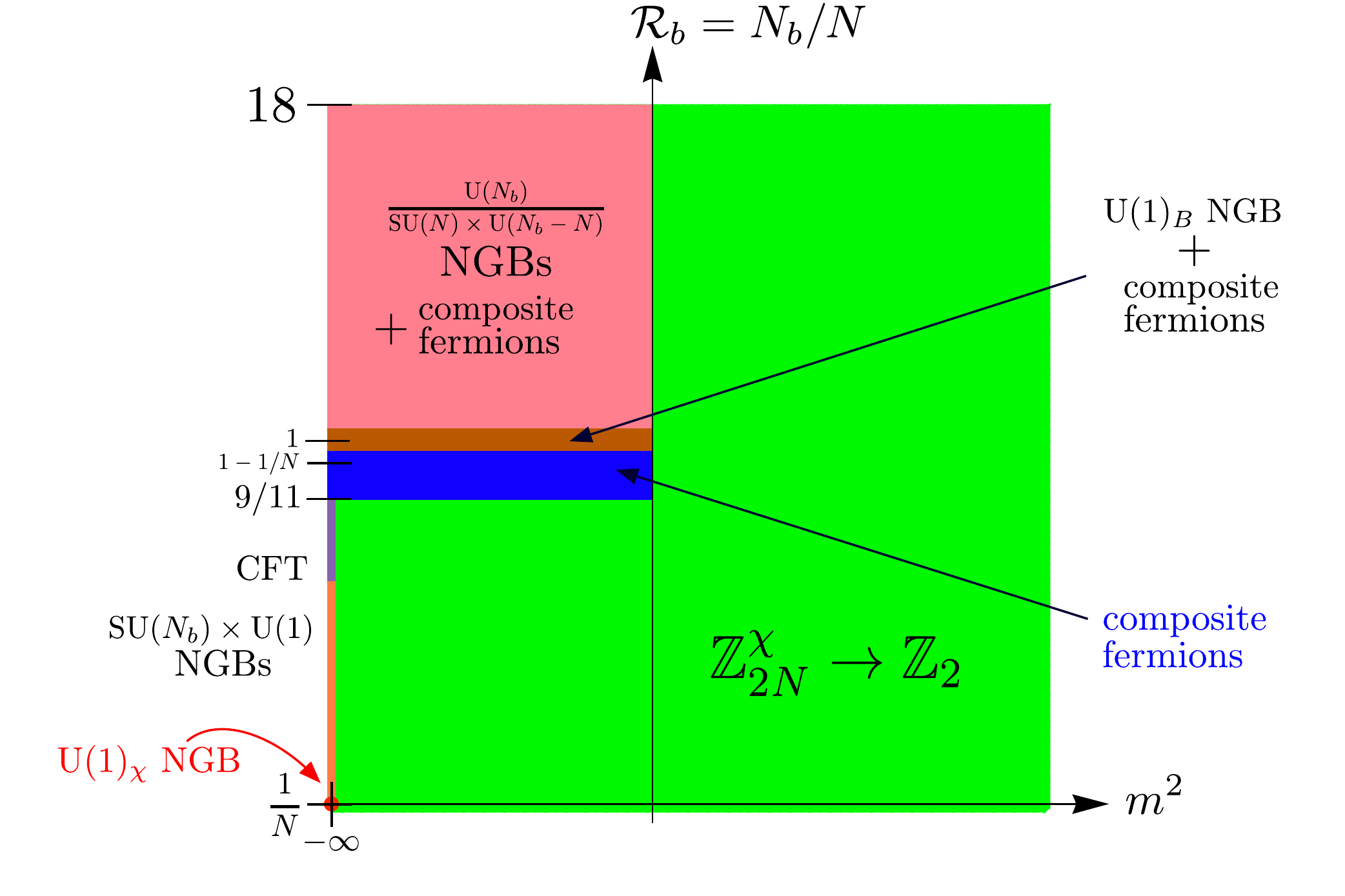}
  \caption{Phase diagram for the $\SU(N)$ QCD (f/adj) with fundamental
    scalars. “Composite fermions” here means the UV adjoint fermion
    dressed with the scalar fields in a gauge-covariant way under the
    gauge group $\SU(N-N_b)$ after higgsing. The theory loses asymptotic
    freedom and is not well-defined in the UV when $N_b/N > 18$.}
  \label{fig:mult-scalars-diag}
\end{figure}

\section{Conclusions and future prospects}\label{sec:conclusions}

In this work, we analyzed the $\SU(N)$ gauge theory with one massless adjoint fermion and massive matter, either bosonic or fermionic. The special case of one fundamental flavor of varying mass was analyzed in detail. The abundance of 't Hooft anomalies involving discrete chiral symmetry $\Z_{2N}^\chi$, combined with Vafa-Witten-Weingarten theorems, restricts the bulk phase quite strongly, which results in spontaneously broken discrete chiral symmetry for any mass of the fundamental fermion or boson, leaving $N$ degenerate vacua in the bulk. Further, the bosonic theory is related to fermionic theory in the sense that when the boson condenses, the theory is higgsed down to $\SU(N-1)$ with one fundamental fermion. So most conclusions about the theory with fundamental bosons can be drawn by studying the fermionic theory.

The massless fundamental fermion limit is particularly interesting, where the discrete chiral symmetry enhances to $\U_{\chi}$, and the domain walls melt into Goldstone bosons. Studying the small fundamental mass regime can be done systematically by perturbing the Goldstone theory. In particular, we studied the domain walls between the $N$ vacua and, with the exception of the one $T$-preserving domain wall, found them all to be trivial. On the other hand, in the opposite limit of infinite fundamental fermion mass, the theory becomes Super Yang-Mills, of which many things are known. The domain walls of Super Yang-Mills are conjectured to hold a TQFT. As the fundamental fermion mass is dialed from small to large, no bulk phase transition occurs, but our analysis shows that a transition must occur on the domain wall. Further, we conjectured that the domain wall theory is IR dual to the corresponding $\SU(N)$ gauge theory with both fundamental and adjoint matter in 3d. Of particular interest is the $T$-preserving domain wall, which exists if $N$ is even. This domain wall cannot be made trivial as it carries a $\U_B-T$ mixed anomaly. The corresponding 3d theories were studied in \cite{Lohitsiri:2022jyz}, where it was proposed that anomalies can be saturated either by composite fermions or spontaneous $T$-breaking, with large $N$ arguments favoring the latter. The 4d domain wall analysis indicates that $T$ breaking is preferred, at least for a small enough mass of the fundamentals. We, however, speculate that there may be a composite fermion phase on the domain wall for some intermediate mass of the fundamentals for $N=2$. We also discuss the decoupling limit of adjoints which results in the usual QCD. 

We generalized the one fundamental flavor case to multi-flavors and discussed the bounds on the conformal window from the a-theorem and the Banks-Zacks 2-loop fixed point. For a sufficiently low number of flavors, we analyze the chiral Lagrangian and map out the phase diagram in the bulk and on the domain wall. As with one flavor, we discuss the decoupling limit of adjoints. 

We end this section by discussing some future prospects. The conjectures about the bulk phase analyticity can be tested on the lattice. More interesting would be to study domain walls on the lattice, or the corresponding 3d gauge theories. The lattice studies of domain walls would require using twisted boundary conditions or spatially varying $\theta$-term, which generically introduce a complex action problem that hinders numerical simulations. Studies of corresponding 3d theories directly also generically require bare Chern-Simons terms and complex fermionic measures, which again hinders lattice studies. Another interesting approach is to study soft SUSY deformations of super QCD setup, and see carefully what happens on the domain walls as supersymmetry is broken.

\acknowledgments{We would like to thank Joe Davighi, Thomas Dumitrescu, Avner Karasik, Zohar Komargodski, Erich Poppitz and David Tong. TS would like to thank the organizers of the Confinement Workshop 2023 in Minneapolis, where interesting discussions took place. This work is supported by the STFC consolidated grant in Particles, Strings and
  Cosmology number ST/T000708/1. TS and NL are also supported by the Royal Society of London. }

\appendix
\section{Majorana and Weyl Fermions in the Real Gauge Group Representation}
\label{app:measure-positive-definite}
\subsection{Converting to the Majorana in 4d}

Consider a Weyl fermion in 4d Lorentz space given by the action
\begin{equation}
\bar \lambda i D\lambda\;,
\end{equation}
where $D=\sigma^\mu D_\mu$ with 
\begin{equation}
\sigma^\mu=(\mathbb I,\tau^1,\tau^2,\tau^3)\;,
\end{equation}
and $\tau^i$ are the Pauli matrices. We will take $D_\mu$ to be the covariant derivative in a real representation of some gauge group, so that $D_\mu^*=D_\mu$. 

We can write the complex Weyl fermion as a real fermion by writing $\lambda=\lambda_1+i \lambda_2$. Then we have that
\begin{equation}
\bar\lambda i D\lambda= \lambda_1^T i\sigma^\mu D_\mu \lambda_1+i\lambda_2^T\sigma^\mu D_\mu \lambda_2-\lambda_1^T\sigma^\mu D_\mu \lambda_2+\lambda_2^T\sigma^\mu D_\mu \lambda_1\;.
\end{equation}
Then we have that $\lambda_1 i\sigma^\mu D_\mu \lambda_1$ and $\lambda_2 i\sigma^\mu D_\mu \lambda_2$ do not get a contribution from the anti-symmetric $\sigma^2$ matrix, while  
 $-\lambda_1\sigma^\mu D_\mu \lambda_2$ and $\lambda_2\sigma^\mu D_\mu \lambda_1$ do not get a contribution from the symmetric matrices $\mathbb I, \sigma^1,\sigma^3$. Now organize $\lambda_1$ and $\lambda_2$ into a column vector $\Psi= \begin{pmatrix}
\lambda_1\\
\lambda_2
\end{pmatrix}$. We can write the above action as
\begin{equation}
\Psi^T i\Gamma^\mu D_\mu \Psi\;,
\end{equation}
where
\begin{align}
&\Gamma^0= \begin{pmatrix}
\mathbb I & 0\\
0& \mathbb I
\end{pmatrix}&&\Gamma^1= \begin{pmatrix}
\sigma^1 & 0\\
0&\sigma^1
\end{pmatrix}\\
&\Gamma^2= \begin{pmatrix}
0&  i\sigma^2\\
-i\sigma^2 & 0
\end{pmatrix}&&\Gamma^3=  \begin{pmatrix}
\sigma^3&0  \\
0& \sigma^3
\end{pmatrix}
\end{align}
Now let us set
\begin{equation}
C = \begin{pmatrix}
0 & i\sigma^2\\
i\sigma^2 & 0
\end{pmatrix}\;.
\end{equation}
Notice that the matrix is real and anti-symmetric. It is also unitary because $C^\dagger = C^{-1}$. We want to set $\Gamma^\mu=C\gamma^\mu$, so that $\gamma^\mu=C^{-1}\Gamma^\mu$. We then have that
 
\begin{equation}
\bar\lambda i D\lambda= i\Psi^TC \gamma^\mu D_\mu \Psi\;,
\end{equation}
with 
\begin{align}
&\gamma^0= \begin{pmatrix}
0 & -i \sigma^2\\
-i\sigma^2 & 0
\end{pmatrix}&&\gamma^1= \begin{pmatrix}
0 &  -\sigma^3\\
-\sigma^3 & 0
\end{pmatrix}\\
&\gamma^2= \begin{pmatrix}
- \mathbb I &  0\\
0 & \mathbb I
\end{pmatrix}&&\gamma^3=  \begin{pmatrix}
0& \sigma^1 \\
\sigma^1& 0
\end{pmatrix}\;,
\end{align}
Note that all the Gamma-matrices are purely real. One can also check that $\gamma^\mu$ as defined above satisfy the Clifford algebra
\begin{equation}
\{\gamma^\mu,\gamma^\nu\}= 2\;\text{diag}(-1,1,1,1)_{\mu\nu}
\end{equation}

\subsection{The Weight of the Dirac and Majorana Fermion in 4d, with real gauge representation}

If we have a Dirac fermion in $4d$ spacetime, we can write its Euclidean action as
\begin{equation}
S=\int d^4x\; \bar\Psi i\slashed D \Psi\;,
\end{equation}
where $\slashed D=\gamma^\mu D_\mu$ is the Dirac operator, and $D_\mu$ is a covariant derivative $D_\mu=\partial_\mu+A_\mu$, where $A_\mu$ is the gauge field in the real representation of some group G, i.e. $A_\mu^*=A_\mu$. 

In Euclidean space $i\slashed D$ is Hermitian, and so its eigenvalues are real. Let $\psi_n$ be the eigenfunctions of $i\slashed D$ with eigenvalues $\lambda_n$. Then we can decompose 
\begin{align}
\Psi=\sum_n \alpha_n \psi_n\;,\\
\bar\Psi= \sum_n \bar\alpha_n \psi_n^\dagger\;,
\end{align}
where $\alpha_n$ and $\bar\alpha_n$ are independent Grassmann numbers, and where\footnote{If there are degeneracies $\lambda_n=\lambda_m$ for $n\ne m$ we can still chose that degenerate eigenstates are orthogonal.} $\int d^4x\; \psi_n^\dagger \psi_m=\delta_{nm}$. We have that the action is given by
\begin{equation}
S=\sum_n \lambda_n \bar\alpha_n \alpha_n\;,
\end{equation}
so if we define the measure of the Grassmann integral as\footnote{the factor of $i$ is there by convention, and is just an overall normalization. In this convention the weight is always positive.} $\prod_n (id\bar\alpha_n d\alpha_n)$ we have that the path-integral weight is
\begin{equation}
\prod_n (i\lambda_n)=\det \slashed D\;.
\end{equation}
Notice that because $\{\slashed D,\gamma^5\}=0$ we have that for every eigenstate $\psi_n$ with eigenvalue $\lambda_n$ there exists an eigenstate $\gamma^5\psi_n$ with the eigenvalue $-\lambda_n$. So we can rewrite (assuming no $\lambda_n=0$)
\begin{equation}
\prod_n (i\lambda_n)= \prod_{n | \lambda_n>0}\lambda_n^2\;.
\end{equation}
so that the weight is positive definite.  In addition we will see that each eigenvalue $\lambda_n$ is twice degenerate because it forms a Kramers doublet. 

Now let us move to Majorana fermions. In this case $\bar\Psi=\psi^T C$ where $C$ is an unitary, anti-symmetric matrix with the property
\begin{equation}
C\gamma^\mu C^{-1}=- (\gamma^\mu)^T\;.
\end{equation}
Now notice that the Dirac operator $i\slashed D$ has a degeneracy, because if $\psi_n$ has an eigenvalue $\lambda_n$, then $C^{-1}\psi_n^*$ has the same eigenvalue. Indeed, since $(\gamma^\mu)^\dagger=\gamma^\mu$, we have
\begin{equation}
i\slashed D C^{-1}\psi_n^*= -C^{-1}i\slashed D^T\psi_n^*= C^{-1} (i\slashed D^\dagger \psi_n)^*=C^{-1} (i\slashed D^\dagger \psi_n)^*=\lambda_n C^{-1}\psi_n^*\;.
\end{equation}
Moreover $C^{-1}\psi_n^*$ is orthogonal to $\psi_n$ by the anti-symmetry of $C$, i.e.
\begin{equation}
(C^{-1}\psi_n^*)^\dagger \psi_n= \psi_n^T C\psi_n=0\;.
\end{equation}
So $i\slashed D$ has at least a double degeneracy of the spectrum.

One can also see this as a Kramers degeneracy \cite{Witten:2015aba}. Indeed if $K$ is a complex conjugation operator, we define $\mathcal T = C^{-1}K$ an operator which commutes with $i\slashed D$. Now $\mathcal T^2= C^{-1}KC^{-1}K=-C^{-1}C=-1$, where we used the unitarity and anti-symmetry of $C$.

Let us hence label $\psi_n^{i}$ with $i=1,2$ labels the Kramers doublet. Now we expand the Majorana fermion fields as
\begin{align}
\Psi= \sum_{n,i} \alpha_n^{i} \psi_n^{i}\;,\\
\bar\Psi=\sum_{n,i}\alpha_n^i (\psi_n^{i})^{T}C\;.
\end{align}
Now notice that
\begin{equation}
 \int d^4x\; (\psi_n^i)^T C i\slashed D \psi_m^j = \lambda_m \int (\psi_n^i)^T C\psi_m
\end{equation}
On the other hand we have that, by partially integrating,
\begin{equation}
\int d^4x\; (\psi_n^i)^T C i\slashed D \psi_m^j = \int d^4x (\slashed D\gamma^\mu \psi_n^i)^T C\psi_m^j= \lambda_n\int d^4x (\psi_n^i)^T C\psi_m^j\;.
\end{equation}
Combining the two expressions we have that\footnote{We will assume that the only degeneracy in the spectrum is the Kramers degeneracy so that $\lambda_n=\lambda_m$ implies $n=m$.}
\begin{equation}
\int d^4x (\psi_n^i)^T C\psi_m^j = 0 \;\quad \text{ if  $n\ne m$}\;.
\end{equation}
On the other hand if $n=m$ then we see that the expression $\int d^4x (\psi_n^i)^T C\psi_n^j$ is anti-symmetric in $i$ and $j$. We use a natural normalization
\begin{equation}
\int d^4x (\psi_n^i)^T C\psi_n^j =\epsilon^{ij}\;.
\end{equation}
Then the action becomes
\begin{equation}
S= \sum_n \lambda_n \alpha_n^i\alpha_n^j\epsilon_{ij}\;.
\end{equation}
Now we define the measure to be
\begin{equation}
\prod_n(id\alpha_n^1d\alpha_n^2)\;,
\end{equation}
so that the weight is
\begin{equation}
\prod_n (i\lambda_n)= \prod_{n | \lambda_n>0} \lambda_n^2\;.
\end{equation}
where the product over $\lambda_n$ is only over one of the Kramers doublet eigenvalue. The above is manifestly positive.

\section{Spectral flow}\label{app:spectral}

Consider a first-order differential operator
\beq
D=\mathbb I\partial_\tau +A(\tau)\;.
\eeq
where $A(\tau)$ and $\mathbb I$ are an $N\times N$ Hermitian and identity matrices respectively. We want to look for the zero modes of the above operator. We solve the differential equation 
\beq
D\psi=0\;.
\eeq
Now let us decompose $\psi=\sum_{n=1}^Nc_n(\tau)\chi_n(\tau)$ into the instantaneous eigenvectors $\chi_n(\tau)$ of $A(\tau)$, i.e.
\beq\label{eq:A_eigeneq}
A(\tau)\chi_n(\tau)=\lambda_n(\tau)\chi_n(\tau)\;.
\eeq
We have that the zero-mode equation becomes equivalent to
\beq
\partial_\tau c_n+ \lambda_n c_n+\sum_{m=1}^NM_{nm}c_m=0
\eeq
where the matrix $M_{nm}$ is an anti-Hermitian matrix given by
\beq
M_{nm}=\chi^\dagger_{m}\partial_\tau \chi_n\;.
\eeq
Let us put the coefficients  $c_n$ into a complex vector $\bm c$, and $\lambda_n$ into a diagonal matrix $A_D$. Then we have that
\beq\label{eq:c_tau_sol}
\bm c(\tau)=Pe^{-\int_0^\tau d\tilde\tau (A_D(\tilde\tau)+M(\tilde\tau))}\bm c(0)\;.
\eeq
Now let us discuss this solution in the adiabatic approximation limit. Namely if we differentiate the equation \eqref{eq:A_eigeneq}, ignoring the derivative of $A$ and the derivative of $\lambda_n$ as small, we have that (sum over $m$ implied)
\beq
\lambda_m M_{mn}-M_{mn}\lambda_n=0 \Leftrightarrow \Lambda M-M A_D=0\;,
\eeq
so $M_{mn}$, commuting with $A_D$ must be diagonal unless $A_D$ has exact degeneracies. Let us assume that this is the case. Then the equation \eqref{eq:c_tau_sol} implies that if we start with $c_n(0)$ not equal to zero for only some $n$ and zero for others, it will stay that way. The diagonal matrix $M$ is just the Barry phase of individual eigenstates. Now notice that only $c_n(\tau)$ for which $\lambda_n$ is positive for $\tau\rightarrow \infty$ and negative for $\tau\rightarrow -\infty$ can be kept if we want normalizable $\psi(\tau)$. Hence we conclude that the operator $D$ has as many zero modes as the net spectral flow. If operator $A(\tau)$ still has some degeneracies, the story is similar because we can always diagonalize $M$ in the subspace of the degeneracies without affecting the discussion.

%%%%%%%%%%%%%%%%%%
\section[Beta-function]{$\beta$-function}
\label{beta function section}
%%%%%%%%%%%%%%%%%%%

The 3-loop $\beta$ function for $n_{R}$ Weyl fermions in representation $R$ of $\SU(N)$ Yang-Mills theory is given by (see \cite{Dietrich:2006cm,Zoller:2016sgq})
\begin{eqnarray}
\begin{split}
\beta(g)=&-\beta_0\frac{g^3}{(4\pi)^2}-\beta_1\frac{g^5}{(4\pi)^4}-\beta_2\frac{g^7}{(4\pi)^6}~,
\\[3pt]
\beta_0=&\frac{11}{6}C_2(G)-\sum_{ R}\frac{1}{3}T_{R}n_{R}~,
\\[3pt]
\beta_1=&\frac{34}{12}C_2^2(G)-\sum_{{ R}}n_{R}\left\{\frac{5}{6} C_2(G)T_{R} + \frac{1}{2}C_2({ R})T_{R}\right\}~,
\\[3pt]
\beta_2=&\frac{2857}{432}C_2^3(G)-\sum_{{R}}\frac{n_{R}T_{ R}}{4} \left[-\frac{C_2^2({ R})}{2}+\frac{205 C_2(G)C_2({R})}{36}+\frac{1415C_2^2(G)}{108} \right]
\\[3pt]
&+\quad\sum_{{R},{R}'} \frac{n_{ R}n_{ R'} T_{R}T_{R'}}{16}\left[\frac{44 C_2({R})}{18}+\frac{158C_2(G)}{54} \right]~,
\end{split}
\label{beta function}
\end{eqnarray} 
where $G$ denotes the adjoint representation.
% and $n_{R}$ is the number of the Weyl flavors in representation ${R}$.
The quadratic Casimir of representation $R$,  $C_2({ R})$, is  
\begin{eqnarray}
t^a_{R}t^a_{ R}=C_2({R})\mathbf{1}_{ R}~,
\label{Casimir}
\end{eqnarray}
and $C_2(G)$ is the quadratic Casimir of the adjoint representation. $T_{R}$ is the Dynkin index, which is defined via
\begin{eqnarray}
\mbox{tr}\left[t^a_{R}t^b_{R}\right]=T_{R}\delta^{ab}~.
\label{trace}
\end{eqnarray}
From Eqs. (\ref{Casimir}) and (\ref{trace}), we obtain the relation 
\begin{eqnarray}
T_{R}\mbox{dim}_G=C_2({ R})\mbox{dim}_{R}~,
\end{eqnarray}
where $\mbox{dim}_{ R}$ is the dimension of $R$. In particular, using the convention $T_{R}=1$ for the fundamental representation $R={\tiny \yng(1)}$, we have $C_2(G)=2N$, $\mbox{dim}_G=N^2-1$.

\bibliographystyle{JHEP} \bibliography{references}

\providecommand{\href}[2]{#2}\begingroup\raggedright\begin{thebibliography}{10}

\bibitem{Kapustin:2014gua}
A.~Kapustin and N.~Seiberg, \emph{{Coupling a QFT to a TQFT and Duality}},
  \href{http://dx.doi.org/10.1007/JHEP04(2014)001}{\emph{JHEP} {\bf 04} (2014)
  001}, [\href{https://arxiv.org/abs/1401.0740}{{\tt 1401.0740}}].

\bibitem{Gaiotto:2014kfa}
D.~Gaiotto, A.~Kapustin, N.~Seiberg and B.~Willett, \emph{{Generalized Global
  Symmetries}}, \href{http://dx.doi.org/10.1007/JHEP02(2015)172}{\emph{JHEP}
  {\bf 02} (2015) 172}, [\href{https://arxiv.org/abs/1412.5148}{{\tt
  1412.5148}}].

\bibitem{Gaiotto:2017yup}
D.~Gaiotto, A.~Kapustin, Z.~Komargodski and N.~Seiberg, \emph{{Theta, Time
  Reversal, and Temperature}},
  \href{http://dx.doi.org/10.1007/JHEP05(2017)091}{\emph{JHEP} {\bf 05} (2017)
  091}, [\href{https://arxiv.org/abs/1703.00501}{{\tt 1703.00501}}].

\bibitem{Gaiotto:2017tne}
D.~Gaiotto, Z.~Komargodski and N.~Seiberg, \emph{{Time-reversal breaking in
  QCD$_{4}$, walls, and dualities in 2 + 1 dimensions}},
  \href{http://dx.doi.org/10.1007/JHEP01(2018)110}{\emph{JHEP} {\bf 01} (2018)
  110}, [\href{https://arxiv.org/abs/1708.06806}{{\tt 1708.06806}}].

\bibitem{Komargodski:2017dmc}
Z.~Komargodski, A.~Sharon, R.~Thorngren and X.~Zhou, \emph{{Comments on Abelian
  Higgs Models and Persistent Order}},
  \href{http://dx.doi.org/10.21468/SciPostPhys.6.1.003}{\emph{SciPost Phys.}
  {\bf 6} (2019) 003}, [\href{https://arxiv.org/abs/1705.04786}{{\tt
  1705.04786}}].

\bibitem{Komargodski:2017smk}
Z.~Komargodski, T.~Sulejmanpasic and M.~\"Unsal, \emph{{Walls, anomalies, and
  deconfinement in quantum antiferromagnets}},
  \href{http://dx.doi.org/10.1103/PhysRevB.97.054418}{\emph{Phys. Rev. B} {\bf
  97} (2018) 054418}, [\href{https://arxiv.org/abs/1706.05731}{{\tt
  1706.05731}}].

\bibitem{Komargodski:2017keh}
Z.~Komargodski and N.~Seiberg, \emph{{A symmetry breaking scenario for
  QCD$_{3}$}}, \href{http://dx.doi.org/10.1007/JHEP01(2018)109}{\emph{JHEP}
  {\bf 01} (2018) 109}, [\href{https://arxiv.org/abs/1706.08755}{{\tt
  1706.08755}}].

\bibitem{Sulejmanpasic:2018upi}
T.~Sulejmanpasic and Y.~Tanizaki, \emph{{C-P-T anomaly matching in bosonic
  quantum field theory and spin chains}},
  \href{http://dx.doi.org/10.1103/PhysRevB.97.144201}{\emph{Phys. Rev. B} {\bf
  97} (2018) 144201}, [\href{https://arxiv.org/abs/1802.02153}{{\tt
  1802.02153}}].

\bibitem{Tanizaki:2018xto}
Y.~Tanizaki and T.~Sulejmanpasic, \emph{{Anomaly and global inconsistency
  matching: $\theta$-angles, $SU(3)/U(1)^2$ nonlinear sigma model, $SU(3)$
  chains and its generalizations}},
  \href{http://dx.doi.org/10.1103/PhysRevB.98.115126}{\emph{Phys. Rev. B} {\bf
  98} (2018) 115126}, [\href{https://arxiv.org/abs/1805.11423}{{\tt
  1805.11423}}].

\bibitem{Tanizaki:2018wtg}
Y.~Tanizaki, \emph{{Anomaly constraint on massless QCD and the role of
  Skyrmions in chiral symmetry breaking}},
  \href{http://dx.doi.org/10.1007/JHEP08(2018)171}{\emph{JHEP} {\bf 08} (2018)
  171}, [\href{https://arxiv.org/abs/1807.07666}{{\tt 1807.07666}}].

\bibitem{Karasik:2019bxn}
A.~Karasik and Z.~Komargodski, \emph{{The Bi-Fundamental Gauge Theory in 3+1
  Dimensions: The Vacuum Structure and a Cascade}},
  \href{http://dx.doi.org/10.1007/JHEP05(2019)144}{\emph{JHEP} {\bf 05} (2019)
  144}, [\href{https://arxiv.org/abs/1904.09551}{{\tt 1904.09551}}].

\bibitem{Anber:2018iof}
M.~M. Anber and E.~Poppitz, \emph{{Two-flavor adjoint QCD}},
  \href{http://dx.doi.org/10.1103/PhysRevD.98.034026}{\emph{Phys. Rev. D} {\bf
  98} (2018) 034026}, [\href{https://arxiv.org/abs/1805.12290}{{\tt
  1805.12290}}].

\bibitem{Anber:2018jdf}
M.~M. Anber and E.~Poppitz, \emph{{Anomaly matching, (axial) Schwinger models,
  and high-T super Yang-Mills domain walls}},
  \href{http://dx.doi.org/10.1007/JHEP09(2018)076}{\emph{JHEP} {\bf 09} (2018)
  076}, [\href{https://arxiv.org/abs/1807.00093}{{\tt 1807.00093}}].

\bibitem{Anber:2019nfu}
M.~M. Anber, \emph{{Self-conjugate QCD}},
  \href{http://dx.doi.org/10.1007/JHEP10(2019)042}{\emph{JHEP} {\bf 10} (2019)
  042}, [\href{https://arxiv.org/abs/1906.10315}{{\tt 1906.10315}}].

\bibitem{Anber:2021lzb}
M.~M. Anber, \emph{{Condensates and anomaly cascade in vector-like theories}},
  \href{http://dx.doi.org/10.1007/JHEP03(2021)191}{\emph{JHEP} {\bf 03} (2021)
  191}, [\href{https://arxiv.org/abs/2101.04132}{{\tt 2101.04132}}].

\bibitem{Sulejmanpasic:2020zfs}
T.~Sulejmanpasic, Y.~Tanizaki and M.~\"Unsal, \emph{{Universality between
  vector-like and chiral quiver gauge theories: Anomalies and domain walls}},
  \href{http://dx.doi.org/10.1007/JHEP06(2020)173}{\emph{JHEP} {\bf 06} (2020)
  173}, [\href{https://arxiv.org/abs/2004.10328}{{\tt 2004.10328}}].

\bibitem{Smith:2021vbf}
P.~B. Smith, A.~Karasik, N.~Lohitsiri and D.~Tong, \emph{{On discrete anomalies
  in chiral gauge theories}},
  \href{http://dx.doi.org/10.1007/JHEP01(2022)112}{\emph{JHEP} {\bf 01} (2022)
  112}, [\href{https://arxiv.org/abs/2106.06402}{{\tt 2106.06402}}].

\bibitem{Anber:2019nze}
M.~M. Anber and E.~Poppitz, \emph{{On the baryon-color-flavor (BCF) anomaly in
  vector-like theories}},
  \href{http://dx.doi.org/10.1007/JHEP11(2019)063}{\emph{JHEP} {\bf 11} (2019)
  063}, [\href{https://arxiv.org/abs/1909.09027}{{\tt 1909.09027}}].

\bibitem{Vafa:1983tf}
C.~Vafa and E.~Witten, \emph{{Restrictions on Symmetry Breaking in Vector-Like
  Gauge Theories}},
  \href{http://dx.doi.org/10.1016/0550-3213(84)90230-X}{\emph{Nucl. Phys. B}
  {\bf 234} (1984) 173--188}.

\bibitem{Bashmakov:2018ghn}
V.~Bashmakov, F.~Benini, S.~Benvenuti and M.~Bertolini, \emph{{Living on the
  walls of super-QCD}},
  \href{http://dx.doi.org/10.21468/SciPostPhys.6.4.044}{\emph{SciPost Phys.}
  {\bf 6} (2019) 044}, [\href{https://arxiv.org/abs/1812.04645}{{\tt
  1812.04645}}].

\bibitem{Poppitz:2009tw}
E.~Poppitz and M.~Unsal, \emph{{Conformality or confinement (II): One-flavor
  CFTs and mixed-representation QCD}},
  \href{http://dx.doi.org/10.1088/1126-6708/2009/12/011}{\emph{JHEP} {\bf 12}
  (2009) 011}, [\href{https://arxiv.org/abs/0910.1245}{{\tt 0910.1245}}].

\bibitem{Anber:2017pak}
M.~M. Anber and L.~Vincent-Genod, \emph{{Classification of compactified
  $su(N_c)$ gauge theories with fermions in all representations}},
  \href{http://dx.doi.org/10.1007/JHEP12(2017)028}{\emph{JHEP} {\bf 12} (2017)
  028}, [\href{https://arxiv.org/abs/1704.08277}{{\tt 1704.08277}}].

\bibitem{DeGrand:2016pgq}
T.~DeGrand, M.~Golterman, E.~T. Neil and Y.~Shamir, \emph{{One-loop Chiral
  Perturbation Theory with two fermion representations}},
  \href{http://dx.doi.org/10.1103/PhysRevD.94.025020}{\emph{Phys. Rev. D} {\bf
  94} (2016) 025020}, [\href{https://arxiv.org/abs/1605.07738}{{\tt
  1605.07738}}].

\bibitem{Ayyar:2017qdf}
V.~Ayyar, T.~DeGrand, M.~Golterman, D.~C. Hackett, W.~I. Jay, E.~T. Neil
  et~al., \emph{{Spectroscopy of SU(4) composite Higgs theory with two distinct
  fermion representations}},
  \href{http://dx.doi.org/10.1103/PhysRevD.97.074505}{\emph{Phys. Rev. D} {\bf
  97} (2018) 074505}, [\href{https://arxiv.org/abs/1710.00806}{{\tt
  1710.00806}}].

\bibitem{Cossu:2019hse}
G.~Cossu, L.~Del~Debbio, M.~Panero and D.~Preti, \emph{{Strong dynamics with
  matter in multiple representations: $\mathrm {SU}(4)$ gauge theory with
  fundamental and sextet fermions}},
  \href{http://dx.doi.org/10.1140/epjc/s10052-019-7137-1}{\emph{Eur. Phys. J.
  C} {\bf 79} (2019) 638}, [\href{https://arxiv.org/abs/1904.08885}{{\tt
  1904.08885}}].

\bibitem{Bergner:2020mwl}
G.~Bergner and S.~Piemonte, \emph{{Lattice simulations of a gauge theory with
  mixed adjoint-fundamental matter}},
  \href{http://dx.doi.org/10.1103/PhysRevD.103.014503}{\emph{Phys. Rev. D} {\bf
  103} (2021) 014503}, [\href{https://arxiv.org/abs/2008.02855}{{\tt
  2008.02855}}].

\bibitem{Shimizu:2017asf}
H.~Shimizu and K.~Yonekura, \emph{{Anomaly constraints on deconfinement and
  chiral phase transition}},
  \href{http://dx.doi.org/10.1103/PhysRevD.97.105011}{\emph{Phys. Rev. D} {\bf
  97} (2018) 105011}, [\href{https://arxiv.org/abs/1706.06104}{{\tt
  1706.06104}}].

\bibitem{Anber:2021iip}
M.~M. Anber, S.~Hong and M.~Son, \emph{{New anomalies, TQFTs, and confinement
  in bosonic chiral gauge theories}},
  \href{http://dx.doi.org/10.1007/JHEP02(2022)062}{\emph{JHEP} {\bf 02} (2022)
  062}, [\href{https://arxiv.org/abs/2109.03245}{{\tt 2109.03245}}].

\bibitem{Hsieh:2018ifc}
C.-T. Hsieh, \emph{{Discrete gauge anomalies revisited}},
  \href{https://arxiv.org/abs/1808.02881}{{\tt 1808.02881}}.

\bibitem{Delmastro:2022pfo}
D.~Delmastro, J.~Gomis, P.-S. Hsin and Z.~Komargodski, \emph{{Anomalies and
  Symmetry Fractionalization}},  \href{https://arxiv.org/abs/2206.15118}{{\tt
  2206.15118}}.

\bibitem{Davighi:2020uab}
J.~Davighi and N.~Lohitsiri, \emph{{The algebra of anomaly interplay}},
  \href{http://dx.doi.org/10.21468/SciPostPhys.10.3.074}{\emph{SciPost Phys.}
  {\bf 10} (2021) 074}, [\href{https://arxiv.org/abs/2011.10102}{{\tt
  2011.10102}}].

\bibitem{Cordova:2019bsd}
C.~C\'ordova and K.~Ohmori, \emph{{Anomaly Obstructions to Symmetry Preserving
  Gapped Phases}},  \href{https://arxiv.org/abs/1910.04962}{{\tt 1910.04962}}.

\bibitem{Cordova:2019jqi}
C.~C\'ordova and K.~Ohmori, \emph{{Anomaly Constraints on Gapped Phases with
  Discrete Chiral Symmetry}},
  \href{http://dx.doi.org/10.1103/PhysRevD.102.025011}{\emph{Phys. Rev. D} {\bf
  102} (2020) 025011}, [\href{https://arxiv.org/abs/1912.13069}{{\tt
  1912.13069}}].

\bibitem{Weingarten1983}
D.~Weingarten, \emph{Mass inequalities for quantum chromodynamics},
  \href{http://dx.doi.org/10.1103/PhysRevLett.51.1830}{\emph{Phys. Rev. Lett.}
  {\bf 51} (Nov, 1983) 1830--1833}.

\bibitem{Cherman:2018jir}
A.~Cherman, S.~Sen and L.~G. Yaffe, \emph{{Anyonic particle-vortex statistics
  and the nature of dense quark matter}},
  \href{http://dx.doi.org/10.1103/PhysRevD.100.034015}{\emph{Phys. Rev. D} {\bf
  100} (2019) 034015}, [\href{https://arxiv.org/abs/1808.04827}{{\tt
  1808.04827}}].

\bibitem{Lohitsiri:2022jyz}
N.~Lohitsiri and T.~Sulejmanpasic, \emph{{Comments on QCD$_{3}$ and anomalies
  with fundamental and adjoint matter}},
  \href{http://dx.doi.org/10.1007/JHEP10(2022)081}{\emph{JHEP} {\bf 10} (2022)
  081}, [\href{https://arxiv.org/abs/2205.07825}{{\tt 2205.07825}}].

\bibitem{Tachikawa:2016cha}
Y.~Tachikawa and K.~Yonekura, \emph{{On time-reversal anomaly of 2+1d
  topological phases}},
  \href{http://dx.doi.org/10.1093/ptep/ptx010}{\emph{PTEP} {\bf 2017} (2017)
  033B04}, [\href{https://arxiv.org/abs/1610.07010}{{\tt 1610.07010}}].

\bibitem{Closset:2012vp}
C.~Closset, T.~T. Dumitrescu, G.~Festuccia, Z.~Komargodski and N.~Seiberg,
  \emph{{Comments on Chern-Simons Contact Terms in Three Dimensions}},
  \href{http://dx.doi.org/10.1007/JHEP09(2012)091}{\emph{JHEP} {\bf 09} (2012)
  091}, [\href{https://arxiv.org/abs/1206.5218}{{\tt 1206.5218}}].

\bibitem{Acharya:2001dz}
B.~S. Acharya and C.~Vafa, \emph{{On domain walls of N=1 supersymmetric
  Yang-Mills in four-dimensions}},
  \href{https://arxiv.org/abs/hep-th/0103011}{{\tt hep-th/0103011}}.

\bibitem{Gomis:2017ixy}
J.~Gomis, Z.~Komargodski and N.~Seiberg, \emph{{Phases Of Adjoint QCD$_3$ And
  Dualities}},
  \href{http://dx.doi.org/10.21468/SciPostPhys.5.1.007}{\emph{SciPost Phys.}
  {\bf 5} (2018) 007}, [\href{https://arxiv.org/abs/1710.03258}{{\tt
  1710.03258}}].

\bibitem{Delmastro:2020dkz}
D.~Delmastro and J.~Gomis, \emph{{Domain walls in 4d$ \mathcal{N} $ = 1 SYM}},
  \href{http://dx.doi.org/10.1007/JHEP03(2021)259}{\emph{JHEP} {\bf 03} (2021)
  259}, [\href{https://arxiv.org/abs/2004.11395}{{\tt 2004.11395}}].

\bibitem{Cordova:2017vab}
C.~Cordova, P.-S. Hsin and N.~Seiberg, \emph{{Global Symmetries, Counterterms,
  and Duality in Chern-Simons Matter Theories with Orthogonal Gauge Groups}},
  \href{http://dx.doi.org/10.21468/SciPostPhys.4.4.021}{\emph{SciPost Phys.}
  {\bf 4} (2018) 021}, [\href{https://arxiv.org/abs/1711.10008}{{\tt
  1711.10008}}].

\bibitem{Hsin:2016blu}
P.-S. Hsin and N.~Seiberg, \emph{{Level/rank Duality and Chern-Simons-Matter
  Theories}}, \href{http://dx.doi.org/10.1007/JHEP09(2016)095}{\emph{JHEP} {\bf
  09} (2016) 095}, [\href{https://arxiv.org/abs/1607.07457}{{\tt 1607.07457}}].

\bibitem{Witten:2015aba}
E.~Witten, \emph{{Fermion Path Integrals And Topological Phases}},
  \href{http://dx.doi.org/10.1103/RevModPhys.88.035001}{\emph{Rev. Mod. Phys.}
  {\bf 88} (2016) 035001}, [\href{https://arxiv.org/abs/1508.04715}{{\tt
  1508.04715}}].

\bibitem{Hsieh:2015xaa}
C.-T. Hsieh, G.~Y. Cho and S.~Ryu, \emph{{Global anomalies on the surface of
  fermionic symmetry-protected topological phases in (3+1) dimensions}},
  \href{http://dx.doi.org/10.1103/PhysRevB.93.075135}{\emph{Phys. Rev. B} {\bf
  93} (2016) 075135}, [\href{https://arxiv.org/abs/1503.01411}{{\tt
  1503.01411}}].

\bibitem{Hason:2020yqf}
I.~Hason, Z.~Komargodski and R.~Thorngren, \emph{{Anomaly Matching in the
  Symmetry Broken Phase: Domain Walls, CPT, and the Smith Isomorphism}},
  \href{http://dx.doi.org/10.21468/SciPostPhys.8.4.062}{\emph{SciPost Phys.}
  {\bf 8} (2020) 062}, [\href{https://arxiv.org/abs/1910.14039}{{\tt
  1910.14039}}].

\bibitem{Tachikawa:2016nmo}
Y.~Tachikawa and K.~Yonekura, \emph{{More on time-reversal anomaly of 2+1d
  topological phases}},
  \href{http://dx.doi.org/10.1103/PhysRevLett.119.111603}{\emph{Phys. Rev.
  Lett.} {\bf 119} (2017) 111603},
  [\href{https://arxiv.org/abs/1611.01601}{{\tt 1611.01601}}].

\bibitem{DiVecchia:2017xpu}
P.~Di~Vecchia, G.~Rossi, G.~Veneziano and S.~Yankielowicz, \emph{{Spontaneous
  $CP$ breaking in QCD and the axion potential: an effective Lagrangian
  approach}}, \href{http://dx.doi.org/10.1007/JHEP12(2017)104}{\emph{JHEP} {\bf
  12} (2017) 104}, [\href{https://arxiv.org/abs/1709.00731}{{\tt 1709.00731}}].

\bibitem{Kitano:2020mfk}
R.~Kitano, N.~Yamada and M.~Yamazaki, \emph{{Is $N = 2$ Large?}},
  \href{http://dx.doi.org/10.1007/JHEP02(2021)073}{\emph{JHEP} {\bf 02} (2021)
  073}, [\href{https://arxiv.org/abs/2010.08810}{{\tt 2010.08810}}].

\bibitem{Witten:2000nv}
E.~Witten, \emph{{Supersymmetric index in four-dimensional gauge theories}},
  \href{http://dx.doi.org/10.4310/ATMP.2001.v5.n5.a1}{\emph{Adv. Theor. Math.
  Phys.} {\bf 5} (2002) 841--907},
  [\href{https://arxiv.org/abs/hep-th/0006010}{{\tt hep-th/0006010}}].

\bibitem{Browder:1962}
W.~BROWDER and E.~THOMAS, \emph{{AXIOMS FOR THE GENERALIZED PONTRYAGIN
  COHOMOLOGY OPERATIONS}},
  \href{http://dx.doi.org/10.1093/qmath/13.1.55}{\emph{The Quarterly Journal of
  Mathematics} {\bf 13} (01, 1962) 55--60},
  [\href{https://arxiv.org/abs/https://academic.oup.com/qjmath/article-pdf/13/1/55/7288473/13-1-55.pdf}{{\tt
  https://academic.oup.com/qjmath/article-pdf/13/1/55/7288473/13-1-55.pdf}}].

\bibitem{Aharony:2013hda}
O.~Aharony, N.~Seiberg and Y.~Tachikawa, \emph{{Reading between the lines of
  four-dimensional gauge theories}},
  \href{http://dx.doi.org/10.1007/JHEP08(2013)115}{\emph{JHEP} {\bf 08} (2013)
  115}, [\href{https://arxiv.org/abs/1305.0318}{{\tt 1305.0318}}].

\bibitem{Anber:2020gig}
M.~M. Anber and E.~Poppitz, \emph{{Generalized \textquoteright{}t Hooft
  anomalies on non-spin manifolds}},
  \href{http://dx.doi.org/10.1007/JHEP04(2020)097}{\emph{JHEP} {\bf 04} (2020)
  097}, [\href{https://arxiv.org/abs/2002.02037}{{\tt 2002.02037}}].

\bibitem{Kapustin:2014dxa}
A.~Kapustin, R.~Thorngren, A.~Turzillo and Z.~Wang, \emph{{Fermionic Symmetry
  Protected Topological Phases and Cobordisms}},
  \href{http://dx.doi.org/10.1007/JHEP12(2015)052}{\emph{JHEP} {\bf 12} (2015)
  052}, [\href{https://arxiv.org/abs/1406.7329}{{\tt 1406.7329}}].

\bibitem{Appelquist:1999hr}
T.~Appelquist, A.~G. Cohen and M.~Schmaltz, \emph{{A New constraint on strongly
  coupled gauge theories}},
  \href{http://dx.doi.org/10.1103/PhysRevD.60.045003}{\emph{Phys. Rev. D} {\bf
  60} (1999) 045003}, [\href{https://arxiv.org/abs/hep-th/9901109}{{\tt
  hep-th/9901109}}].

\bibitem{Dietrich:2006cm}
D.~D. Dietrich and F.~Sannino, \emph{{Conformal window of SU(N) gauge theories
  with fermions in higher dimensional representations}},
  \href{http://dx.doi.org/10.1103/PhysRevD.75.085018}{\emph{Phys. Rev. D} {\bf
  75} (2007) 085018}, [\href{https://arxiv.org/abs/hep-ph/0611341}{{\tt
  hep-ph/0611341}}].

\bibitem{Zoller:2016sgq}
M.~F. Zoller, \emph{{Four-loop QCD $\beta$-function with different fermion
  representations of the gauge group}},
  \href{http://dx.doi.org/10.1007/JHEP10(2016)118}{\emph{JHEP} {\bf 10} (2016)
  118}, [\href{https://arxiv.org/abs/1608.08982}{{\tt 1608.08982}}].

\end{thebibliography}\endgroup
\end{document}